\documentclass [11pt]{article}
\usepackage{amsmath,amsthm,amsfonts,amscd,eucal,latexsym,amssymb}
\def\sp{\hskip -5pt}
\def\spa{\hskip -3pt}
\def\sdelta{\not\spa\Delta}                                           

\def\past{{\,\prec\sp\prec\,}}

\newsymbol\bt 1202           %%% \boxtimes

\def\bC{{\mathbb C}}           %%%  complex numbers and so on

\def\bI{{\mathbb I}}

\def\bN{{\mathbb N}}

\def\bR{{\mathbb R}}

\def\b1{\bI}
\def\ine{\:\varepsilon\:}

\newsymbol\rest 1316         %%% restriction symbol

\def\gA{{\mathfrak A}}       %%% Ghotic

\def\gC{{\mathfrak C}}

\def\gF{{\mathfrak F}}
\def\gG{{\mathfrak G}}
\def\gH{{\mathfrak H}}

\def\gL{{\mathfrak L}}

\def\gP{{\mathfrak P}}

\begin{document}

%%%%%%%%%%%%%%  Preprint number %%%%%%%%%%%%%%%%%%%%%%%%%%

\hfill{\sl preprint - UTM 615}
\par
\bigskip
\par
\rm

%%%%%%%%%%%%%   Title %%%%%%%%%%%%%%%%%%%%%%%%%%

\par
\bigskip
\LARGE
\noindent
{\bf  Aspects of noncommutative Lorentzian geometry for globally hyperbolic spacetimes.}
\bigskip
\par
\rm
\normalsize

%%%%%%%%%%%%%%%%%%%%%%%%%%%%%%%%%%%%%%%%%%%%%

%%%%%%%%%%%% Author %%%%%%%%%%%%%%%%%%%%%%%%%%%

\large
\noindent {\bf Valter Moretti}

\large
\smallskip

\noindent
Department of Mathematics of the University of Trento,\\
I.N.d.A.M, Istituto Nazionale di Alta Matematica ``F.Severi'',  unit\`a locale  di Trento,\\ 
 I.N.F.N., Istituto Nazionale di Fisica Nucleare,  Gruppo Collegato di Trento,\\
via Sommarive 14,
I-38050 Povo (TN), Italy\\
E-mail: moretti@science.unitn.it\\

\large
\smallskip

\rm\normalsize

%%%%%%%%%%%%%%%%%%%%%%%%%%%%%%%%%%%%%%%%%%%

%%%%%%%%%%%% Date %%%%%%%%%%%%%%%%%%%%%%%%%%

\bigskip
\par
\hfill{\sl  Revised version, September  2003}
\par
\medskip
\par\rm

\noindent {\bf Abstract:}    Connes'  functional formula of the Riemannian distance is generalized to the Lorentzian case  using the so-called Lorentzian distance, the d'Alembert 
operator and the causal functions of a globally hyperbolic spacetime.
As a step of the presented machinery,  a proof of the almost-everywhere smoothness of the Lorentzian distance 
considered as a function of one of the two arguments is given.
Afterwards, using a $C^*$-algebra approach, the spacetime causal structure and the Lorentzian distance are generalized  into noncommutative structures giving rise to a Lorentzian version
of part of Connes' noncommutative geometry.   The generalized noncommutative 
spacetime consists of a direct set   of Hilbert spaces 
and a related class of $C^*$-algebras of operators. In each algebra a convex cone made of self-adjoint elements is selected
which generalizes the class of causal functions. The generalized events, called {\em loci}, are realized as the elements of the inductive
limit of the spaces of the algebraic states on the $C^*$-algebras. A partial-ordering relation between pairs of 
loci generalizes the causal order relation in spacetime. A generalized Lorentz distance of loci  is  
defined by means of a class of densely-defined operators which play the r\^ole of a  Lorentzian metric. 
Specializing back the formalism to the usual globally hyperbolic spacetime, it is found that compactly-supported
probability measures give rise to a non-pointwise extension of the concept of events. 

%%%%%%%%%%%%%%%%  Article text %%%%%%%%%%%%%%%%%%%%%%%

\section{Introduction.}

\noindent {\bf 1.1.}
{\em Some aspects of Connes' Riemannian non-commutative geometry.}
Connes' noncommutative geometry is a very impressive coherent  set of mathematical theories which encompass parts of mathematics
born by  very far and different contexts \cite{Connes}. On the physical ground, applications of Connes' noncommutative geometry
include general relativity, quantum field theory and many other research areas \cite{Connes,Landi,GBVF}.
As regards the content of this paper we are interested in the approach of chapter VI of \cite{Connes} (see also Chapter 6 of \cite{Landi}).
The basic ingredient introduced by Connes to develop the analogue of differential calculus for noncommutative algebras is
given by a so-called  {\em spectral triple}, $({\cal A},{\cal H},D)$. ${\cal A}$ is a unital algebra which is a
subalgebra of the natural $C^*$-algebra  of bounded operators on  a Hilbert space ${\cal H}$.
$D : {\cal D}(D) \to {\cal H}$ is a self-adjoint operator on ${\cal H}$, ${\cal D}(D)\subset {\cal H}$ being
a dense linear manifold, such that the resolvent $(D-\lambda I)^{-1}$ is compact for each $\lambda \not \in \bR$. 
$[D,a]$ must be  well defined at least as a quadratic form (see VI.1 in \cite{Connes}) and bounded for every $a\in {\cal A}$. \\
Every smooth compact $n$-dimensional Riemannian manifold $M$ equipped with a (Euclidean) spin structure
determines a natural {\em commutative} (i.e., ${\cal A}$ is commutative) spectral triple.
In that case ${\cal A}$ is the normed commutative unital involutive (the involution being the usual complex conjugation)
algebra  of  Lipschitz\footnote{I.e., for some $K_f\geq0$, it holds
 $|f(p)-f(q)|\leq K_f {\bf d}_E(p,q)$ for every  $p,q\in M$, ${\bf d}_E$ being the distance  in $M$.}
maps $f : M \to \bC$, the norm being  the usual sup-norm $||\cdot||_\infty$.
${\cal H}$  is the space $L^2(M,S)$ of the square integrable
sections of the irreducible $\bC^{2^{[\scriptsize \mbox{dim} D/2]}}$-spinor bundle over $M$ with measure $\mu_{\bf g}$ associated to the metric ${\bf g}$ on $M$.
The positive Hermitean  scalar product used to define $L^2$ reads
$$(\psi,\phi) := \int_M \psi^{\dagger}(x) \phi(x) \: d\mu_{\bf g}(x)\:.$$
This scalar product induces an operator norm which we denote by $||\cdot||_{{\bf L}(L^2(M,S))}$.
Finally $D$ is the Dirac operator associated with the Levi-Civita connection. It turns out that if $f\in {\cal A}$ is seen as a multiplicative operator,
$||f||_\infty= ||f||_{{\bf L}(L^2(M,S))}$, $f^* = \overline{f}$, $1=I$, where $1: M\to \bC$
is the constant map $1(x)=1$. Therefore ${\cal A}$   is a subalgebra  of the $C^*$-algebra of the bounded operators on $L^2(M,S)$ as it must be. \\
Remarkably, one can realize the topological and metric structure of the manifold in terms
of the spectral triple only (see propositions 6.5.1 and  10.1.1 in \cite{Landi}). Let us summarize this result.
In the following  $\overline{\cal A}$ denotes  the  (unital) $C^*$-algebra given by the completion of ${\cal A}$.
 $M$ is  homeomorphic to
the space of (the classes of unitary equivalence of) irreducible representations of
the $C^*$-algebra  $\overline{\cal A}$, equipped with
the topology of the pointwise convergence (also said Gel'fand's or  $*$-weak topology).   In the commutative case, the irreducible representations
are unidimensional and coincide with  the {\em pure algebraic states} on $\overline{\cal A}$. In this sense the points of $M$ are pure
algebraic states. All that  is essentially due \cite{Connes,Landi,GBVF} to the  well-known ``commutative Gel'fand-Naimark theorem''\cite{Simon}.
In practice, $\overline{\cal A}$ turns out to be  nothing but the $C^*$-algebra of the complex-valued
continuous functions on $M$, $C(M)$ with the norm $||\cdot ||_\infty$, and the pure state associated to any $p\in M$ trivially acts as
$p(f) := f(p)$ for every $f\in C(M)$.
As regards the metric,   one has the functional formula
\begin{eqnarray}
{\bf d}_E(x,y) = \sup \left\{\left|f(x)-f(y)\right| \:\: \left|\:\:\: f\in {\cal A},\:\: 
\left|\left|\left[D, f\right]\right|\right| \leq 1\right\}\right. \:, \label{d}
\end{eqnarray}
where ${\bf d}_E$ is the distance in the manifold which is induced by the metric.
Notice that there is no reference to paths in the manifold, despite the left-hand side is defined as the infimum of the length
of the paths from $p$ to $q$:
 \begin{eqnarray}
 {\bf d}_E(p,q) := \inf_{\Omega_{p,q}}\{ L(\gamma)\}\:, \label{ddef}
 \end{eqnarray}
 where $\Omega_{p,q}$ is the class of all continuous piecewise-smooth curves jointing  $p$ and $q$ and $L(\gamma)$
 is the Riemannian  length of $\gamma\in \Omega_{p,q}$.
 As remarked by Connes \cite{Connes}, this fact is interesting on a pure physical ground. Indeed the path of quantum
particles do not exist: wave functions exists but  one must assume the existence of geometrical structures also discussing quantum particles.
There is an analogous formula for the
integration of functions $f\in {\cal A}$ over $M$ based on the {\em Dixmier trace} $\mbox{tr}_\omega$
(below $c(n)$ is a coefficient depending on the dimension $n$ of the manifold $M$ only)  \cite{Connes,Landi,GBVF},
\begin{eqnarray}
\int_M f(x)\: d\mu_{\bf g}(x) = c(n)\:\: \mbox{tr}_\omega \left(f|D|^{-n}\right) \label{tr}\:.
\end{eqnarray} 
Whenever the algebra ${\cal A}$ of a spectral triple is taken noncommutative  (\ref{d}) can be re-interpreted as defining  a distance
in the space of pure states \cite{Connes,Landi,GBVF} and generalized interpretations are possible for (\ref{tr}).
Similar noncommutative generalizations can be performed concerning much of differential
and integral calculus finding out very interesting and useful mathematical structures giving rise to a remarkable  interplay between mathematics and
 theoretical physics \cite{Connes,Landi,GBVF}.
It is worth noticing that, for most applications, the Dirac operator $D$ can be replaced by the Laplace-Beltrami one $\Delta$ as suggested in
\cite{FG,FGR} (see also \cite{Landi}) and this is the way we follow within the present work.

 Most physicists interested in quantum  gravity believe that the Planck-scale geometry may reveal a  structure very different from the geometry at macroscopic scales.
This is a strong motivation for developing further any sort of noncommutative geometry. However, physics deals with {\em Lorentzian} spacetimes 
rather that {\em Euclidean}\footnote{We use "Euclidean" as synonym of "Riemanniann" throughout.} spaces. To this end, the principal aim of this paper is the attempt to find the Lorentzian analogue of (\ref{d}).
Actually, we shall see that this is nothing but the first step in order to develop  a
noncommutative approach of  the spacetimes causality.\\

\noindent {\bf 1.2.}
{\em The Lorentzian  puzzle.}  
The Lorentzian geometry, i.e. the geometry of  spacetimes, is more complicated than the Euclidean
one due to the presence of, local and global,
{\em causal structures}. These take  temporal and causal relations among  events into account.    The local, metrical and causal, structure
is given by the Lorentzian metric. A physically relevant global causal structure is involved in the definition of  a {\em globally hyperbolic}
spacetime.  Roughly speaking, a globally hyperbolic spacetime is a time-oriented Lorentzian manifold (that is a spacetime)
which admits spacelike surfaces, called  Cauchy surfaces,
such that the assignment of Cauchy data on those surfaces determines the evolution 
of any field everywhere in the manifold if  the field satisfies, for instance, Klein-Gordon 
equation.   A globally hyperbolic spacetime seems to be the natural {\em scenario}
where one  represents the theory on the matter content of the universe, including
(quantum)  fields, elementary interactions and all that \cite{Wald84,Wald94}.
In order to built up a Lorentzian noncommutative geometry, a generalization of the (local and global) {\em causal structure}
of a spacetime is necessary.  To make contact with Connes' program a natural question arises: What is the Lorentzian analogue of ${\bf d}_E$ to be used 
to generalize (\ref{d}) in Lorentzian manifolds?
An interesting object defined in either Euclidean 
and Lorentzian manifolds is  the so-called {\em Synge world function} $\sigma$ (see the Appendix A) which is related with the function ${\bf d}_E$ in Euclidean 
manifolds:
Any smooth, either Riemannian or Lorentzian, manifold
is locally endowed with a smooth  function $\sigma : N\times N \to \bR$
where $N$ is any convex normal neighborhood.
$\sigma$ maps $x,y\in N$ into one half the (signed) squared length
of the unique geodesic segment, which joints $x$ and $y$, contained in  $N$.
In Riemannian manifolds $\sigma \geq 0$. In Lorentzian manifolds,
the sign is positive if and only if  $x,y$ are spatially separated, negative if and only if $x,y$
are time-like related, and $\sigma(x,y)=0$ for  either $x=y$ or when $x,y$ are null related. It is known
that $\sigma$   completely determines the metric at each point of the spacetime. 
In Euclidean manifolds  ${\bf d}_E=\sqrt{2\sigma}$ holds
 whenever $x,y$ belong to a common  convex normal neighborhood, so, at least locally,
	 it is possible to define ${\bf d}_E$ in terms of $\sqrt{2\sigma}$.     However, any attempt to generalize (\ref{d}) in Lorentzian manifolds  by means of any analogue of 
   ${\bf d}_E$ built up by means of $\sigma$ faces the basic issue of the
indefiniteness of the Lorentzian world function.   ${\bf d} := \sqrt{2\sigma}$ would be complex-valued and so useless to restore some identity similar to (\ref{d}).
  One could try to define ${\bf d}$  for spatially separated events  only
 by taking the squared root of $2\sigma$ in that case.
An immediate  drawback is that  the definition would not  work whenever $x$ and $y$ are too far from each other since $\sigma$ is not well defined
outside convex normal neighborhoods.
To avoiding the problem, one may try to  use  (\ref{ddef}) for $x,y$ spatially separated with
$\Omega_{x,y}$ now denoting the class of {\em space-like}  continuous piecewise  smooth curves jointing  $x$ and $y$.
This is not a nice idea too, because it would entail ${\bf d}(x,y)= 0$ (and thus also ${\bf d}(x,y) \neq \sqrt{2 \sigma(x,y)}$)
at least for $x$ and $y$ sufficiently close to each other and spatially separated. This is  because, in convex normal neighborhoods,  one may arbitrarily
approximate null piece-wise smooth curves by means of piecewise smooth space-like curves  with the same
endpoints.

Actually several other problematic issues are related to the indefiniteness of ${\bf d}^2$. For instance,
if $D$ indicates the Dirac operator, the identity $$||[D, f]|| = {\mbox{ess}\sup}_M |{\bf g}(d f,d f)|\:,$$ necessary to  give rise
to (\ref{d}) (e.g.,   see
\cite{Connes,Landi,GBVF}),
fails to be fulfilled.  This is because the left-hand side is not well-defined as a Hilbert-space operator norm since, in Minkowski spacetime
(but this generalizes to any Lorentzian manifold equipped with a spin structure),
the natural  Lorentz invariant
scalar product of spinors turns out to be indefinite.
We do not address to these issues in the present work because we shall employ the Laplace-Beltrami-D'Alembert operator instead of
the Dirac one (see \cite{Strohmaier01} for another approach based on the Dirac operator and Krein spaces).

Another problematic technical issue related to the indefiniteness of the metric  is the failure of the Lipschitz condition  to define a valuable
background algebra of functions ${\cal A}$. Indeed,  in the Euclidean case ${\bf d}_E(p,\cdot)$
 cannot  be everywhere smooth but
it turns out to be Lipschitz because of the triangular inequality (false in the Lorentzian case).
The Lipschitz condition  plays a relevant r\^ole in proving (\ref{d}) and in the choice of
the algebra ${\cal A}$ which contains ${\bf d}_E(p,\cdot)$.
We remark  that also the compactness of the manifold has to be dropped in the Lorentzian case because a compact spacetime
contains a closed timelike curve (proposition 3.10 in \cite{BEE}) and thus fails to be physical.
The failure of the compactness  gives rise to  problems in the Euclidean case.
However approaches to noncommutative Euclidean  geometry exist in some cases \cite{Gayral}.
If $M$ is a Hausdorff locally compact space but it is not compact,
there is a homeomorphism from $M$ onto the space of complex homomorphism of the nonunital $C^*$-algebra of the complex functions on $M$
which vanish at infinity, $C_0(M)$, equipped with the pointwise-convergence topology \cite{FE-DO}.  So the points of $M$ can be thought
as multiplicative functionals on the $C^*$-algebra $\overline{\cal A} := C_0(M)$, and ${\cal A}$ can be taken as the algebra
of complex continuous compactly-supported functions in $M$, $C_c(M)$. However, in the noncompact case  (\ref{d}) cannot be re-stated as it stands.

In the Lorentzian case,  possible attempts to  solve all these problems (also connected with Hamiltonian formulation of field theories
including the gravitational field)
 \cite{Kopf1,Kopf2,Kalau,Hawkins} are based on the
foliation of the manifold by means of space-like hypersurfaces. On these hypersurfaces, provided they are compact (and endowed with
spin structures), one can restore Connes' standard non-commutative
approach referring to the Euclidean  distance induced by the Lorentzian background metric.   However, barring globally {\em static} spacetimes,
any choice of the foliation is quite arbitrary.
 Moreover the relation between
spatial spectral triples and  causality seems to be  quite involved. Finally, a classical background spacetime cannot completely
eliminated through this way reducing possible attempts to formulate approaches to quantum gravity.  Another  approach to
 noncommutative Lorentzian geometry is presented in \cite{Strohmaier01}  in terms of Krein spaces. However the issue of the generalization
 of  (\ref{d}) is not investigated, but attention is focused on the generalization of (\ref{tr}) and the noncommutative differential calculus.\\

\noindent {\bf 1.3}. {\em A natural Lorentzian approach.}
 In this paper, first of all we show that  there  is a possible  generalization of (\ref{d}) in any physically well-behaved spacetime (see 1.5 for more details on the used definitions).  In fact, in every {\em globally hyperbolic}
  spacetime $M$ (i.e., a connected time oriented Lorentzian manifold which admits {\em Cauchy surfaces})
a functional identity similar to (\ref{d}) arises
 which uses the so-called
 {\em Lorentzian distance} ${\bf d}(x,y)$ \cite{BEE,O'Neill}, the class of almost-everywhere  smooth {\em causal functions}  and
the Laplace-Beltrami-d'Alembert operator, locally  $\Delta = \nabla^\mu \nabla_\mu$, associated to the Levi-Civita connection derivative
$\nabla$. 
(Actually the same result holds working  with a vector fiber bundle $\gF \to M$ and  more complicate second-order hyperbolic operators,
see remark 1 after Theorem 3.1 below).
The original idea to express the Lorentzian distance by a functional formula using the metric Laplacian was formulated  by Parfinov and Zapatrin in  \cite{PaZa} where
part of the approach developed in the first part of  this work was presented into a more elementary form
without  the requirement of global hyperbolicity.\\
Let us illustrate the ingredients pointed out above. Take $p,q\in M$. First suppose  that $p\neq q$ and  $p\preceq q$ which means
that $q$ belongs to  the {\em causal future} of $p$
(i.e., the subset of $M$ of the events $r$ such that there is a causal future-directed curve 
from $p$ to $r$).
In that case, the {\em Lorentzian distance} from  $p$ to $q$  is defined as
 ${\bf d}(p,q) := \sup\{L(\gamma) \:|\: \gamma\in \Omega_{p,q}\}\:,$
 $\Omega_{p,q}$ denoting  the set of all  causal future-directed curves  from $p$ to $q$ and $L(\gamma) \geq 0$ is the length of
 $\gamma$. ${\bf d}(p,q) := 0$ if either $p=q$ or  $p\not \preceq q$.
  ${\bf d}$ enjoys an inverse triangular inequality if $p\preceq q \preceq r$:
${\bf d}(p,r)\geq {\bf d}(p,q)+{\bf d}(q,r)$.
${\bf d}$ is a natural object in {\em time oriented} Lorentzian manifolds, i.e., spacetimes,
and it turns out to be  continuous in {\em globally hyperbolic} spacetimes. ${\bf d}$  plays a crucial r\^ole in Lorentzian geometry \cite{BEE,O'Neill}
 because one can re-built  the  topology,  the differential structure, the metric tensor and the
time orientation of the spacetime by using ${\bf d}$ only, as we shall see shortly.
If $N\subset M$ is open,  a {\em causal functions} on ${N}$
 is a continuous map  $f:{N} \to \bR$ which  does not  decrease along every causal future-directed curve contained in ${N}$.
${\cal C}_{[\mu_{\bf g}]}({N})$
 denotes the class of causal functions on $N\subset M$
which are smooth almost everywhere in $N$.
${\cal X}$ denotes the class of all regions $I$ in the spacetime $M$
which are open, causally convex (i.e., if $p,q$ belong to such a region, also every future-directed causal curve
from $p$ to $q$ lies in the region) and such that $\overline{I}$ is compact, causally convex and $\partial I$ has measure zero.\\
The Lorentzian equation which corresponds to (\ref{d}) reads, in a globally hyperbolic spacetime $M$, for $p,q\in M$
with $q$ in the causal future of $p$,
\begin{eqnarray}
{\bf d}(p,q) =
 \inf\left\{ \langle f(q)-f(p) \rangle \:\:\left|\:\:  f\in {{\cal C}_{[\mu_{\bf g}]}(\overline{I})}, \:\: I \in  {\cal X},\:\: p,q\in \overline{I},
 \:\: \left|\left| \left[f,\left[f,\sdelta\right]\right]^{-1}
 \right|\right|_{I} \leq  1  \right.\right\}\:,
 \label{dl}
 \end{eqnarray}
 where $\langle \alpha\rangle := \max\{0,\alpha\}$ for $\alpha\in \bR$,
$2\spa$ ${\sdelta} :=  \Delta$, the latter being the Laplace-Beltrami-d'Alembert operator.
 $||\cdot||_{S}$ denotes the uniform norm of operators $A : L^2(S, \mu_{\bf g}) \to L^2(S, \mu_{\bf g})$ where
$\mu_{\bf g}$ being the  measure on $S\subset M$ naturally induced by the  metric
${\bf g}$ of the spacetime. The restriction to a suitable class of  compact sets $\overline{I}$ is useful
to realize the events of the spacetime as pure states of unital $C^*$-algebras of functions containing
the causal functions. It holds despite the manifold is not compact and these functions, in general,
 are not bounded on the whole manifold.\\
Afterwards we analyze, from the point of view of the $C^*$-algebras, the ingredients
above showing that noncommutative generalizations are  possible. In particular we  introduce,  in
suitable algebraic context, the generalizations of the causal ordering relation
$\preceq$  and of the Lorentzian distance. Specializing back to the commutative case, these generalizations give
rise to a non-pointwise concept of event (compactly-supported probability measures on globally hyperbolic spacetimes)
preserving the notion of causal ordering relation and Lorentzian distance.\\

\noindent {\bf 1.4}.
{\em Structure of the work}.
This paper is organized as follows. The remaining part of Section 1 
contains basic definitions, notations and conventions used 
throughout the paper.
In Section 2 we introduce the Lorentzian distance and the {\em causal functions} on a spacetime. More precisely  (a) we present the basic properties
of ${\bf d}$, (b) we show that it completely determines
the structure of the spacetime and (c) we prove some
propositions  necessary to generalize (\ref{d}).
 In particular we prove a theorem concerning the almost-everywhere smoothness of 
${\bf d}$ in globally hyperbolic spacetimes. Section 3  is devoted to prove (\ref{dl}).
Section 4 contains an algebraic analysis of the introduced mathematical structures and several
generalizations. In particular (a) we introduce the concept of {\em locus} which generalizes the concept of event
(or point in noncompact Euclidean  manifolds) and we prove that (b) loci reduce to compactly supported
(regular Borel) probability measures in the commutative case.
Finally we show (c) that $\preceq$ and ${\bf d}$ can be extended into analogous mathematical objects related to the space of the loci
which give rise to a noncommutative causality.\\

\noindent {\bf 1.5}.
{\em Basic definitions, notations and conventions.}
Throughout the work  ``iff'' means  ``if and only if'' and ``smooth'' means $C^\infty$.  Concerning differentiable manifolds we assume usual definitions. 
More precisely,   a ($n$-dimensional) differentiable manifold  $M$ is a connected,   Hausdorff, second countable topological space which is locally homeomorphic
to $\bR^n$ and is  equipped with  a $C^\infty$-differentiable structure.
Concerning differentiable functions in nonopen sets we give the following definition.
If  $M$ is a differentiable manifold, $U\subset M$ is  open and nonempty, and $V\subset \partial U$,
$C^\infty(U\cup V)$
 denotes the set of functions $f : U\cup V \to \bR$ such that $f\sp \rest_U \in C^\infty(U)$ and, for every $y\in V$,
each derivative of any order, computed in a coordinate patch in some open neighborhood $U_y$ of $y$,
can be extended into a continuous function in $U_y\cap (U\cup V)$.
We assume that the reader knows basic definitions and properties of manifolds 
equipped with (Lorentzian or Riemannian) metrics, Levi-Civita connection and geodesical flux.  A.1 in Appendix A contains   definitions and properties of
the exponential map and the related mathematical machinery  (convex normal neighborhoods).  \\
 We address the reader
to \cite{O'Neill,BEE,Wald84,Penrose72,HE} as general reference textbooks on spacetime structures.
Let us summarize  basic definitions, further definitions used in the paper 
will be given before relevant statements in the text.  Appendix A contains a complete summary.  \\
 A (smooth) {\bf Lorentzian manifold} $(M,{\bf g})$ is a  $n\geq 2$-dimensional
smooth manifold $M$ with a smooth Lorentzian metric ${\bf g}$ (with signature $(-,+,\cdots,+$)). 
  We use the following terminology concerning the classification of vectors and co-vectors. A vector
$T\in T_xM$, $T\neq 0$, is said to be {\bf space-like}, {\bf time-like} or {\bf null}
if, respectively, ${\bf g}_x(T,T)>0$,  ${\bf g}_x(T,T)<0$, ${\bf g}_x(T,T)=0$.   $T$ is said
to be  {\bf causal} if it is either time-like or null.
The same terminology is used for
co-vectors $\omega \in T^*_xM$ referring to $\uparrow\spa \omega \in T_xM$, where ${\bf g}_x(\uparrow\spa \omega, \:\cdot\:) = \omega$.\\
 We remind the reader that 
a  Lorentzian manifold $(M,{\bf g})$ is said to be  {\bf time orientable} if it admits a smooth non vanishing vector field $Z\in TM$ which is
everywhere time-like. Afterwards a {\bf time orientation}, ${\cal O}_t$, is the choice of one of the two
equivalence classes of smooth time-like vector fields $Z$ with respect to the equivalence relation
$Z\sim Z'$ iff ${\bf g}(Z,Z') <0$ everywhere.
For each point $p\in M$, an orientation determines an analogous equivalence  class of time-like vectors of $T_pM$, ${\cal O}_{tp}$.   With the given definitions, a causal vector (co-vector )
 $T\in T_pM$ ($\omega\in T^*_pM$) is said to be {\bf future directed} if
${\bf g}_{p}(Z(p),X) <0$ (${\bf g}_{p}(Z(p),\uparrow\spa\omega) <0$) and  {\bf past directed} if 
${\bf g}_{p}(Z(p),X) >0$ (${\bf g}_{p}(Z(p),\uparrow\spa\omega) >0$).\\
A {\bf spacetime} $(M, {\bf g}, {\cal O}_t)$ is a  Lorentzian manifold $(M, {\bf g})$ which is time orientable and  equipped with
 a time orientation ${\cal O}_t$; the points of $M$  are also called {\bf events}. \\
To conclude we give the definition of causal curves. In  spacetime $M$,  a piecewise $C^1$ curve 
 (see A.5 for the detailed definition of  {\bf piecewise $C^k$}  {\bf curve}
 used in this work)
$\gamma$ is said to be 
 {\bf time-like}, {\bf space-like}, {\bf null}, {\bf causal} if its tangent vector  $\dot\gamma$  is
respectively  time-like, space-like, null, causal. Moreover, the curve  
is said to be  {\bf  future} ({\bf past}) {\bf directed} 
if its tangent vector  $\dot\gamma$  is {\bf  future} ({\bf past}) {\bf directed}.

\section{Lorentzian distance and causal functions.}

In this section we collect and review important notions and results in Lorentzian geometry, in particular  focusing on the r\^ole of  Lorentzian distance. Part of these results are well known 
but spread in the literature. A relevant result proven in  Section 2.1 concerns the 
almost-everywhere smoothness  of the Lorentzian distance (Theorem 2.1). In Section 2.2 is investigated the 
interplay of the Lorentzian distance and the notion of causal function in a spacetime. Finally, a preliminary
formulation of the functional formula of the  Lorentzian distance is presented (Theorem 2.2)
using the built up machinery. \\

\noindent {\bf 2.1.}  {\em The r\^ole  and the properties of the  Lorentzian distance in spacetimes.}
To define the Lorentzian distance it is necessary to recall the notion of  {\bf Lorentzian length} of a (causal) curve. As is well known,  the  Lorentzian length   $L(\gamma)$ of  a piecewise $C^1$ curve $\gamma: [a,b] \to M$    is  
  \begin{eqnarray}
 L(\gamma):= \int_a^b \sqrt{|{\bf g}_{\gamma(t)}(\dot{\gamma}(t),\dot{\gamma}(t))|} \:dt \:. \label{length}
 \end{eqnarray}
Obviously the definition does not depend on the used parametrization. \\
It is convenient to extend the definition of Lorentzian length to  {\em continuous causal curves} 
because several definitions and results of Lorentzian geometry  found in the literature require the use of  continuous causal curves.
A continuous curve  $\gamma : I \to M$ is said to be a 
{\bf  continuous future-directed causal curve}  (see A.6) if the following requirement is fulfilled.  For each $t\in I$, there is a neighborhood of $t$, $I_t$ and
a  convex normal neighborhood of $\gamma(t)$, $U_t$, such that  the following requirements are fulfilled. For $t'\in I_t\setminus \{t\}$, one has $\gamma(t')\neq \gamma(t)$ and
(a) there is a  future-directed causal (smooth) geodesic segment  $\gamma' \subset U_t$  from $\gamma(t)$ to $\gamma(t')$
if $t'>t$ or (b) 
there is a  future-directed causal (smooth) geodesic segment $\gamma' \subset U_t$  from $\gamma(t')$ to $\gamma(t)$
if $t'<t$.       Similar definitions hold concerning  {\bf continuous future-directed timelike curves}, by replacing
``causal'' with ``timelike'' in the definitions above.
The definition of
$L(\gamma)$  is extended as follows (\cite{HE} p.214) to continuous future-directed causal curves
$\gamma$.  Suppose that $\gamma$, from $p$ to $q$, is such that, for every open neighborhood $U_\gamma$ of $\gamma$,
 there is a  future-directed timelike piecewise $C^1$ curve $\gamma'$ from $p$ to $q$, then define $L_{U_\gamma}(\gamma):= \sup{L(\gamma')}$
 varying $\gamma'$ in $U_\gamma$ as said. Then $L(\gamma) := \inf L_{U_{\gamma}}(\gamma)$ where $U_\gamma$ varies in the class
 of all open neighborhoods of $\gamma$.  If $\gamma$ does not fulfill the initial requirement then $\gamma$ must be an unbroken null geodesic
 (see \cite{HE} p.215) and thus one defines
 $L(\gamma):=0$\footnote{A maybe equivalent definition can
 be given noticing that a continuous future directed causal curve $\gamma$ satisfies a local Lipschitz condition
(with respect to the coordinates of a sufficiently small neighborhood of each point of $\gamma$) and thus it is almost everywhere differentiable.
 So,  one defines $L(\gamma)$ using (\ref{length}) too  (see \cite{BEE} p. 136).}.\\
 
 \noindent {\bf Remark}. From now on a,
 either future-directed or past-directed,  causal curve is supposed to be a continuous, respectively  future-directed or past-directed, causal curve. Moreover  continuous curves $\gamma :I\to M$
and $\gamma' : I' \to M$ are identified if there is an increasing  homomorphism $h : I\to I'$
and $\gamma'\circ h =\gamma$.\\
 
\noindent Let us give the definition of Lorentzian distance.
We remind the reader that,  in a spacetime $(M, {\bf g}, {\cal O}_t)$, if $p,q\in M$,  $p\preceq q$ means that either $p=q$ or
 there is a  future-directed causal  curve from $p$ to $q$,  whereas $p\prec q$   means that $p\preceq q$
 and $p\neq q$,  and finally
 $p\past q$ means that there is a  future-directed time-like curve from $p$ to $q$. 
($\past$ and $\preceq$ are clearly transitive relations moreover,  if $p,q,r\in M$,
$p\past q$ and $q\preceq r$ entail $p\past r$,  similarly $p\preceq q$ and $q\past r$ entail
$p\past r$  \cite{Penrose72}.)  \\

 \noindent {\bf Definition 2.1}.
 {\em Let $(M,{\bf g}, {\cal O}_t)$ be a spacetime. If $p,q \in M$ and
$\Omega_{p,q}$ denotes the class of the  future-directed causal
  curves from $p$ to $q$, the {\bf Lorentzian distance from $p$ to $q$}, ${\bf d}(p,q) \in [0,+\infty)\cup \{+\infty\}$
 is \cite{O'Neill,BEE}
 \begin{eqnarray}
 {\bf d}(p,q) := \left\{
\begin{array}{cl}
\sup\{L(\gamma)\:|\: \gamma\in \Omega_{p,q}\} & \mbox{ if $p\prec q$, }\\
0   & \mbox{ if $p\not \prec q$. }  \label{delta}
\end{array}
\right.
 \end{eqnarray}}\\

\noindent {\bf Remarks} {\bf (1)} By the given definition of $L(\gamma)$,  ${\bf d}(p,q)=\sup\{L(\gamma)\}$ attains the same value if
one restricts the range of $\gamma$ to the piecewise $C^1$ curves of $\Omega_{p,q}$.\\
{\bf (2)} Differently from the Euclidean case, in general $\Omega_{p,q} \neq \Omega_{q,p}$, and thus ${\bf d}(p,q) \neq {\bf d}(q,p)$.\\

 \noindent The Lorenz distance enjoys several relevant properties which
 will be useful later.  Proposition 2.1 below presents the elementary properties 
 of the Lorentzian distance in relation with the {\em causal sets} of a spacetime.
  From now on we use the following definitions of causal sets in a spacetime $(M, {\bf g}, {\cal O}_t)$.  The topological and causal  properties of these  sets which are employed in the work  are presented in  A.9, A.11, A.12 and A.23. If $S\subset M$,
  \begin{align}
J^{+}(S) := &\{ q\in M\:|\: p \preceq q \mbox{ for some $p\in S$}\}   \mbox{  is  the {\bf causal future} of $S$ } \nonumber \:,   \\
J^{+}(S) := &\{ q\in M\:|\: q \preceq p \mbox{ for some $p\in S$}\}   \mbox{  is  the {\bf causal past} of $S$ }\nonumber\:,\\
I^{+}(S)  :=  &\{ q\in M\:|\: p \past q \mbox{ for some $p\in S$}\}  \mbox{  is  the {\bf chronological future} of $S$ } \nonumber\:,   \\
I^{-}(S)  :=  &\{ q\in M\:|\: q \past p \mbox{ for some $p\in S$}\}   \mbox{  is  the {\bf chronological past} of $S$ }  \nonumber \:.
\end{align}
Moreover $I(p,q) := I^+(p) \cap I^-(q)$  and   $J(p,q) := J^+(p) \cap J^-(q)$.
$p,q\in M$ are said to be  {\bf time related}, if either $I^+(p) \cap I^-(q)\neq \emptyset $ or  $I^-(p) \cap I^+(q) \neq \emptyset$,
 {\bf causally related} if either $J^+(p) \cap J^-(q) \neq \emptyset$ or $J^-(p) \cap J^+(q)\neq \emptyset$.
  Causally related events $p,q\in M$, $p\neq q$, which are not time related are called {\bf null related}.
$S,S' \subset M$ are said to be {\bf spatially separated} if  $(J^+(S) \cup J^-(S)) \cap S' =\emptyset$
(which is equivalent to $(J^+(S') \cup J^-(S')) \cap S =\emptyset$).\\
We remind the reader that 
 a set $S$ of a spacetime $M$ is  {\bf causally convex}  when
 $J(p,q) \subset S$ if $p,q\in S$ (see A.11 and  for properties of causally convex sets and strongly causal spacetimes).  A spacetime
is {\bf strongly  causal} when every event admits a fundamental set of open neighborhoods
consisting of {causally convex} sets.
 A spacetime is called  {\bf chronological} if  there are no events $p,q$ such that $p\past q\past p$ (equivalently, it  does not contain any closed future-directed timelike curve).
Finally, a {\bf globally hyperbolic} spacetime (see also A.16-A.23 and  the end of 8.3 in \cite{Wald84} about possible equivalent definitions)
is a strongly-causal spacetime $(M, {\bf g}, {\cal O}_t)$ such that every  $J(p,q)$ is either empty or compact  for each pair $p,q\in M$ (see A.12-A.15 for further definitions and properties,  here we remind the reader only that a globally hyperbolic spacetime is both strongly causal and chronological).\\

\noindent  {\bf Proposition 2.1.}
 {\em If $(M,{\bf g}, {\cal O}_t)$ is a spacetime and $p,q,r\in M$:\\
 {\bf (a)} $I^+(p) = \{ q\in M \:|\: {\bf d}(p,q)>0\}$. Moreover, if $p\neq q$ and both ${\bf d}(p,q)$ and ${\bf d}(q,p)$ are
  finite, then either ${\bf d}(p,q)=0$ or ${\bf d}(q,p)=0$;\\
 {\bf (b)} if $p \preceq q \preceq r$,  the {\bf inverse triangular inequality} holds, that is
 \begin{eqnarray}
 {\bf d}(p,r) \geq {\bf d}(p,q) + {\bf d}(q,r) \label{antitriangle}\:;
 \end{eqnarray}
 {\bf (c)} ${\bf d}$ is lower semicontinuous on $M\times M$;\\
 {\bf (d)} if $U_p$ is a, sufficiently small, convex normal neighborhood of $p$, ${\bf d}(p, \cdot)\sp\rest_{U_p \cap J^+(p)}$
 is finite, belongs to the class $C^\infty(U_p \cap J^+(p))$ and, for all $q \in U_p \cap J^+(p)$,
  \begin{eqnarray}
 \sigma(p,q) = -\frac{1}{2}{\bf d}(p,q)^2 \label{delta-sigma}\:;
 \end{eqnarray}
 where $\sigma(p,q)$ is one half the squared geodesic distance from $p$ to $q$, also called Synge's world function,
 defined by using the exponential map (see A.0 in Appendix A).\\
 {\bf (e)} if $p \preceq q$ and there is a curve $\gamma \in \Omega_{p,q}$ with
$L(\gamma) = {\bf d}(p,q)$ (i.e., $\gamma$ is {\bf maximal}),
then $\gamma$ can be re-parametrized to be a smooth geodesic.

 If  $(M,{\bf g}, {\cal O}_t)$ is globally hyperbolic it also holds:  \\
 {\bf (f)} $J^+(p) = \overline{\{q\in M\:|\: {\bf d}(p,q)>0\}}$;\\
 {\bf (g)} ${\bf d}$ is finite; \\
 {\bf (h)}   ${\bf d}$ is continuous on $M\times M$; \\
{\bf (i)} if $p \prec q$, there is a causal geodesic from $p$ to $q$, $\gamma$ with $L(\gamma) = {\bf d}(p,q)$.}\\

\noindent {\em Proof.} Items {\bf (a)},{\bf (c)},{\bf (e)},{\bf (g)},{\bf (h)} are proven in Section 4.1 of \cite{BEE}.
{\bf (b)} is a trivial consequence of the definition of ${\bf d}$.
Concerning {\bf (d)},   everything is a consequence
of  the smoothness of $\sigma$ and of (\ref{delta-sigma}).  The latter can be proven noticing that the length from $p$
of causal geodesic segments through $p$, in a convex normal neighborhood  is maximal
(proposition 4.5.3 in Section 4.5 of \cite{HE}) and using theorem 4.27 in \cite{BEE}. {\bf (f)} is a consequence of (a) and A.12.
The proof of  {\bf (i)} can be found in \cite{O'Neill} p.411. $\Box$\\

 \noindent A very remarkable result of Lorentzian geometry is that the Lorentzian distance determines the whole, local and global (topological, differential, metric), structure of a spacetime 
 as  summarized in Proposition 2.2. \\

 \noindent {\bf Proposition 2.2.}
{\em  Let  $(M,{\bf g}, {\cal O}_t)$  be a spacetime with Lorentzian distance ${\bf d}$ and  $n:=\mbox{\em dim} M$.\\
{\bf (a)} If $M$ is strongly causal (in particular if $M$ is globally hyperbolic), its topology is generated by
the sets $\{x\in M\:|\: {\bf d}(p,x) \cdot {\bf d}(x,q)>0\}$ for all pairs $p,q\in M$ with $p\past q$
(we assume that $0\cdot \infty=\infty \cdot 0 =0$). \\
{\bf (b)} There is an atlas of  $M$, $\{(U_p,\varphi_p)\}_{p\in M}$, $U_p$ being an open neighborhood of $p$
with coordinate maps given by  $\varphi_p :q\mapsto ({\bf d}(p_1,q),\ldots, {\bf d}(p_n,q))\in \bR^n$,
$p_1,p_2,\ldots,p_{n}$ being suitable events about $p$.\\
{\bf (c)} For every pair of smooth vector fields $X,Y$ and every event $p\in M$ it holds
 \begin{eqnarray}
{\bf g}_p(X_p,Y_p) = -\frac{1}{2}\lim_{p\past q\to p} X_q(Y_q({\bf d}(p,q)^2)) \label{rebuilt}\:.
\end{eqnarray}
{\bf (d)} If $M$ is chronological (in particular if $M$ is globally hyperbolic),  $T_p\in T_pM$ is  timelike future-directed iff
${\bf d}(p,\exp_p(tT_p))>0$, $t\in (0,u]$ for some $u>0$. \\
{\bf (e)} Let $(M',{\bf g}', {\cal O}_t')$ be another spacetime with Lorentzian distance ${\bf d}'$.
If $M$ is strongly causal (in particular if $M$ is globally hyperbolic) and $f: M \to M'$  (not necessarily continuous)
is surjective and ${\bf d}'(f(p),f(q)) = {\bf d}(p,q)$ for all $p,q\in M$,
then $f$ is a diffeomorphism (and thus {\em a fortiori} a  homeomorphism), preserves the metric, i.e. $f^*{\bf g}' ={\bf g}$,
and preserves the time orientation.}\\

\noindent {\em Proof.}
{\bf (a)} See the end of A.11.
{\bf (b)} Let $n:= \mbox{dim}M$. Fix $p\in M$ and a sufficiently small  convex normal neighborhood $U$ of $p$. Take a basis of $T^*_pM$ made of
future directed co-vectors $\omega_k$, $k=1,\ldots,n$,  considers $n$ geodesics $\gamma_k$
through $p$, with respectively tangent vectors $\uparrow \spa \omega_k$ and  take $n$ events $p_k\in \gamma_k\cap U\cap I^-(p)$.
The maps $x\mapsto {\bf d}(p_k,x)$
are smooth in a neighborhood of $p$ by (d) of Proposition 2.1. Using that proposition
and (\ref{AA}) one gets  $d {\bf d}(p_k,x)|_p = \beta_k \omega_k$ (there is no summation over $k$)
for some reals $\beta_k\neq 0$. Since the co-vectors $\omega_k$ are linearly independent,
such a requirement is preserved  by the vectors $d {\bf d}(p_k,x)$ in a neighborhood of $p$ and
the maps $x\mapsto {\bf d}(p_k,x)$ define an admissible coordinate map about $p$.
{\bf (c)} In a Riemannian normal coordinate system centered on $p$,
$\sigma(p,q)= (1/2)g_{ab}(p)x_q^ax_q^b$. Hence
${\bf g}_p(X_p,Y_p) = \lim_{q\to p} X_q(Y_q(\sigma(p,q)))$
by direct computation. The limit does not depend on the used curve because
$q\mapsto X_q(Y_q(\sigma(p,q)))$ is continuous about $p$.
Using  $\gamma$ from $p$ to some $q_0\in I^+(p)$
with $\gamma\setminus \{p\} \subset I^+(p)$, (d) of Proposition 2.1 implies (\ref{rebuilt}). 
{\bf (d)} If $T_p$ is time-like and future-directed,  $t\mapsto\exp_p(tT)$ is a timelike
future directed curve, thus $\exp_p(tT)\in I^+(p)$ if  $t>0$ and the thesis is a consequence of
(a) of Proposition 2.1. Conversely, if  $T_p\in T_pM$ and ${\bf d}(p,\exp_p(tT_p))>0$ when $t\in (0,u]$ for some
$u>0$ then $exp_p(tT_p) \in I^+(p)$ in that interval for (a) of Proposition 2.1. Taking $t_0<u$, $t_0>0$ sufficiently small,
there is a convex normal neighborhood $U_p$ containing either $p$,
$q:= \exp_p(t_0T_p)$ and $\exp_p(tT_p)$ for $t\in (0,t_0]$.
Theorem 8.1.2 in \cite{Wald84} implies that the unique geodesic in $U_p$ from $p$ to $q$ must be timelike and thus
$T_p$ is such. If $T_p$ were past directed, $t\mapsto \exp_p(tT_p)$ would be such
giving  $I^+(p)\cap I^-(p) \neq \emptyset$ which violates the chronological condition.
{\bf (e)} One has to prove the injectivity of $f$ only, because the proof of the remaining items  is a direct consequence of  (a)-(d).
The preservation of the Lorentz distance implies that $p\past q$ in $M$ iff  $f(p)\past f(q)$ in $M'$.
Then suppose $p\neq q$ in $M$ and $f(p)=f(q)$. Let $V$ be an open causally convex  neighborhood of $p$ with $q\not \in V$. Take
$q_1,q_2\in V$ with $q_1 \past p \past q_2$. It holds $I^+(q_1)\cap I^-(q_2) \subset V$ and thus  $q\not\in I^+(q_1)\cap I^-(q_2)$.
However $f(q_1)\past f(p) = f(q) \past f(q_2)$ implies $q_1\past q\past q_2$ and  $q\in I^+(q_1)\cap I^-(q_2)$ which is a contradiction.
$\Box$\\

\noindent  {\bf Remark}.   The item (e) can be made stronger (see theorem 4.17 in \cite{BEE}) proving that
if  $(M',{\bf g}', {\cal O}_t')$ is another spacetime with Lorentzian distance ${\bf d}'$,
$(M,{\bf g}, {\cal O}_t)$ is strongly causal and $f: M \to M'$
(not assumed to be continuous) is surjective and for some constant $\beta>0$,
 ${\bf d}'(f(p),f(q)) = \beta{\bf d}(p,q)$ for all $p,q\in M$,
then $f$ is a diffeomorphism and satisfies $f^*{\bf g}' =\beta{\bf g}$. \\

\noindent  We can state the first important technical result of this section in Theorem 2.1. The theorem concerns
some features of the structure of the {\em cut locus} in Lorentzian geometry and establishes 
 that   the Lorentzian distance is almost-everywhere smooth if considered as a function
 of one of the two arguments.
These properties, in turn, will be used to prove the functional formula of the Lorentzian distance
(in particular they are useful to prove Proposition 2.3).  \\
To understand the statement of the theorem  we remind the reader that
a subset $X$ of a manifold $M$ is said to {\bf have measure zero}
if for every local  chart $(U, \phi)$, the set $\phi(U\cap X) \subset \bR^{\mbox{dim}(M)}$ has Lebesgue measure zero.
When $M$ is endowed with a nondegenerate smooth metric ${\bf g}$, it turns out that $X\subset M$
has measure zero if and only of it has measure zero with respect to the positive complete Borel measure 
$\mu_{\bf g}$ induced by ${\bf g}$ on $M$. \\
Some further preliminary definitions and results concerning the {\em nonspace-like cut locus} are necessary.
We use notations and definitions
 in chapter 9 of \cite{BEE}.
 Consider $p\in M$, with $M$ globally hyperbolic. Let ${\bf h}$ be a complete Riemannian metric\footnote{It exists on any differentiable
 Hausdorff second-countable manifold as proven by Nomizu and Ozeki, {\em Proc. Amer. Math. Soc.} {\bf 12}, 889-891.} on $M$.
 Define $$UM:=\{ v\in TM \:|\: {\bf h}(v,v) =1,\:\: {\bf g}(v,v) \leq 0,
 \:\: \mbox{$v$ is future directed} \},\:\:\:\:\:\:\:\: UM_p := \{ v\in UM \:|\: \pi(v)= p\}\:.$$
If $v\in UM$, $t\mapsto c_v(t)$, with $t\in [0, a)$, denotes the unique geodesic starting from $p= \pi(v)$ with initial tangent vector given by
 $v$ and maximal domain.  Finally define, for $v\in UM$, $$s_1(v) := \sup\{t > 0\:|\: \:\:c_v(t) \:\: \mbox{is maximal form $p$ to $c_v(t)$}\}\:.$$  ``$c_v(t)$
 is maximal from $p$ to $c_v(t)$''  means   \cite{BEE}
 that   $L(c_v\sp \rest_{[0,t]}) = {\bf d}(p, c_v(t))$.
Using (b) of Proposition 2.1, it  arises that if a future-directed causal geodesic segment $\gamma: [a,b] \to M$ is maximal, then $\gamma\spa
\rest_{[a',b']}$ is so for $a\leq a'<b'\leq b$.
Notice that $s_1(v)>0$ in strongly causal spacetimes and thus in globally hyperbolic spacetimes
 because, in these spacetimes,  every geodesic is maximal in a  convex normal neighborhood containing the initial point \cite{BEE}.
  It is known (see proposition 9.33 in \cite{BEE}) that $s_1$ is lower semicontinuous in globally hyperbolic spacetimes
  and,  if (1) the spacetime
 is globally hyperbolic, (2) $s_1(v)$ is finite and (3) $c_v$ extends
 to $[0,s_1(v)]$, then $s_1$ is continuous in $v$.      Finally define
 $$\Gamma_{ns}^+(p) := \{s_1(v)v \:|\: v\in UM_p, s_1(v)< +\infty,  \:\:\mbox{$c_v$ extends to $[0,s_1(v)]$} \}\:\:\:
 \mbox{and}\:\:\: C^+(p):= \exp({\Gamma_{ns}^+(p)})\:. $$
   The second definition is consistent because $c_v$ extends to $[0,s_1(v)]$ iff it is defined in some maximal domain
$[0,s_1(v)+\epsilon)$, $\epsilon>0$
 and this is equivalent to saying that  $c_{s_1(v)s}$ is defined in some maximal domain $[0, 1 +\frac{\epsilon}{s_1(v)})$.
 Therefore  if $v\in UM_p$,  ``$s_1(v)< +\infty$ and  $c_v$ extends to $[0,s_1(v)]$''  is equivalent to ``$s_1(v)v \in U_p$'' and so
  $$\Gamma_{ns}^+(p) = \{s_1(v)v \:|\: v\in UM_p, s_1(v)v \in U_p \}\:.$$
  $C^+(p)$ is a subset of $J^+(p)$ by construction  and
it  is called the {\bf future nonspace-like cut locus of $p$}. If $s_1(v)v\in \Gamma_{ns}^+(p)$,   $\exp(s_1(v)v)$ is called the
   {\bf future cut  point of $p$} along $c_v$. The {\bf past nonspace-like cut locus} is defined similarly, with the obvious changes.
Everything  can be re-stated for the past nonspace-like cut locus with the necessary obvious replacements.
By theorem 9.35  of \cite{BEE}, in globally hyperbolic spacetimes,  $C^+(p)$ is closed (and thus $J^+(p)\setminus C^+(p)$ is the union
of the open set $I^+(p)\setminus C^+(p)$ and $\partial I^+(p)\setminus C^+(p)\subset \partial\left(I^+(p)\setminus C^+(p)\right)$) .\\

\noindent {\bf Theorem 2.1}.
 {\em Let $(M,{\bf g}, {\cal O}_t)$ be a globally hyperbolic spacetime and take any $p\in M$.
The following statements hold.\\
{\bf (a)} $\partial I^+(p) = \partial J^+(p) = J^+(p)\setminus I^+(p)$ and $C^+(p) \subset J^+(p)$ are  closed, without internal points, with measure zero;    \\
{\bf (b)} $J^+(p) \setminus \left(C^+(p) \cup \partial J^+(p) \right) = I^+(p) \setminus C^+(p)$
is  open and homeomorphic to $\bR^{\mbox{dim}(M)}$;\\
{\bf (c)} $\exp_p$ defines a diffeomorphism onto $I^+(p)\setminus C^+(p)$ with domain given
by an open subset of $T_pM$ of the form
$${\cal A}_p = \{X\in T_pM\:|\: X \mbox{is timelike and future directed}, 0 < |{\bf g}_p(X,X)| < \lambda_X \mbox{for some $\lambda_X >0$} \}\:;$$
{\bf (d)}
${\bf d}(p,\cdot)^2$ belongs  to
$C^\infty\left(J^+(p)\setminus C^+(p)\right)$
and
${\bf d}(p,\cdot)$ belongs  to
$C^\infty\left(I^+(p)\setminus C^+(p)\right)$;\\
{\bf (e)}  ${\bf d}(p,\cdot)$ satisfies the {\bf timelike eikonal equation} for $q\in I^+(p)\setminus C^+(p)$,
\begin{center}${\bf g}_q(\uparrow\spa d_q {\bf d}(p,q),\uparrow \spa d_q {\bf d}(p,q)) = -1$.\end{center}}

\noindent {\em Proof.} See the Appendix C.
  $\Box$  \\

\noindent  {\bf Remarks}.
{\bf (1)} The statement and the proof of the item  (b) are known in the literature \cite{BEE}.\\
{\bf (2)}
 $C^\infty\left(J^+(p)\setminus C^+(p)\right)$ is valid in the sense of 1.5.
Indeed since  $C^+(p)$ is  closed, $I^+(p)$ is open
and  $J^+(p) = \overline{I^+(p)}$ (see A.12), one has  that  $J^+(p)\setminus C^+(p) =  (I^+(p) \setminus C^+(p)) \cup (\partial I^+(p) \setminus C^+(p))$
where $I^+(p) \setminus C^+(p)$ is open and  $\partial I^+(p) \setminus C^+(p)\subset \partial  (I^+(p) \setminus C^+(p))$.\\
{\bf (3)} Due to the possibility of  reversing the time orientation preserving the globally hyperbolicity,
it turns out that, fixing the latter argument of ${\bf d}(p,q)$ and varying the former, one gets a function in $C^\infty\left(J^-(q)\setminus C^-(q)\right)$ and the analogues of items (a)-(e) above hold.  Finally
$q\in C^+(p)$ iff $p\in C^-(q)$ as a consequences of theorems 9.12
and 9.15 in \cite{BEE}.\\

\noindent {\bf 2.2.}
{\em Causal functions and  Lorentzian distance.}  We introduce a lemma and a proposition
necessary to generalize (\ref{d}) to Lorentzian manifolds in
terms of the Lorentzian distance.    To this end, we have to give some introductory definitions
in particular concerning so-called {\em causal functions}. The introduced machinery, together with
the results achieved in Section 2.1 will make us able to present a preliminary version 
of the formula of the Lorentzian distance in globally hyperbolic spacetimes (Theorem 2.2). \\

\noindent {\bf Definition 2.2}.
 {\em Let $(M, {\bf g}, {\cal O}_t)$ be  a spacetime. Let  $N\subset M$ such that $N= A\cup B$ where $A$ is open and
 $B\subset \partial A$.
 A continuous function $f : N \to \bC$ is said to be  {\bf essentially smooth on} $N$ if there is a closed  set $C_f\subset N$ with
measure zero, such that $f\spa\rest_{{N}\setminus C_f}$ is smooth. ${\cal E}_{[\mu_{\bf g}]}(N)$
indicates the class of such functions.} \\

 \noindent{\bf Definition 2.3}.
 {\em Let $(M, {\bf g}, {\cal O}_t)$ be  a spacetime. Let $N\subset M$.
 A continuous function $f : N\to \bR$ is either  a {\bf causal function} or a {\bf time function} on $N$ if, respectively, it is  non-decreasing
 or increasing along every  future-directed causal curve contained in $N$.
 ${\cal C}(N)$ and  ${\cal T}(N)$ respectively denote  the class of  causal functions and the class of time functions
 on $N$. If $N$ is taken as in Def. 2.2, ${\cal C}_{[\mu_{\bf g}]}(N) :=
{\cal E}_{[\mu_{\bf g}]}(N) \cap {\cal C}(N)$,  ${\cal T}_{[\mu_{\bf g}]}(N) :=
{\cal E}_{[\mu_{\bf g}]}(N) \cap {\cal T}(N)$.}  \\

\noindent {\bf Remark}.   Notice that ${\cal T}(N) \subset {\cal C}(N)$. Moreover, if $N\subset M$ is taken as in Def. 2.2
and $M$ is globally hyperbolic, ${\cal T}(N)\cap C^\infty(N)\neq \emptyset$
because a  smooth time function exists on the whole manifold $M$
(see A.13 and A.15). In general spacetimes ${\cal C}(N)\cap C^\infty(N)\neq \emptyset$ because the constant functions are causal
functions. \\

 \noindent The following technical lemma and a proposition are useful in generalizing   (\ref{d}). 
 The proposition  states that, in suitable domains, ${\bf d}$ defines a natural causal/time function which is also essentially
smooth.  \\
 
 \noindent {\bf Lemma 2.1}.
  {\em In a globally hyperbolic spacetime $(M, {\bf g}, {\cal O}_t)$ with Lorentzian distance ${\bf d}$, take an open causally convex (A.11) set
  $I\subset M$ such that  $\partial I$ has measure zero.
  If  $f \in {\cal C}_{[\mu_{\bf g}]}(\overline{I})$ then,  $df$ is either $0$ or causal and past directed 
  in an open set  $J\subset \overline{I}$ with $\mu_{\bf (g)}(J) = \mu_{\bf g}(I) (= \mu_{\bf g}(\overline{I}))$ and
  \begin{eqnarray}
      \mbox{\em ess}\inf \{|d_z f|\:|\:z\in \overline{I}\} \leq  \inf \left\{ \left.\frac{f(y)-f(x)}{{\bf d}(x,y)} \:\:\right|\: \: x,y \in {\overline{I}},\: x
\past y\right\}
  \label{lemma1}\:.
  \end{eqnarray}
  Above, $\leq$ can be replaced by $=$ if $f \in {\cal T}_{[\mu_{\bf g}]}(\overline{I})$.}\\

 \noindent {\em Proof.} See the Appendix C. $\Box$ \\

 \noindent {\bf Proposition 2.3}.
{\em Let $(M,{\bf g}, {\cal O}_t)$ be a globally hyperbolic spacetime  let ${\bf d}$ indicate the corresponding
Lorentzian distance and, for each $p\in M$, define the functions  $f_p(\cdot) := {\bf d}(p,\cdot)$ and    $h_p(\cdot) := -{\bf d}(\cdot,p)$.
 It holds\\
{\bf (a)} $f_p,h_p\in {\cal E}_{[\mu_{\bf g}]}(M)$; \\
{\bf (b)} $f_p\spa\rest_{I^+(p)} \in {\cal T}_{[\mu_{\bf g}]}(I^+(p))$ and  $h_p\spa\rest_{I^-(p)} \in {\cal T}_{[\mu_{\bf g}]}(I^-(p))$; \\
{\bf (c)} $f_p\spa\rest_{N} \in {\cal C}_{[\mu_{\bf g}]}(N)$ and  $h_p\spa\rest_{N} \in {\cal C}_{[\mu_{\bf g}]}(N)$ for every $N\subset M$
as in Def. 2.2.}    \\

\noindent {\em Proof.} We prove the thesis for $f_p$, the other case is analogous.    {\bf (a)} is a direct consequence of  Theorem 2.1.
and the fact that $f_p(x) = 0$ if $x\not \in J(p)$.
{\bf (b)}
Let $\gamma \subset I^+(p)$ be a causal  future-directed curve. Take  $x,y\in \gamma$ with $x=\gamma(t), y=\gamma(t')$
and  $t'>t$. We want to show that it holds $f_p(x) <  f_p(y) $, i.e., ${\bf d}(p,y) \leq {\bf d}(p,x)$ is not possible.
Notice that $y\neq x$ because the spacetime is globally hyperbolic and thus causal, in fact we have $p \past x \prec y$ (and thus $p\past y$).
Suppose that  ${\bf d}(p,x) \geq {\bf d}(p,y)$.
By (b) of Proposition 2.1 it must also  hold
${\bf d}(p,y) \geq {\bf d}(p,x) +{\bf d}(x,y)$. Putting together and using ${\bf d}(x,y)\geq 0$ one gets
$$0\leq {\bf d}(x,y) \leq {\bf d}(p,y) - {\bf d}(p,x) \leq 0\:.$$
 The only chance is ${\bf d}(x,y) = 0$ and ${\bf d}(p,y) ={\bf d}(p,x)$.
Since the spacetime is globally hyperbolic, there must be a future-directed maximal null geodesic $\gamma_2$ from $x$ to $y$ by (i)
of Proposition 2.1. By the same item there must be a time-like maximal future-directed geodesic $\gamma_1$ from $p$ to $x$.
$\gamma_1 * \gamma_2$ is a causal future-directed  curve from $p$ to $y$. Moreover it holds  $L(\gamma_1 * \gamma_2)= {\bf d}(p,x) + 0 =  {\bf d}(p,y)$.
By (e) in Proposition 2.1, $\gamma_1 * \gamma_2$ can be re-parametrized into a maximal geodesic from $p$ to $y$
which must be time-like, since ${\bf d}(p,y)>0$, $y$ being in $I^+(p)$. This is impossible since $\gamma_2$ is  null.
{\bf (c)} If $N\cap J^+(p) =\emptyset$ 
the proof is trivial since $f_p$ is constant on $N$. Suppose that $N\cap J^+(p) \neq \emptyset$ and 
that $\gamma \subset N$ is a future-directed causal curve with $\gamma(u) \in J^+(p)$ for some $u$, the remaining cases being trivial.
In these hypotheses $\gamma(u') \in J^+(p)$ for $u'>u$  because of A.7.
Then there are  various cases to be analyzed for $t<t'$ where we use the fact that  $f_p$ vanishes outside $I^+(p)$ by Proposition 2.1.
 (1) if $\gamma(t), \gamma(t')\not \in J^+(p)$, the thesis holds because $0= f_p(\gamma(t)) \leq f_p(\gamma(t')) = 0$.
(2) If $\gamma(t) \not \in J^+(p)$ and $\gamma(t') \in J^+(p)$ the thesis holds because $0=f_p(\gamma(t)) \leq f_p(\gamma(t')) \geq 0$.
 (3) If $\gamma(t),\gamma(t') \in I^+(p)$, the thesis holds by (a). (4) $\gamma(t),\gamma(t') \in \partial I^+(p)=\partial J(p)$.
In that case $f_p(\gamma(t)) = f_p(\gamma(t'))=0$ by (a) and (f) of Proposition 2.1. (4) $\gamma(t)\in \partial I^+(p)$ and $\gamma(t') \in I^+(p)$, in that case
$0= \gamma(t) < \gamma(t')$ by (a) and (f) of Proposition 2.1. The case    $\gamma(t')\in \partial I^+(p)$ and $\gamma(t) \in I^+(p)$  is forbidden
because $p \past \gamma(t) \preceq \gamma(t')$ implies $p \past \gamma(t')$ by the remark in A.7.
 $\Box$\\

 \noindent The last technical proposition necessary to state 
 the preliminary version of the functional formula of the Lorentzian distance concerns
 the interplay of  relatively-compact causally-convex sets in globally hyperbolic spacetimes
 and essentially smooth causal functions. \\
 In A.16-A.23 of Appendix A the definition of {\em Cauchy surface} and the relevant properties 
 of these surfaces are given.  An important results of Lorentzian geometry (see A.20) states that: {\em  a spacetime $(M, {\bf g}, {\cal O}_t)$ is globally hyperbolic iff it admits a
Cauchy surface}. This statement  can be adopted as an equivalent definition of a globally hyperbolic spacetime
(see remark in the end of 8.3 in \cite{Wald84} for a proof of  equivalence of the various definitions of globally hyperbolicity). In a globally hyperbolic spacetime  $M$, if $S\subset M$ is a smooth Cauchy surface
and $p\in J^+(S)$, $I(S,p)$ and $J(S,p)$ respectively denote $I^{-}(p) \cap I^+(S)$ and  $J^{-}(p) \cap J^+(S)$.  One can straightforwardly  prove that 
 $I(s,p)$  is not empty iff $p\in I^+(p)$.  It is not very difficult to show that $I(S,p)$ and $J(S,p)$ are causally convex. A.8 implies that $I(S,p)$ is open and  $I(S,p) \subset J(S,p)$. The sets 
$I(p,S)$ and $J(p,S)$ enjoy analogous properties. \\

\noindent {\bf Proposition 2.4}.
{\em Let $(M,{\bf g}, {\cal O}_t)$ be  a globally hyperbolic spacetime and let ${\cal X}$ denote the class of open, nonempty,
causally-convex   subsets of $M$, $I$, such that  $\overline{I}$ is  compact, causally convex  and $\partial I$ has measure zero. The following
statements hold.\\
{\bf (a)}  The class ${\cal X}$
is a {\bf covering} of $M$, i.e., $\bigcup {\cal X} = M$,
and  defines a {\bf direct set}  with respect to the set-inclusion partial-ordering relations,
i.e., if $A,B \in {\cal X}$ there is $C\in {\cal X}$ such that $A\cup B\subset C$.\\
{\bf (b)} If $A\in {\cal X}$, ${\cal T}_{[\mu_{\bf g}]}(A) \neq \emptyset$ and ${\cal C}_{[\mu_{\bf g}]}(\overline{A}) \neq \emptyset$. \\
{\bf (c)} If $p,q\in M$, $p\preceq q$ iff $f(p) \leq f(q)$ for all $f\in {\cal C}_{[\mu_{\bf g}]}(\overline{I})$ and $I\in {\cal X}$ such that $p,q\in \overline{I}$. \\
{\bf (d)} (i) If $S\subset M$ is a smooth Cauchy surface for $M$ and  either $p\in I^+(S)$ or $p\in I^-(S)$, respectively $I(S,p) \in {\cal X}$ and
$\overline{I(S,p)}= J(S,p)$
or  $I(p,S) \in {\cal X}$ and  $\overline{I(p,S)}= J(p,S)$ .

(ii) If $p\in M$, there is a fundamental system of neighborhoods of $p$ made of sets $I(r,s)\in {\cal X}$ with  $\overline{I(r,s)}= J(r,s)$}.\\

\noindent {\em Proof}. See the Appendix B. $\Box$.\\

 \noindent Now we are able to  state and prove the second important theorem  of this section which is nothing but a preliminary version of  the functional formula of the Lorentz distance.\\
 {From} now on we use the following notation:
If $T\in T_pM$, $|T| := \sqrt{|{\bf g}_p(T,T)|}$, similarly, if $\omega \in T^*_pM$,   $|\omega| := \sqrt{|{\bf g}_p(\uparrow\spa \omega, \uparrow\spa
 \omega)|}$.\\

\noindent {\bf Theorem 2.2}.
 {\em Let $(M,{\bf g},{\cal O}_t)$ be a globally hyperbolic spacetime and $p, q \in M$.
Defining $\langle \alpha \rangle := \max\{0,\alpha\}$ for all $\alpha \in \bR$, it holds
 \begin{eqnarray}
 {\bf d}(p,q) =
 \inf \{\langle f(q)-f(p)\rangle \:\:|\:\:
 f \in {\cal C}_{[\mu_{\bf g}]}(\overline{I}),\:\: I\in {\cal X},\:\: p,q\in \overline{I},
\:\: {\mbox{\em ess}\inf}_{\overline{I}} |d f| \geq 1\} \:.    \label{step1}
 \end{eqnarray}}

  \noindent {\em Proof.}
Define $\mu(p,q) :=   \inf \{\langle f(q)-f(p)\rangle \:\:|\:\: f \in {\cal C}_{[\mu_{\bf g}]}(\overline{I}),\:\: I\in {\cal X}, \:\: p,q\in \overline{I},
\:\: {\mbox{ess}\inf}_{\overline{I}} |d f| \geq 1\} \:.$\\
We want to show that $\mu(p,q) = {\bf d}(p,q)$.
First consider the case $p\preceq q$. \\
To this end consider the
map $f_{p}:  x \mapsto {\bf d}(p,x)$, where $x\in \overline{I_{f_p}}$ with $I_{f_p}= I(p,S)$, $S$ being a smooth Cauchy surface with $p,q\in I^-(S)$. Theorem 2.1 and Proposition 2.3
say that such a $f_p$ can be used to evaluate $\mu(p,q)$ because all of the necessary requirements are fulfilled.
We trivially have $0\leq  {\bf d}(p,q) = f_{p}(q) - f_{p}(p) = \langle f_{p}(q) - f_{p}(p) \rangle$
 and thus
$\mu(p,q) \leq {\bf d}(p,q)$.
To conclude, it is sufficient to show that
 $\mu(p,q) \geq  {\bf d}(p,q)$.
   By Lemma 2.1, if $I\in {\cal X}$, $f\in {\cal C}_{[\mu_{\bf g}]}(\overline{I})$ and  ${\mbox{ess}\inf}_{\overline{I}} |d f|
  \geq 1$,
we have
     $$\inf \left\{ \left.\frac{f(y)-f(x)}{{\bf d}(x,y)} \:\:\right|\: \: x,y \in \overline{I},\: x\past y\right\} \geq 1\:.$$
   Therefore, in $\overline{I}$, $x\past y$ entails
     $\langle f(y)-f(x)\rangle  \geq {\bf d}(x,y)$.
The inequality holds also if
$x\preceq y$ because, by (a) and (f) of Proposition 2.1,  if $x\preceq y$ and $x \past y$ is false, it must be ${\bf d}(p,q)=0$. In that case
$\langle f(y)-f(x)\rangle  \geq {\bf d}(x,y)$ is trivially true. In particular, if $p,q\in \overline{I}$ and $p\preceq q$, then
$0 \leq {\bf d}(p,q) \leq  \langle f(q)-f(p)\rangle$.
By the definition of $\mu$, this implies $\mu(p,q)  \geq {\bf d}(p,q)$.\\
Let us consider the case $q \preceq p$. Similarly to above, take $f_{q}:  x \mapsto {\bf d}(q,x)$ in some $J(q,S)$ with $p, q\in I^-(S)$.
$f_q$ can be used to compute $\mu(p,q)$
obtaining  $f_q(q)-f_q(p) \leq 0$ which implies
$\langle f_q(q)-f_q(p)\rangle = 0$ and thus $\mu(p,q) =0$ because $0\leq \mu(p,q)\leq \langle f_q(q) - f_q(p)\rangle$ by definition.\\
Finally consider the case of $p$ and $q$ spatially separated.  In that case it is possible to find  (see below)
two, sufficiently small, regions $I(x,y), I(x',y')$ with $p\in I(x,y)$, $q\in I(x',y')$  and such that $\overline{I(x,y)}=J(x,y)$ and $\overline{I(x',y')}=J(x',y')$
are spatially separated. We conclude that $A := I(p,y)\cap I(q,y') \in {\cal X}$. Then $x\mapsto f(x) := {\bf d}(p,x)+{\bf d}(q,x)$   defines an element of
${\cal C}_{[\mu_{\bf g}]}(\overline{A})$  and satisfies
${\bf g}(df,df)=-1$ a.e. by construction, hence it can be used to evaluate  $\mu(p,q)$ producing
$\mu(p,q)=0 ={\bf d}(p,q)$ because $\langle f(q) - f(p)\rangle =\langle 0-0\rangle =0$ and $0\leq \mu(p,q)\leq \langle f(q) - f(p)\rangle$.
 Let us prove the existence of   $I(x,y), I(x',y')$ with the properties above.  Since $\{q\} \cap (J^+(p)\cup J^-(p)) =\emptyset$ and
 $J^+(p)\cup J^-(p)$  is closed (A.12), there is a neighborhood of $q$, $V$ which satisfies $V\cap (J^+(p)\cup J^-(p)) =\emptyset$. As the spacetime
 is strongly causal,
 $V$ can be fixed with  the form $I(x',y')$. By a suitable restriction (A.8) it is possible to fix $J(x',y')$ such that $q\in I(x',y')$
 and $J(x',y') \cap (J^+(p)\cup J^-(p)) =\emptyset$. This is equivalent to $\{p\} \cap (J^+(J(x',y'))\cup J^-(J(x',y')))=\emptyset$.
 A.12 implies that  $J^+(J(x',y'))\cup J^-(J(x',y'))$ is closed because, since the spacetime is globally hyperbolic, $J(x',y')$ is compact.
 Using the same way followed above, one can find $I(x,y)$ such that $p\in I(x,y)$ and $J(x,y) \cap (J^+(J(x',y'))\cup J^-(J(x',y')))=\emptyset$.
 We have proven that  there are
two regions $I(x,y), I(x',y')$ with $p\in I(x,y)$, $q\in I(x',y')$  and $J(x,y)$, $J(x',y')$ are spatially separated.
$\Box$

\section{The functional formula of the Lorentzian distance.}

\noindent {\bf 3.1.} {\em Laplace-Beltrami-d'Alembert operator and the net of Hilbert spaces.} The  results achieved in Section 2  allow us 
to generalize (\ref{d}) in
a globally hyperbolic spacetime $(M,{\bf g},{\cal O}_t)$ using  the Lorentzian distance ${\bf d}$.
The procedure consists of a translation of the statement of Theorem 2.2, Eq.(\ref{step1}) in particular, in terms of operators.
To this end, a  preliminary discussion on the remaining  ingredients (operators) which appear in (\ref{dl}) is 
necessary. \\
Consider the class of Hilbert spaces   $L^2(\overline{I},\mu_{\bf g})$, ${I}\in {\cal X}$.
These spaces are naturally isomorphic to closed subspaces of $L^2(M,\mu_{\bf g})$.
$||\cdot ||_{{\bf L}(L^2(\overline{I}))}$ denotes the uniform norm operator  in the corresponding $L^2(\overline{I},\mu_{\bf g})$.  In those spaces  three classes of useful operators can be defined:
the operators $\Delta_I$ which are obtained by means of  a suitable restriction of the Laplace-Beltrami-d'Alembert operator,  the functions $f\in {\cal C}_{[\mu_{\bf g}]}(\overline{I})$
viewed as multiplicative operators and the commutators  $[f,[h,\Delta_I]]$.\\

\noindent{\bf Definition 3.1}.
 {\em Let  $(M, {\bf g}, {\cal O}_t)$ be a globally hyperbolic spacetime.
Referring to the notations above, the {\bf Laplace-Beltrami-d'Alembert operator} on $L^2(M,\mu_{\bf g})$, is
$$\Delta : C^{\infty}_0(M) \to L^2(M,\mu_{\bf g})\:,$$ with
 $\Delta := \nabla_\mu\nabla^\mu$ in local coordinates,
$\nabla$ denoting the Levi-Civita covariant derivative.
$\Delta_I$ denotes the restriction of $\Delta$ to the, dense in $L^2(\overline{I},\mu_{\bf g})$, linear manifold 
$C^\infty(\overline{I})$, $I\in {\cal X}$}.\\

\noindent As a general remark we notice that $\Delta$  is densely defined, symmetric and 
 admit self-adjoint extensions because
it commutes with the complex conjugation, conversely every $\Delta_I$ is not symmetric 
because it is not Hermitean in the considered domain because of nonvanishing boundary terms.\\
Let us pass to  consider the causal functions and commutators.
Every $f\in {\cal C}_{[\mu_{\bf g}]}(\overline{I})$ (${I}\in {\cal X}$)
can be seen as a multiplicative (self-adjoint) operator in $L^2(\overline{I},\mu_{\bf g})$
with domain given by the whole space $L^2(\overline{I},\mu_{\bf g})$.
The commutator  $[f,[h,\Delta_I]]$ is well-defined as an operators in $L^2(\overline{I},\mu_{\bf g})$ with
the domain and the properties stated below.  
A remarkable step  which permits to translate (\ref{step1}) into (\ref{dl}) is the identity established  by 
Eq. (\ref{dc})
in the item (b) below.\\
 In the following, if $A$ is an operator in a  Hilbert space  with
scalar product $(\cdot,\cdot)$, $A \leq \alpha I$ means $\alpha I-A \geq 0$, i.e., $(\Psi,(\alpha I-A) \Psi) \geq 0$ for all $\Psi$ in the domain of $A$. \\

\noindent {\bf Lemma 3.1}.
{\em  In a globally hyperbolic spacetime $(M,{\cal O}_t,{\bf g})$ take ${I}\in {\cal X}$ and $f,h\in  {\cal C}_{[\mu_{\bf g}]}(\overline{I})$.
Let ${\cal D}_{I,f,g} :=  C^\infty_0(I\setminus (S_{f}\cup S_h))$, $S_{t}$
being the set  of singular points of $t\in  T_{[\mu_{\bf g}]}(I)$. \\
{\bf (a)} ${\cal D}_{I,f,g}\subset L^2(\overline{I},\mu_{\bf g})$ is a dense linear manifold, invariant  for  either
$f$, $h$, $\Delta_I$.\\
{\bf (b)}  $\Delta_I$ and $[f,[h,\Delta_I]]$ are symmetric on ${\cal D}_{I,f,g}$,  the latter operator is also essentially self-adjoint on ${\cal D}_{I,f,g}$  and
\begin{eqnarray}
[f,[h,\Delta_I]] = 2 {\bf g}(\uparrow \spa df, \uparrow \spa dh)  \:\:\: \mbox{almost everywhere in $\overline{I}$}\label{dc} \:.
\end{eqnarray}
{\bf (c)} The following equivalent relations hold

(i)  $\sigma({\overline{[f,[h,\Delta_I]]}})\subset (-\infty,0]$,

(ii) ${{[f,[h,\Delta_I]]}} \leq 0$ on ${\cal D}_{I,f,g}$,

(iii)  ${\overline{[f,[h,\Delta_I]]}} \leq 0$.}\\

\noindent {\em Proof.}  {\bf (a)} It is obvious that  ${\cal D}_{I,f,g}$ is a linear manifold in $L^2(\overline{I},\mu_{\bf g})$. It is also dense
therein because, as $S_{f}\cup S_h$ is closed, $I\setminus (S_{f}\cup S_h)$ is open and $C^\infty_0(I\setminus (S_{f}\cup S_h))$ is dense in
$L^2(I\setminus (S_{f}\cup S_h),\mu_{\bf g})$ which coincides with $L^2(\overline{I},\mu_{\bf g})$ because $S_{f}\cup S_h\cup \partial I$ has measure zero.
 The invariance properties can be proven by direct inspection.  {\bf (b)} 
 $\Delta_I$ restricted to the  linear manifold ${\cal D}_{I,f,g}$ is Hermitean by construction
(notice that $I\setminus (S_{f}\cup S_h)$ is open)
and thus it is symmetric too because the domain is dense.
 (\ref{dc}) can be proven by direct inspection
on $I\setminus (S_{f}\cup S_h)$. 
$[f,[h,\Delta_I]] = 2{\bf g}(\uparrow \sp df, \uparrow \sp dh)$
entails the Hermiticity (and thus the symmetry, the domain being dense) because   ${\bf g}(\uparrow \sp df, \uparrow \sp dh)$ is a real measurable function which
acts as a multiplicative operator. However the symmetry also follows form standard properties of the commutator and the symmetry
of the operators $f,h, {\Delta_I}$.   The essentially self-adjointness of   $[f,[h,\Delta_I]]$ on ${\cal D}_{I,f,g}$  is assured by Nelson's theorem \cite{Simon}
proving that ${\cal D}_{I,f,g}$ is made by analytic vectors. The proof is immediate using the fact that   ${\bf g}({\uparrow \sp df, \uparrow \sp dh})$ is   smooth
and thus bounded when restricted to any compact set contained in   $I\setminus (S_{f}\cup S_h)$. {\bf (c)}
By Lemma 2.1, $df$ and $df$ are almost everywhere causal and past directed where they do not  vanish, therefore
 $[f,[h,\Delta_I]]= {\bf g}(\uparrow \sp df, \uparrow \sp dh) \leq 0$ almost everywhere. In turn it entails (ii), namely 
 $(\Psi, [f,[h,\Delta_I]] \Psi)\leq 0$  for all
 $\Psi \in {\cal D}_{I,f,g}$. Let us prove the equivalence of (i),(ii) and (iii). The unique self-adjoint extension of   $[f,[h,\Delta_I]]$ 
coincides with the closure of the same operator and thus (ii) implies (iii). Moreover  (iii)  implies  (ii)
trivially. Using the spectral measure of
  $\overline{[f,[h,\Delta_I]]}$ one trivially see that (i)
   is equivalent to (iii).
 $\Box$\\

 \noindent{\bf 3.2.} {\em The functional formula of the Lorentz distance.}
To conclude, we can state and prove the formula  (\ref{dl}) which generalizes (\ref{d}) in  globally hyperbolic
spacetimes.\\

 \noindent {\bf Theorem 3.1}.
 {\em Let $(M,{\bf g},{\cal O}_t)$ be a globally hyperbolic spacetime
with Lorentzian distance ${\bf d}$ and define  ${\sdelta}_{I} := \frac{1}{2} \Delta_{I}$ and  $\langle \alpha\rangle := \max\{0,\alpha\}$ if $\alpha \in \bR$.
The Lorentzian distance of $p,q\in M$ can be computed as follows
  \begin{eqnarray}
{\bf d}(p,q) =
 \inf\spa\left\{ \spa \langle f (q)-f(p)\rangle \spa\left|\:  f\in {{\cal C}_{[\mu_{\bf g}]}(\overline{I})},\: I\in {\cal X},\: p,q\in \overline{I},
 \: \left|\left| \overline{\left[f,\left[f,{\sdelta}_{I}\right]\right]}^{-1}
 \right|\right|_{{\bf L}(L^2(\overline{I}))} \sp \leq  1 \spa \right.\right\},
\label{fine1}
\end{eqnarray}
 where   $|| \overline{[f,[f,{\sdelta}_{I}]]}^{-1}
 ||_{{\bf L}(L^2(\overline{I}))} \leq  1$ (which includes the requirement on the
 existence of $\overline{[f,[f,{\sdelta}_{I}]]}^{-1}$)  can be replaced by one of the following equivalent requirements
\begin{eqnarray}
    [f,[f,  \Delta_{I}]]  &\leq& -I \:\:\: \: \mbox{(on ${\cal D}_{I,f,f}$)} \label{one}\:,\\
    \overline{[f,[f,  \Delta_{I}]]}  &\leq& -I  \label{two} \:,\\
    \sigma(\overline{[f,[f,  \Delta_{I}]]}) &\subset& (-\infty, -1] \label{three}\:.
\end{eqnarray}}
\noindent {\em Proof.}  First  we show that under the assumption  $[f,[f,{\sdelta}_{I}]]  \leq 0$ (which holds by (c) of Lemma 3.1 as
$f\in {\cal C}_{[\mu_{\bf g}]}(I)$),
the four requirements  (\ref{one}),  (\ref{two}), (\ref{three}) and
 (R): ``$\overline{[f,[f,{\sdelta}_{I}]]}^{-1}$ exists and  $||\overline{[f,[f,{\sdelta}_{I}]]}^{-1}||_{{\bf L}(L^2(\overline{I}))}  \leq  1$'',
are equivalent.  The proof of the equivalence of   (\ref{one}),  (\ref{two}), (\ref{three}) is essentially the same
 used to prove the equivalence of the analogous three conditions in (c) of Lemma  3.1, we leave the details to the reader.
 Using the spectral representation of $\overline{[f,[f,{\sdelta}_{I}]]}$,
 and viewing $\overline{[f,[f,{\sdelta}_{I}]]}^{-1}$  as a spectral function of the former,
  (\ref{three}) implies (R) straightforwardly. On the other hand, using the spectral theorem for $\overline{[f,[f,{\sdelta}_{I}]]}^{-1}$,
  (R) implies that $\sigma(\overline{[f,[f,{\sdelta}_{I}]]}) \subset (-\infty,-1] \cup [1, +\infty)$  (Use $\sigma(A)\subset [-||A||,||A||]$ and
   $\sigma(A^{-1})\subset
  \overline{\{1/\lambda \:|\: \lambda \in \sigma(A)\setminus\{0\}\}}$ provided $0\not \in \sigma_p(A)$ this being our case when
  $A=\overline{[f,[f,{\sdelta}_{I}]]}^{-1}$ because $A$ admits inverse by construction). Then $[f,[f,{\sdelta}_{I}]]  \leq 0$,  which is equivalent to
  $\sigma(\overline{[f,[f,{\sdelta}_{I}t]]})\subset (-\infty,0]$ by (c) of Lemma 3.1,   entails  $\sigma(\overline{[f,[f,{\sdelta}_{I}]]}) \subset (-\infty,-1]$ which is (\ref{three}). 
To conclude and prove (\ref{fine1}) we  reduce to the expression for ${\bf d}$ given in Theorem 2.2. The condition
 $\mbox{ess}\inf_{\overline{I}}|d f| \geq 1$ which appears  in the thesis of Theorem 2.2 is equivalent to  $\mbox{ess}\inf_{\overline{I}}|d f|^2 \geq 1$ which, in turn,
 is equivalent to  \begin{eqnarray} {\mbox{ess}\sup}\left\{|{\bf g}_x(\uparrow\spa d f,\uparrow\spa d f)|^{-1}\:|\: x\in \overline{I}\right\} \leq 1\:.
\label{dc-1} \end{eqnarray}
Using  the function ${\bf g}_x(\uparrow\spa d f,\uparrow\spa d f)^{-1} = - |{\bf g}_x(\uparrow\spa d f,\uparrow\spa d f)|^{-1}$
as a multiplicative (self-adjoint) operator on the whole space $L^2(\overline{I}, \mu_{\bf g})$, (\ref{dc-1}) can  equivalently be re-written
\begin{eqnarray} || {\bf g}_x(\uparrow\spa d f,\uparrow\spa d f)^{-1} ||_{{\bf L}(L^2(\overline{I}))} \leq 1\:.
\label{dc-2} \end{eqnarray}
 On the other hand,  holding $ {\bf g}_x(\uparrow\spa d f,\uparrow\spa d f)^{-1} \cdot {\bf g}_x(\uparrow\spa d f,\uparrow\spa d f) =1$ a.e.,
 and ${\bf g}_x(\uparrow\spa d f,\uparrow\spa d f) = [f,[f, {\sdelta}_{I}]]$  a.e., we also  have
\begin{eqnarray}
[f,[f, {\sdelta}_{I}]] \circ {\bf g}_x(\uparrow\spa d f,\uparrow\spa d f)^{-1} &=& I_{L^2(\overline{I})}\:,\\
{\bf g}_x(\uparrow\spa d f,\uparrow\spa d f)^{-1} \circ [f,[f, {\sdelta}_{I}]]  &=& I_{{\cal D}_{I,f,f}}\:.
 \end{eqnarray}
Notice that  the closure of $[f,[f, {\sdelta}_{I}]]$ is an operator because $[f,[f, {\sdelta}_{I}]]$ is essentially 
self-adjoint ((b) of Lemma 3.1), moreover ${\bf g}_x(\uparrow\spa d f,\uparrow\spa d f)^{-1}$ is bounded by (\ref{dc-2}). These two
facts together imply that
 $[f,[f, {\sdelta}_{I}]]$ can be replaced by $\overline{[f,[f, {\sdelta}_{I}]]}$   in
both the identities above (also replacing $I_{{\cal D}_{I,f,f}}$ with the identity operator on the domain of 
$\overline{[f,[f, {\sdelta}_{I}]]}$). Then, the uniqueness of the inverse operator implies that
  (\ref{dc-2})  is nothing but $||\overline{[f,[f,{\sdelta}_{I}]]}^{-1}||_{{\bf L}(L^2(\overline{I}))}  \leq  1$. $\Box$\\

\noindent  {\bf Remark}.
 Theorem 3.1 holds if replacing $M$  with a vector fiber bundle $\gF \to M$ equipped with a positive Hermitean fiber-scalar product,
and using  a second-order differential
 operator working on compactly-supported almost-everywhere smooth sections,  locally given by
$$\sdelta^{(X,V)} =\frac{1}{2}\left[ (\nabla^\mu - iX^\mu)(\nabla_\mu-iX_\mu) + V\right]\:.$$
$X$ is  any smooth Hermitean $SU(N)$-connection field
$V$ defining a Hermitean linear map $V_x : {\cal F}_x \to {\cal F}_x$ on each fiber ${\cal F}_x$, $x\in M$.
This is because the identity (\ref{dc}) is preserved
\begin{eqnarray}
[h,[f,\sdelta^{(X,V)}]]
= {\bf g}(\uparrow\spa d h,\uparrow\spa d f)I \label{dc3}\:,
\end{eqnarray}
where $I$ is the fiber identity.

\section{The algebraic point of view: generalizations towards a Lorentzian noncommutative geometry.}

As found in  Section 3, a generalization of the functional identity for the Riemannian distance exists
in globally hyperbolic spacetimes.   Here, we shall not attempt
to give a complete investigation  of noncommutative Lorentzian causal structures  but we try to extract the algebraic content
from the  structure involved in  the commutative case obtaining  generalizations of  the causal structure  in both the commutative
and  noncommutative case. In particular we present a set of five axioms on {\em noncommutative causality} which give a straightforward
generalization of the causal structure of globally hyperbolic spacetimes. We stress that there is no guarantee for the minimality of the
presented set of axioms. \\

\noindent {\bf 4.1}.{\em Algebraic ingredients.} Assume that $(M, {\cal O}_t, {\bf g})$ is a globally hyperbolic spacetime
and adopt all the notations and definition  given in Section 1, 2 and 3 (including Appendix A).
In  particular we focus attention on the ingredients used to write down (\ref{dl}) from the point 
of view of $C^*$-algebra theory.
A relevant mathematical object is the net of Hilbert subspaces,
$\gH = \{ {\cal H}_I\:|\: I\in {\cal X}\}$,
where ${\cal H}_I= L^2(\overline{I},\mu_{\bf g})$.
$\gH$ enjoys several properties induced by the properties of the class of subsets
${\cal X}$ defined in Proposition 2.4.
In the following  $\leq$, used between elements of ${\cal X}$, indicates the partial ordering relation on ${\cal X}$  given by the set-inclusion relation.
$({\cal X}, \leq)$ is a direct set as shown in (a) of  Proposition 2.4.  We have a consequent trivial proposition concerning the elements of $\gH$ .  \\

\noindent {\bf Proposition 4.1}.
{\em Referring to the given definitions and notations,\\
{\bf (a)} for any pair  $I,J\in {\cal X}$ with $I\leq J$,   ${\cal H}_I \subset {\cal H}_J$. More precisely,
there is a  Hilbert-space isomorphism from  ${\cal H}_I$ onto a (closed)  subspace of  ${\cal H}_J$;\\
{\bf (b)} $\gH$ is a direct set with respect that inclusion relation. More precisely, for any pair $I,J\in {\cal X}$ there is $K\in {\cal X}$
with $I,J\leq K$ such that $\overline{{\cal H}_I + {\cal H}_J} \subset  {\cal H}_K$.}\\

\noindent  A second set of relevant mathematical objects is given as follows.
An elementary computation proves that  if $f\in C(\overline{I})$,
$C(\overline{I})$ denoting the commutative unital $C^*$-algebra of the continuous complex functions on $\overline{I}$,
$|| f ||_\infty = ||O_f||_{{\bf L}({\cal H}_I)}$, where $O_f$ is the multiplicative operator 
$O_f h := f\cdot h$ for all $h\in  {\cal H}_I$.
Moreover the involution in $C(\overline{I})$, i.e. the complex conjugation $\overline{\:\cdot\:}$, is equivalent to the
involution in ${\bf L}({\cal H}_I)$, that is the Hermitean conjugation $\cdot^*$.  Therefore   $C(\overline{I})$ can
 be viewed as a subalgebra of the $C^*$-algebra of all bounded operators on  ${\cal H}_I$,
${\bf L}({\cal H}_I)$.\\
From now on  we use the following notation $\gA_0 := \{{\cal A}_{I}\}_{I\in {\cal X}}$,
where   ${\cal A}_{I}$  denotes the commutative unital normed $*$-algebras
containing all  of multiplicative operators $O_f$, $f\in {\cal E}_{[\mu_{\bf g}]}(I)$.  Moreover
${\gA} := \{\overline{{\cal A}_{I}}\}_{I\in {\cal X}}$  where  $\overline{{\cal A}_{I}}$  indicates
the $C^*$-algebra given  by  the Banach completion of ${\cal A}_{I}$.   \\

\noindent {\bf Lemma 4.1}
{\em Referring to the given definitions and notations, if $I\in {\cal X}$,
$\overline{{\cal A}_{I}}$ is (isometrically) isomorphic to the $C^*$-algebra of the continuous functions on $\overline{I}$, $C(\overline{I})$.}\\

 \noindent {\em Proof.} $C^\infty(\overline{I})\subset {\cal A}_{I}$.
$C^\infty(\overline{I})$ is  $||\cdot ||_\infty$-dense in
$C(\overline{I})$ by Stone-Weierstrass' approximation theorem
because ${C^\infty(\overline{I})}$ and thus the closed sub $*-algebra$ of $C(\overline{I})$,  $\overline{C^\infty(\overline{I})}$,
separates the points of  $\overline{I}$
and so  $\overline{C^\infty(\overline{I})}$ must coincide
with the algebra $C(\overline{I})$ it-self. $\Box$\\

\noindent {\bf Proposition 4.2}.
{\em Referring to the given definitions and notations, for $I,J \in {\cal X}$ and
$I\leq J$, define $\Pi_{I,J}(a) := a\sp\rest_{{\cal H}_I}$, $a\in \overline{{\cal A}_{J}}$, then\\
{\bf (a)} $\Pi_{I,J}({\cal A}_J) \subset {\cal A}_I$ and thus $\Pi_{I,J}\spa \rest_{{\cal A}_J}: {\cal A}_J \to {\cal A}_I$ is  a  continuous (norm decreasing)  unital
$*$-algebra
homomorphism; \\
{\bf (b)} $\overline{\Pi_{I,J}({\cal A}_{J})} =\overline{{\cal A}_{I}}$, in other words
$\Pi_{I,J} : \overline{{\cal A}_J} \to \overline{{\cal A}_I}$ is a surjective continuous unital $C^*$-algebra homomorphism.}\\

\noindent {\em Proof.} {\bf (a)} can be proved by direct inspection using the fact that, in the sense of Lemma 4.1, $a\spa\rest_{\overline{I}} = a\spa\rest_{{\cal H}_I}$
where $a \in C(\overline{J})$ in the left-hand side is viewed as a function and $a \in {\bf L}({\cal H}_J)$ in the right-hand side is viewed as a multiplicative operator. Let us prove {\bf (b)}.
$\overline{\Pi_{{I},J}({\cal A}_{J})}
 = \overline{{\cal A}_{I}}$ and the surjectivity on $\overline{{\cal A}_I}$
 of $\Pi_{{I},{J}}$ to $\overline{{\cal A}_{J}}$ are trivially equivalent because $\Pi_{{I},{J}}$ is continuous.
We directly prove the surjectivity. Using Lemma 4.1, it is sufficient to show that, for every $f\in C(\overline{I})$ there is $g\in C(M)$
such that $g\spa\rest_{\overline{I}}=f$. Since $M$ is Hausdorff, locally compact and  $\overline{I}$ is compact, the existence of $g$
follows from the Tietze extension theorem \cite{Rudin70}.  $\Box$\\

\noindent We have an immediate corollary: \\

\noindent {\bf Corollary}.
{\em In the hypotheses of Proposition 4.2, for $I,J,K \in {\cal X}$, $\Pi_{I,I}=Id$
and  $I\leq J \leq K$ entails $\Pi_{{I},K} = \Pi_{{I},J}\circ \Pi_{J,K}$.}\\

\noindent The third ingredient is given by the  class of  causal functions. It  takes the causal structure of the spacetime into account. Let us examine this ingredient
 from the  algebraic point of view.
First of all, notice that  $I\in {\cal X}$ entails  ${\cal C}_{[\mu_{\bf g}]}(\overline{I})\subset {\cal A}_I$.
From now on we use the notation $Co_I := {\cal C}_{[\mu_{\bf g}]}(\overline{I})$ and $\gC := \{ Co_{I} \}_{I\in {\cal X}}$.
$Co_I$ is called the {\bf causal cone} in ${\cal A}_I$.\\

\noindent{\bf Proposition 4.3}.
{\em Referring to the given definitions and notations, for $I,J\in {\cal X}$:\\
{\bf (a)} $Co_{I}$  is a convex  cone containing the origin (i.e.
$\alpha t+\beta t'\in Co_{I}$  for $\alpha,\beta\in [0,+\infty)$ and  $t,t'\in Co_{I}$) whose elements are  self-adjoint
(i.e., $t\in Co_{I}$ implies $t^*=t$).\\
{\bf (b)} $[Co_I] := \{t_1-t_2 +i(t_3 -t_4) \:\:|\:\:  t_k\in Co_I, k=1,2,3,4 \}$
 is a dense sub $*$-algebra of $\overline{{\cal A}_I}$;\\
{\bf (c)} $I\leq J$ entails  $\Pi_{I,J}(Co_{J}) \subset Co_{{I}}$.}\\

\noindent {\em Proof}. The only nontrivial statement is {\bf (b)}, let us prove it.  First notice  that   ${[Co_I]}$ is closed with respect to the 
algebra operations and $\bI \in {[Co_I]}$, $\bI$ being the unit of ${\cal A}_I$. 
Indeed, by the given definitions,  $u,v \in {[Co_I]}$ entails $\alpha u+\beta v \in {[Co_I]}$ for all
$\alpha, \beta\in \bC$ and $u\in {[Co_I]}$ entails $u^*\in {[Co_I]}$.  Then
 notice that $\bI$  is nothing but  the constant map
  $x \mapsto 1$ which   is an element of
  ${\cal C}_{[\mu_{\bf g}]}(\overline{I})= Co_I$.
Moreover if $t\in Co_I$, since $\overline{I}$ is compact and $t$ is continuous, there is $\alpha>0$ 
such that if $t_\alpha := t+\alpha\bI$, $t_\alpha (x) >0$ for all $x\in \overline{I}$.
So take   $t,t'\in Co_I$ and define $t_\alpha,t'_{\alpha'} >0$ as said. 
It is clear that $t_\alpha\cdot t'_{\alpha'} \in Co_I$ because the product of positive non-decreasing
functions is a non-decreasing increasing  
function. $t_\alpha\cdot t'_{\alpha'} \in Co_I$ means $t\cdot t' + \alpha \alpha'\bI + \alpha t' +\alpha' t \in {[Co_I]}$, therefore
the definition of $[Co_I]$ implies $t\cdot t' \in {[Co_I]}$.
That result trivially generalizes to any pair $u,u' \in {[Co_I]}$. We have proven that $[Co_I]$ is a  sub $*$-algebras of $\overline{{\cal A}_I}$.
Now we prove that ${[Co_I]}$
separates the points of $\overline{I}$ and hence the closed  algebra $\overline{[Co_I]}$
coincides with $\overline{{\cal A}_I} = C(\overline{I})$ because of Stone-Weierstrass' theorem.
Let us show that, if $p,q\in \overline{I}$, there is $t_1\in Co_I(\subset [Co_I])$ such that $t_1(p)\neq t_1(q)$.
Indeed, if $p \prec q$, take $p' \past p$, fix any $t\in Co_I$ and define
$t_1 := t+ {\bf d}(p',\cdot)
\spa\rest_{\overline{I}}$. By (c) of Proposition 2.3,
  $t_1\in Co_I$. Then $t_1(p) <t_1(q)$ by construction because ${\bf d}(p',p)< {\bf d}(p',q)$ by (b) of Proposition 2.3.
If $p,q\in \overline{I}$ are spatially separated there is  $p'\past p$ with $q \not \in J^+(p')$ (see the proof of Theorem 2.2).
If $t\in Co_I$, take $\alpha \in [0, +\infty)$ with $t(p) + \alpha {\bf d}(p',p) > t(q)$. Then $t_1 := t + \alpha {\bf d}(p',\cdot)\spa\rest_{\overline{I}}\in Co_I$
and  $t_1(p)\neq t_1(q)$. $\Box$\\

 \noindent {\bf Remark}. Since nontrivial causal functions cannot have compact support, we are forced to consider the unital normed $*$-algebras
 ${\cal A}_{I}$, as natural objects instead of the {\em nonunital} normed
$*$-algebras $C_c(I)$
(the compactly-supported continuous functions on the open set $I$) if we want that some time function as ${\bf d}(p,\cdot)$ 
belongs to ${\cal A}_{I}$, as it results necessary from the proof of Proposition 4.3.
On the physical ground this is related to the fact that a physical spacetime cannot be compact.
A consequence of such a choice is that the class of
$C^*$-algebras $\{\overline{{\cal A}_{I}}\}$ is not  a {\em net} of $C^*$-algebras in the sense used in Quantum Field Theory
\cite{Haag} and it is not possible to define an overall  $C^*$-algebra given by  the  inductive limit of the net.\\

\noindent The last ingredient we go  to introduce is the class of
densely-defined operators used in 3.2, $\gG :=\{{\bf G}_I\}_{I\in {\cal X}}$, where ${\bf G}_I := {\sdelta}_I : {\cal D}_I \to {\cal H}_I$   and
${\cal D}_I := C^\infty(\overline{I})$, ${\cal H}_I = L^2(\overline{I}, \mu_{\bf g})$. ${\bf G}_I$ will be said the {\bf causal operators}
on ${\cal H}_I$. \\

\noindent {\bf Proposition 4.4}.
{\em  Referring to the given definitions and notations: \\
{\bf (a)} for every $J\in {\cal X}$, $f,g\in Co_J$, there is a linear manifold ${\cal D}_{J,f,g} \subset {\cal D}_J$ such that:

(i) ${\cal D}_{J,f,g}$  is dense in ${\cal H}_J$ and invariant with respect to $f,g,{\bf G}_I$;

(ii) if $K\in {\cal X}$, $K\leq J$, and $\Psi\in {\cal D}_{K,\Pi_{K,J}(f),\Pi_{K,J}(g)}$,
\begin{eqnarray}
\left[f,\:\left[g\:,\: {\bf G}_J\right]\right] \Psi =  \left[\Pi_{K,J}(f)\:,\:\left[\Pi_{K,J}(g)\:,\: {\bf G}_K\right]\right] \Psi      \label{rest}\:;
\end{eqnarray}

(iii)  $[f\:,\:[g\:,\:{\bf G}_J]]$ is essentially self-adjoint  on ${\cal D}_{J,f,g}$;

(iv) if $\alpha,\beta >0$,  it holds    \begin{eqnarray}
{\cal D}_{J,f,f} \cap {\cal D}_{J,g,g} \cap {\cal D}_{J,f,g} \cap {\cal D}_{J,g,f} \subset {\cal D}_{J,\alpha f + \beta g,\alpha f + \beta g}\:,
\label{core}
\end{eqnarray}
and    ${\cal D}_{J,f,f} \cap {\cal D}_{J,g,g} \cap {\cal D}_{J,f,g} \cap {\cal D}_{J,g,f}$ is a core for
$[\alpha f+\beta g, \: [\alpha f+\beta g,\: {\bf G}_J]]$;

(v)    $[f\:,\:[g\:,\:{\bf G}_J]] \leq 0$  on ${\cal D}_{J,f,g}$.\\
{\bf (b)} $Co_{{\bf G}_K} := \{ f\in Co_K \:\:|\:\:\: [f\:,\:[f\:,\:{\bf G}_K]] \leq -\gamma I\:\:\:\:\mbox{for some $\gamma >0$}  \}$ is not empty.} \\

\noindent {\em Proof}. {\bf (a)} (\ref{rest}) and (\ref{core}) can be proven by direct inspection,
 ${\cal D}_{J,f,f} \cap {\cal D}_{J,g,g} \cap {\cal D}_{J,f,g} \cap {\cal D}_{J,g,f}$ is a core for
$[\alpha f+\beta g,\: [\alpha f+\beta g,{\bf G}_J]]$ because that operator is essentially self-adjoint on that domain. This fact  can straightforwardly be
shown by following that  way, based on Nelson's theorem, used
in  the proof of   Lemma 3.1.
The remaining statements of the thesis are
 parts of the thesis of Lemma 3.1. {\bf (b)}  As the spacetime is globally hyperbolic there is a smooth time function $t$ with $dt$ everywhere timelike (see A.13).
Therefore the smooth function ${\bf g}(\uparrow\spa dt,\uparrow\spa dt)$ is strictly negative on the compact $\overline {K}$ and thus posing
$-\gamma := \max_{\overline K} {\bf g}(\uparrow\spa dt,\uparrow\spa dt)$,  one has
$\gamma >0$ and  $f:= t\sp\rest_{\overline{K}} \in Co'_K$ because, by (\ref{dc}), $[f,[f,{\bf G}_K]] = {\bf g}(\uparrow\spa dt,\uparrow\spa dt) \leq -\gamma I$.
$\Box$\\

\noindent {\bf Corollary}.
 {\em With the hypotheses of Proposition 4.4,
 $t\in Co_{{\bf G}_K}$ entails  $f+ \alpha t \in Co_{{\bf G}_K}$ for
every $f\in Co_K$ and $\alpha >0$, in particular $Co_{{\bf G}_K} $ is a convex cone.}    \\

\noindent {\em Proof}.  Take $t\in Co_{{\bf G}_K}$. 
By (iv) in (a) of Proposition 4.4, if $A:={\cal D}_{K,t,t}\cap {\cal D}_{K,f,f}\cap {\cal D}_{K,f,t}\cap {\cal D}_{K,t,f}$
$$\overline{[f+ \alpha t,[f+\alpha t, {\bf G}_K]]\spa\rest_{A}} =
\overline{[f+ \alpha t,[f+\alpha t, {\bf G}_K]]}\:.$$
In  $A$, it also holds
$[f+ \alpha t,[f+\alpha t, {\bf G}_K]]= [f,[f {\bf G}_K]] + \alpha [f,[t, {\bf G}_K]]+ \alpha [t,[f, {\bf G}_K]] +\alpha^2 [t,[t, {\bf G}_K]]$. Finally (v) in (a)
implies $[f+ \alpha t,[f+\alpha t, {\bf G}_K]]\leq -\gamma I$ in the considered domain and so $\overline{[f+ \alpha t,[f+\alpha t, {\bf G}_K]]} \leq -\gamma I$.
In particular it holds in
${\cal D}_{K,f+\alpha t,f+\alpha t}$. $\Box$ \\

\noindent Putting all together we can state  the following  {\bf general algebraic hypotheses} 
which are fulfilled in the commutative case. However it is worthwhile stressing {\em they do not 
require the commutativity it-self explicitly} and thus they could be used in noncommutative generalizations.  A fifth axiom will be introduced shortly. \\

\noindent {\em {\bf (AH1)}  $\gH =\{{\cal H}_I\}_{I\in {\cal X}}$ is a class of Hilbert spaces
labeled in a partially-ordered direct set $({\cal X}, \leq)$ such that (a), (b) of Proposition 4.1 is fulfilled. \\
 {\bf (AH2)} $\gA = \{{\cal A}_I\}_{I\in {\cal X}}$
is a class of unital sub $*$-algebras of the 
$C^*$-algebra of the bounded operators on ${\cal H}_I$, ${\bf L}({\cal H}_I)$.   $\overline{{\cal A}_I}$ denotes the unital $C^*$-algebra
obtained as the Banach completion of ${\cal A}_I$ and we assume that (a), (b) of Proposition 4.2 holds (and thus its corollary holds too) with
 $\Pi_{I,J}$ defined therein. \\
{\bf (AH3)} $\gC= \{Co_I\}_{I\in {\cal X}}$, with $Co_I\subset {\cal A}_I$, fulfills (a), (b) an (c) of Proposition 4.3.\\
{\bf (AH4)}
 $\gG =\{{\bf G}_I\}_{I\in {\cal X}}$, with ${\bf G}_I : {\cal D}_I \to {\cal H}_I$ and
${\cal D}_I\subset {\cal H}_I$, is a class of densely-defined operators satisfying (a) and (b) of  Proposition 4.4 (and thus its corollary).}\\

\noindent {\bf 4.2}. {\em Events, loci and causality.}  Let us examine  how the events of $M$ and its topology arise in the algebraic picture introduced above.
In particular, we show that  the presented approach gives rise to a generalization of the concept of event in a spacetime, preserving the causal
relations.
When a manifold is compact, its points can be realized in terms of pure algebraic states on the $C^*$-algebra of continuous functions on the manifold
\cite{Connes,Landi,GBVF}. If a manifold is {\em only} locally compact the construction is more complicated and involves irreducible $\bC$-representations of the
{\em nonunital} $C^*$-algebra
of the functions which vanish at infinity\cite{Landi,FE-DO}.  Here we want to develop an alternative procedure, involving pure states,
 which is useful from
a metric point of view.
 We remind  the reader that a linear functional $\omega : {\cal A} \to \bC$ where ${\cal A}$ is a $C^*$-algebra, is said to be  {\bf positive} if
$\omega(a^*a)\geq 0$ for all $a\in {\cal A}$.  If ${\cal A}$ is unital, $\omega$ is said to be {\bf normalized}
if $\omega(\b1) = 1$, $\b1$ being the unit element of ${\cal A}$.  In  unital  $C^*$-algebras, the positivity
of a linear functional $\omega$   implies (a) the boundedness of $\omega$ and (b) $||\omega||= \omega(\bI)$
(see proof of theorem 5.1 in \cite{Simon}) so the normalization condition
can be equivalently stated by requiring that $||\omega||=1$.
A positive normalized linear functional on a unital $C^*$-algebra is called  ({\bf algebraic}) {\bf state}.
Concerning the GNS theorem we address the reader to
\cite{Simon}   (theorem 5.1 in \cite{Simon}) where a concise proof of that theorem is provided.
In particular we remark that, by the GNS theorem,
a positive normalized linear functional on a unital $C^*$-algebra is {\bf real}, i.e., $\overline{\omega(a)} = \omega(a^*)$,
and this implies that $\omega(a) \in \bR$ if $a$ is self-adjoint $a=a^*$.
A state is said to be  {\bf pure} when it is an extremal element of the convex set of states.
As is  known, a state is pure iff  it admits an irreducible  GNS representation \cite{Simon}.  \\

\noindent {\bf Proposition 4.5}.
{\em In our general algebraic hypotheses, let ${\cal S}_{I}$ denote the convex set of algebraic states 
$\lambda_{I}$ on $\overline{{\cal A}_{I}}$, $I\in {\cal X}$
and let ${\cal S}_{p{I}} \subset {\cal S}_{I}$ denote the subset of pure states.
Define the maps $J_{J,I} : {\cal S}_{I} \to {\cal S}_{J}$ with $J_{J,I} : \lambda_{I} \mapsto \lambda_{I}\circ \Pi_{{I}, {J}}$,
where $I,J\in {\cal X}$ and  $I\leq J$. Then,\\
{\bf (a)} $J_{{I},{I}} = Id$  and  $J_{{J},{I}}$ is injective if  $I \leq J$.\\
{\bf (b)} $J_{K,I} = J_{K,J} \circ J_{J,I}$ provided $I\leq J\leq K$.}\\

 \noindent {\em Proof.} Everything is a trivial consequence of Proposition 4.2 and its corollary. $\Box$ \\

 \noindent As ${\cal X}$ is a direct set and Proposition 4.4 holds, it is natural to consider the {\em inductive limit} of the spaces ${\cal S}_{I}$ with respect to the maps
$J_{I,J}$ and give the following definition. The definition of causal ordering $\unlhd$  given below is a direct generalization of  (c) in Proposition 2.4.  After the introduction of generalization of the notion of
event in terms of the notion of  {\em locus} (also given in the definition below),
it will be clear that  $\unlhd$ is nothing but a generalization of the causal partial
ordering on the spacetime.\\

\noindent {\bf Definition 4.1}.
{\em   In our general algebraic hypotheses and using the notation introduced above,\\
{\bf (1)} a {\bf locus} on $\gA$, $\Lambda$, is an element of the inductive limit of the class $\{{\cal S}_{I}\}_{{I}\in {\cal X}}$,
with respect to the class of maps
$\{J_{{I},{J}}\}$.
That is, $\Lambda$ is an equivalence class of  states in $\cup_{I\in {\cal X}} {\cal S}_{I}$ with respect to the equivalence relation
\begin{eqnarray}
\lambda_{I} \sim \lambda_{J}    \:\:\:\:\:\mbox{iff there is $K\geq I,J$ in ${\cal X}$ with}\:\:\:\: J_{{K,{I}}}(\lambda_{I})
=J_{{{K},{J}}}(\lambda_{J})     \label{equivalence}\:;
\end{eqnarray}
${\gL}$   denotes the space of loci , i.e., the inductive limit of the class $\{{\cal S}_{I}\}_{{I}\in {\cal X}}$.\\
{\bf (2)} $\Lambda \in \gL$ is said to be  {\bf pointwise} iff there is some pure state $\lambda_{{I}_0}\in \Lambda$.  $\gL_p\subset \gL$ indicates the space of pointwise loci on $\gA$;\\
{\bf (3)} $\Lambda \in \gL$ is said to  {\bf belong} to $I\in {\cal X}$, and we write $\Lambda \ine I$, iff  ${\cal S}_I \cap \Lambda \neq \emptyset$. In that case
we define   $\Lambda(f) := \lambda_I(f)$ for every $f\in \overline{{\cal A}_I}$ and $\lambda_I \in \Lambda\cap {\cal S}_I$;\\
{\bf (4)} For $\Lambda, \Lambda' \in \gL$, we say that $\Lambda'$ {\bf causally follows} $\Lambda$, and we write $\Lambda \unlhd\Lambda'$, iff $\Lambda(f) \leq \Lambda'(f)$
for every $I\in {\cal X}$ with $\Lambda,\Lambda' \ine I$ and every $f\in Co_I$.}\\

\noindent {\bf Remark}. Definition 4.1 is consistent, i.e.  the equivalence relation preserves positivity and normalization.
Indeed, for  $I\leq K$, $\lambda_{I}$ is respectively positive/normalized iff
$J_{{K},{I}}(\lambda_{I})$
is respectively such.  We leave the trivial proof to the reader, based on the fact that $ \Pi_{{I},{K}}$    is a homomorphism
of unital $C^*$-algebras. The well-definedness of $\Lambda(f)$ is proven in (c) below.   \\

\noindent {\bf Proposition 4.6}.
{\em  In our general algebraic hypotheses and using the notation introduced above, assuming  $\Lambda,\Lambda',\Lambda''\in \gL$ and
$I,J,K\in {\cal X}$, we have the following statements.\\
{\bf (a)} If $\Lambda\ine I$, there is only one   $\lambda_{I} \in \Lambda\cap {\cal Lambda}_I$. \\
{\bf (b)}  $\Lambda \ine I$ and $I\leq J$
 entail  $\Lambda \ine J$. In that case
$\lambda_{J}:=J_{J,I}(\lambda_{I}) \in \Lambda \cap {\cal S}_{J}$ for $\lambda_I\in \Lambda \cap {\cal S}_I$
and $\Lambda(f) = \Lambda(\Pi_{I,J}(f))$ as $f\in \overline{{\cal A}_J}$.\\
{\bf (c)} $\Lambda \in \gL_p$ iff  every $\lambda_{I} \in \Lambda$ is a pure state.
Hence $\gL_p$ is the inductive limit of the class $\{{\cal S}_{p{I}}\}_{{I}\in {\cal X}}$ with respect to the class of the above-defined
maps $\{J_{J,I}\}$, $I\leq J$ in ${\cal X}$.\\
{\bf (d)} $\Lambda \unlhd \Lambda'$, $\Lambda'\unlhd \Lambda''$ imply that $\Lambda(f) \leq \Lambda''(f)$
for every $f\in Co_I$  such that $\Lambda,\Lambda',\Lambda''\ine I$.}\\

 \noindent {\em Proof.} {\bf (a)}  The thesis is a direct consequence of the injectivity of the maps $J_{I,K}$. {\bf (b)}
$\lambda_{J} \sim \lambda_{I}$  by construction. The remaining part  is a direct consequence of the given definitions.
Let us pass to prove {\bf (c)}.  If every $\lambda_{I}\in \Lambda$ is pure, $\Lambda \in \gL_p$ by definition, so consider
the other case. Suppose there is a pure state $\lambda_{I}\in \Lambda$, we want to show that all the remaining states $\lambda_{J}\in \Lambda$
 are pure too. By definition of locus there must be $K\in {\cal X}$ with $I,J\leq K$
and $\lambda_{J}\circ \Pi_{J,K} = \lambda_{{I}}\circ \Pi_{{I},K} =: \lambda_{K} $.
GNS theorem (theorem 5.1 in \cite{Simon}) and the surjectivity
 of $\Pi_{{I},K}$ imply that if   $\langle H, \pi, \Omega \rangle$ is a GNS triple for $\overline{{\cal A}_{I}}$ associated to $\lambda_{I}$,
 $\langle H,\pi\circ\Pi_{{I},{K}}, \Omega \rangle$ is a
GNS triple for $\overline{{\cal A}_{K}}$ associated to $\lambda_{K}$, and $\pi\circ\Pi_{{I},K}$ is irreducible iff
$\pi$ is irreducible. Similarly   if   $\langle H', \pi', \Omega' \rangle$ is a GNS triple for $\overline{{\cal A}_{J}}$
associated to $\lambda_{J}$,
 $\langle H',\pi'\circ\Pi_{J,K}, \Omega' \rangle$ is another
GNS triple for the same algebra $\overline{{\cal A}_{K}}$ associated to the same state $\lambda_{K}$ and $\pi'\circ\Pi_{{I},K}$ is irreducible iff
$\pi'$ is irreducible. Since (by GNS theorem) all GNS triples for an algebra ($\overline{{\cal A}_{K}}$) referred to a state ($\lambda_{K}$)
 are unitarily equivalent and the irreducibility is unitarily
invariant, we conclude that $\pi$ is irreducible iff $\pi'$ is irreducible. This is the thesis.
The proof of {\bf (d)} is immediate by the given definitions and the item (b). $\Box$\\

\noindent The relationship between points on $M$ and pointwise loci is established by the following theorem which does not require  either
the spacetime structure or a differentiable manifold structure. The only requirement is that $M$ is a Hausdorff locally-compact topological space.
More generally, the theorem shows that there is a bijection  between  loci on $M$ and compactly-supported
regular  Borel probability measures $\mu$ with compact support on $M$.
Such a bijective function reduces to a homeomorphism when restricted to the space of  pointwise loci equipped  with a suitable topology.
We remind  the reader that the support of a regular Borel measure is the complement of the largest open  set with measure zero.
Below  $\int_M f\: d\mu$ is well defined by posing $f\equiv 0$ outside $\overline{J}$ since $\mbox{supp}(\mu) \subset \overline{J}$.\\

\noindent {\bf Theorem 4.1}.
{\em Let  $M$ be a locally-compact  Hausdorff topological space and ${\cal X}$ a covering of  $M$
made of open relatively compact subsets  and defining  a direct set with respect to the set-inclusion relation.
Define  $\gA :=\{\overline{{\cal A}_I}\}_{I\in {\cal X}}$ with $\overline{{\cal A}_I} := C(\overline{I})$,
${\cal S}_I$, $\gL$ and $\gL_p$ as done in Def. 4.1, $\Pi_{I,J}(a) := a\spa\rest_{\overline{I}}$  and  $J_{J,I}$ as  in prop.4.3.
Finally, denote   the space of compact-support  regular Borel probability measures
on $M$ by $\gP$.
Consider the map  $F : \gP \to \gL$, such that  for $\mu\in \gP$,
$$F(\mu) := \left\{\lambda^{(\mu)}_{J} \in {\cal S}_{J} \:\left|\:   J\in {\cal X}  \:\:\mbox{with}\:\:
 \mbox{supp}(\mu) \subset \overline{J}, \:\:  \lambda_{J}^{(\mu)}(f) := \int_M f\: d\mu \:\: \mbox{for}\:\: f\in \overline{{\cal A}_{J}}\right\} \right.\:.$$
{\bf (a)} $F$ is well-defined, i.e.,  $F(\mu)$ is a locus for every $\mu\in \gP$. Moreover  $F(\mu) \ine I\in {\cal X}$ 
iff  $\mbox{supp}(\mu)\subset \overline{I}$.\\
{\bf (b)}  $F$ is bijective onto the set of the loci $\gL$.  \\
{\bf (c)}  $F$ restricted to the space of Dirac measures $\{\delta_x\}_{x\in M}$ gives rise to
 a homeomorphism from $M$  onto $\gL_p$  equipped with the inductive-limit topology,
 every ${\cal S}_{{I}}$, $I\in {\cal X}$, being endowed with Gel'fand's topology.}\\

 \noindent {\em Proof.} See the Appendix B. $\Box$.\\

\noindent The following theorem proves that the relation $\unlhd$ among loci  is nothing but a generalization of the causal partial
ordering on the spacetime.  \\

\noindent{\bf Theorem 4.2}.
{\em  In the hypotheses of  theorem 4.1, also assume that $(M, {\bf g}, {\cal O}_t)$ is
 a globally hyperbolic spacetime and ${\cal X}$ is defined as in Proposition 2.4,
$\gC = \{Co_I\}_{I\in {\cal X}}$ with $Co_I := T_{[{\bf g}]}(I)$.
Consider the relation $\unlhd$ defined in $\gL$ by Def. 4.1. and $\Lambda,\Lambda',\Lambda'' \in \gL$, then \\
 {\bf (a)} $\Lambda \unlhd\Lambda'$ and $\Lambda' \unlhd \Lambda$ together entail $\Lambda =\Lambda'$;\\
 {\bf (b)} if $\Lambda\unlhd\Lambda'$ and $\Lambda'\unlhd\Lambda''$ then, $\Lambda,\Lambda'' \ine I \in {\cal X}$ entails
 $\Lambda'\ine I$ and thus $\unlhd$ is transitive and defines a partial ordering relation on $\gL$;\\
 {\bf (c)} if $F$ is that in Theorem 4.1, for every  pair $x,y\in M$, $F(\delta_x)\unlhd F(\delta_y)$ iff $x\preceq y$.}\\

\noindent {\em Proof.} See the Appendix B. $\Box$.\\

\noindent Actually most of the content of Theorem 4.2 can be generalized  using  the general algebraic hypotheses as
well as a further {\bf causal convexity} axiom:\\

\noindent {\bf (AH5)}  {\em For $\Lambda,\Lambda',\Lambda'' \in \gL$,   if
 $\Lambda \unlhd \Lambda'$, $\Lambda'\unlhd\Lambda''$ and  $\Lambda,\Lambda'' \ine I \in {\cal X}$, then
 $\Lambda'\ine I$.}\\

\noindent Notice that (AH5)  is fulfilled in the globally-hyperbolic-spacetime case by (b) of Theorem 4.2.\\

\noindent{\bf Theorem 4.3}.
{\em  In the general algebraic hypotheses, including the causal convexity axiom (AH5), and employing notations  above,
 $\unlhd$ is a partial-ordering relation in $\gL$.}\\

 \noindent {\em Proof.}  $\Lambda\unlhd \Lambda$ is a trivial consequence of the  definition of $\unlhd$. The fact that
 $\Lambda \unlhd\Lambda'$ and  $\Lambda' \unlhd \Lambda$   together
 entail $\Lambda = \Lambda'$ can be proven as in Theorem 4.2  where we have not used the spacetime structure.
 The transitivity of $\unlhd$ follows from (AH5) and (d) of Proposition 4.6. $\Box$  \\

\noindent {\bf 4.3}. {\em Lorentzian distance.}  We conclude by presenting a generalization of the Lorentzian distance in the general case.    The following definition is
very natural and can also be used in the generalized commutative case in the hypotheses of Theorem 4.2 concerning compactly
supported probability
measures on a spacetime. Notice that the definition makes sense by (AH3) and (AH4) which assure the existence of some function satisfying
$[t, [t, {\bf G}_{I}]] \leq -I$ below. \\

\noindent {\bf Definition 4.2}. {\em  In the general algebraic hypotheses including the causal convexity axiom (AH5) and employing notations  and
conventions above, the {\bf Lorentzian distance} of $\Lambda,\Lambda' \in \gL$ is
 \begin{eqnarray}
{\bf D}(\Lambda,\Lambda') =
 \inf \left\{ \langle \Lambda'(t) -\Lambda(t)\rangle \left|\:  t\in Co_{I},\:\:\: \Lambda,\Lambda'\ine I \in {\cal X},
 \:\:\: \left[ t, \left[ t, {\bf G}_{I}\right]\right] \leq -I
 \right.\right\},
\label{fineF}
\end{eqnarray}
where  $\langle \alpha\rangle := \max\{0,\alpha\}$ if $\alpha \in \bR$.}\\

\noindent The item  (iii) of (a) in (AH4) implies the following result,
the proof being the same given for the corresponding part of Theorem 3.1.\\

\noindent {\bf Proposition  4.7}.
{\em  In definition 4.2 the condition $[t,[ t, {\bf G}_{I}]] \leq -I$
can be replaced by one of the three 
following  conditions:
\begin{align}
 \sigma(\overline{[t,[ t, {\bf G}_{I}]]}) &\subset (-\infty, -1]\:,\\ 
\overline{[t,[ t, {\bf G}_{I}]]} &\leq -I \label{D2}\:,\\
\overline{[t,[ t, {\bf G}_{I}]]}^{-1} &\:\:\:\mbox{exists and}\:\:\:\: \left|\left|\overline{[t,[ t, {\bf G}_{I}]]}^{-1}
\right|\right|_{{\bf L}({\cal H}_I)}\leq 1\:.
\end{align}}
 We have  a conclusive theorem.\\

\noindent {\bf Theorem 4.4}.
{\em In the general algebraic hypotheses including the causal convexity axiom (AH5), employing notations  and
conventions above the Lorentzian distance enjoys the following properties for $\Lambda,\Lambda', \Lambda''\in \gL$.\\
{\bf (a)}  In the hypotheses of Theorem 4.2 and assuming ${\bf G}_I = {\sdelta}_I$ (defined in Theorem 3.1 for  $I\in {\cal X}$),
${\bf D}(F(\delta_p),F(\delta_q)) = {\bf d}(p,q)$ for every pair $p,q\in M$.\\
{\bf (b)}  $0 \leq {\bf D}(\Lambda,\Lambda') < +\infty$. In particular, ${\bf D}(\Lambda,\Lambda')= 0$ if either $\Lambda=\Lambda'$ or $\Lambda \not \spa\unlhd \: \Lambda'$.\\
{\bf (c)}  If $\Lambda \unlhd\Lambda' \unlhd \Lambda''$ then ${\bf D}(\Lambda,\Lambda'') \geq {\bf D}(\Lambda,\Lambda') + {\bf D}(\Lambda',\Lambda'')$.}\\

\noindent {\em Proof}. {\bf (a)} The right-hand side of  the definition of ${\bf D}(F(\delta_p),F(\delta_q))$ in (\ref{fineF}) can be re-written as the right-hand side
of (\ref{fine1}) in Theorem 3.1.  So the proof of the thesis is obvious.
{\bf (b)} The set in the right-hand side of (\ref{fineF}) is not empty because, if  $\Lambda\in \gL$, there is some $I\in {\cal X}$ with
$\Lambda \ine \gL$  by definition of locus, moreover (AH4) implies that there is some $f\in Co_{{\bf G}_I} \neq \emptyset$ and thus    $t=\alpha f \in Co_I$ and
$\left[ t, \left[ t, {\bf G}_{I_t}\right]\right] \leq -I$ for some $\alpha>0$.
 Then positivity and boundedness  of ${\bf D}$ hold by definition. $\Lambda=\Lambda'$ implies ${\bf D}(\Lambda,\Lambda')= 0$  by the definition of
 ${\bf D}$.
  Finally  suppose $\Lambda {\not \spa\unlhd} \: \Lambda'$.
 In that case there must exists $f\in Co_I$ for some $I\in {\cal X}$ such that
$\Lambda,\Lambda'\ine I$ and  $\Lambda(f) - \Lambda'(f)= \epsilon >0$.  Define $f_\nu := \nu f$, $f_\nu \in Co_I$ for all $\nu >0$ because $Co_I$ is a
convex cone and $\Lambda(f_\nu) - \Lambda'(f_\nu)= \nu \epsilon$. Then
Take  $t_\gamma\in Co_{{\bf G}_I}$ (which exists by (AH4)) with  $\gamma>0$ such that
$[t_\gamma, [t_\gamma, {\bf G}_I]] \leq -\gamma I$.
Therefore,   by (AH4) (and (iv), (v)  of (a) in Proposition 4.4 and its corollary
in particular),
$t_\nu := f_\nu + (1/\sqrt{\gamma}) t_\gamma$ is in $Co_I$ as before and  satisfies
$[t_\nu, [t_\nu, {\bf G}_I]] \leq -I$. Finally
$ \Lambda'(t_\nu) -\Lambda(t_\nu) = -\nu\epsilon + (1/\sqrt{\gamma}) ( \Lambda'(t_\gamma) -\Lambda(t_\gamma)) <0$ if $\nu>0$
is sufficiently large.   Then the definition of ${\bf D}(\Lambda,\Lambda')$  gives  ${\bf D}(\Lambda,\Lambda')=0$.
{\bf (c)} Take $I\in {\cal X}$ with $\Lambda,\Lambda''\ine I$. $\Lambda \unlhd\Lambda' \unlhd \Lambda''$ and (AH5) entail
$\Lambda' \ine I$ and thus $\Lambda''(f) - \Lambda(f) = (\Lambda''(f) - \Lambda'(f)) + (\Lambda'(f) - \Lambda(f))$ makes sense. In the given hypotheses, by (b),
the identity can also be written $\langle\Lambda''(f) - \Lambda(f) \rangle = \langle\Lambda''(f) - \Lambda'(f)\rangle + \langle\Lambda'(f) - \Lambda(f)\rangle$.
Using the definition of ${\bf D}$ it entails $\langle\Lambda''(f) - \Lambda(f) \rangle \geq {\bf D}(\Lambda,\Lambda') + {\bf D}(\Lambda',\Lambda'')$.
Finally, since $f$ is arbitrary, it implies the thesis. $\Box$ \\

\noindent It is possible to define relations analogous  to $\past$ and $\prec$ respectively, which we  denote by
 $\lhd\sp\lhd$ and $\lhd$. $\Lambda {\lhd\sp\lhd} \Lambda'$ means ${\bf D}(\Lambda,\Lambda') >0$,
 and   $\Lambda \lhd \Lambda'$  means  $\Lambda \unlhd \Lambda'$ and
 $\Lambda \neq \Lambda'$ together. The final corollary shows that the content of A.7 can be restated
 in the general context without using causal path. \\

\noindent{\bf Corollary}. {\em In the hypotheses of Theorem 4.4 and with the given definitions: \\
{\bf (a)} ${\lhd\sp\lhd}$ and $\lhd$ are transitive and $\Lambda {\lhd\sp\lhd} \Lambda'$ implies $\Lambda \lhd \Lambda'$\:;\\
{\bf (b)}   either $\Lambda  {\lhd\sp\lhd} \Lambda'$ and $\Lambda'  {\unlhd} \Lambda''$,
 or $\Lambda  {\unlhd} \Lambda'$ and $\Lambda'  {\lhd\sp\lhd} \Lambda''$   implies $\Lambda {\lhd\sp\lhd} \Lambda''$.}\\

 \noindent {\em Proof}. {\bf (a)}  By (b) of Theorem 4.4,
 $\Lambda {\lhd\sp\lhd} \Lambda'$  entails $\Lambda \unlhd \Lambda'$ and $\Lambda \lhd \Lambda'$, hence, by (c), ${\lhd\sp\lhd}$ is transitive.
    $\lhd$ is transitive too because of Theorem 4.3 and  the definition of $\unlhd$. {\bf (b)} is a direct consequence of (c) in Theorem 4.4 and
    ${\bf D}\geq 0$. $\Box$

\section{Open issues and outlook.}

This paper shows that a generalization of part of the noncommutative Connes' program is possible in order to
encompass Lorentzian and causal structures of (globally hyperbolic) spacetimes.   However several relevant issues
remain open.   Obviously,  first of all concrete models of the  presented generalized formalism   should be presented
in the non-commutative case, moreover the minimality of the proposed axioms should be analyzed.
An important point which should be investigated is  the interplay between the topology  of the space of loci
and  ${\bf D}$. In the  commutative case and considering the events of a globally hyperbolic spacetime,
${\bf d}$ turns out to be continuous with respect to the topology of the manifold. Presumably a natural
topology of the space of  loci, in the general case, could be the inductive limit topology,
each space ${\cal S}_I$ being equipped with the $*$-weak topology. One expects that ${\bf D}$ is
continuous with respect such a topology. Another point is the following. We have focused attention on the Lorentzian generalization
of (\ref{d}) avoiding to tackle difficulties
involved in possible generalizations of (\ref{tr}) which, presumably,  should require a careful analysis of the spectral properties
of the metric operators ${\bf G}_I$ introduced above. Such an analysis could reveals contact points with the content of \cite{Strohmaier01}
in spite of the evident differences of the presented approach and obtained results. Another important question which should be 
investigated concerns possible physical applications of the presented mathematical structure.   \\

\noindent {\bf Acknowledgments}. I am grateful to A.Cassa, S.Delladio,  G.Landi,  M.Luminati  and F.Serra Cassano
for useful mathematical suggestions and comments.  I would like to thank R.M.Wald for a short but useful technical discussion and
H.Thaler who pointed out ref \cite{Strohmaier01} to me.\\

 \section*{Appendix A. Exponential map, Synge's world function, Spacetimes.}
 
{\bf A.0}. 
{\em  Exponential map, Synge's world function}.
Let  $(M,{\bf g})$ be a smooth Riemannian or Lorentzian manifold. $\pi : TM \to M$  denotes the natural
projection of $TM$ onto $M$ and, if $v\in TM$, $v_{\pi(v)}$ is the vector of $T_{\pi(v)}M$ associated to $v$.
If $v\in TM$ and $\lambda \in \bR$, $\lambda v$ is the element of $TM$ with $\pi(\lambda v)= v$ and $(\lambda v)_{\pi(\lambda v)}
= \lambda v_{\pi(v)}$.\\
Consider the map $(t,v) \mapsto {\gamma}(t,v)\in M$, where
$\gamma(.,v)$ is the unique geodesic  which starts from
$\pi(v)$ at $t=0$ with initial tangent vector $v_{\pi(v)}$ and $t$ belongs to the
{\em maximal domain}  $(a_v,b_v)$ ($a_v<0<b_v$). From known theorems on maximal solutions of
(first order) differential equations on manifolds ($TM$) \cite{O'Neill}, the domain of $\gamma$, $\cup_{v\in TM} (a_v,b_v)\times\{v\}$ is
open in $\bR\times TM$
and  $\gamma$ is smooth therein. Then pick out the set $U\subset TM $ of elements  $v$, such that
$1\in (0,b_v)$. It is possible to show  that $U$ is open.
Notice that for each $v \in TM$,  there is a sufficiently small $\lambda>0$ such that $1\in (0,b_{\lambda v})$,
because of the identity $\gamma(\lambda t,v)= \gamma(t, \lambda v)$.
From that identity one trivially proves that $U$ is  {\bf starshaped}, i.e., if $v\in U$ then $\lambda v \in U$ for $\lambda \in [0,1]$.
The {\bf exponential map}, $\exp : U \to M$, is defined as $\exp(v) := \gamma(1,v)$ \cite{KN}. Notice that
$\exp \in C^\infty(U)$.
If $p\in M$, $\exp_p$ denotes $\exp\sp \rest_{T_pM}$ and the open neighborhood of $0$, $U_p:= \{v\in U\:|\: \pi(v)=p\} \subset T_pM$, its natural domain.
By direct inspection, one finds that $d\exp |_{v} \neq 0$ if $v$ belongs to the zero section of $TM$. This entails
that if  one shrinks each $U_p$ sufficiently about $p$ to some starshaped and open neighborhood of $p$, $V_p\subset U_p$,
 $\exp_p\spa\rest_{V_p}$ defines  a diffeomorphism from  $V_p$ to $\exp_p(V_p)$ which is  open too.
 If $\{e_\alpha|_p\}\subset T_pM$ is a basis,
  $(t^1,\ldots t^D) \mapsto exp_p(t^\alpha e_\alpha|_p)$, $t = t^\alpha e_\alpha|_p\in V_p$, defines a {\bf  normal
  coordinate system centred on $p$}.
 An open set $C\subset M$
 is called a ({\bf geodesically}) {\bf convex normal}  neighborhood
 if there is an open and {\em starshaped} set $W \subset TM$, with $\pi(W) = C$ such that $exp\sp\rest_W$ is a diffeomorphism
onto $C\times C$. It is clear that $C$ is connected and  there is only one geodesic segment
 joining any pair $q,q'\in C$ which is {\em completely contained in $C$}, that is  $t\mapsto exp_q(t ((exp_q)^{-1}q'))$
 $t\in [0,1]$.  It is possible to take $C$ diffeomorphic to an open ball in $\bR^{\mbox{dim}M}$ \cite{KN}.
 Moreover if $q\in C$, $\{e_\alpha|_q\}\subset T_qM$ is a basis,
  $(t^1,\ldots, t^D) \mapsto exp_q(t^\alpha e_\alpha|_q)$, $t = t^\alpha e_\alpha|_q  \in W_q$ defines a {\em global} normal
  coordinate system onto $C$ centred on $q$. The class of the convex normal neighborhood of a point $p\in M$ is not empty and
   defines a fundamental system of
 neighborhoods of $p$ \cite{KN,doCarmo,O'Neill,Friedlander}.\\
 In $(M,{\bf g)}$ as above, $\sigma(x,y)$ indicates one half the squared geodesic distance of $x$ from $y$, also known as
 {\bf Synge's world function}:
 $\sigma(x,y) := \frac{1}{2}{\bf g}_x(exp_x^{-1}y,exp_x^{-1}y)$ \cite{Friedlander}.  By definition
 $\sigma(x,y)=\sigma(y,x)$ and $\sigma$ turns out to be  smoothly defined on $C\times C$ if $C$ is a convex normal neighborhood.
 With the signature $(-,+,\cdots,+)$, we have  $\sigma(x,y) > 0$ if the events are space-like separated,
 $\sigma(x,y)<0$ if the events are time related and $\sigma(x,y)=0$ if the events belong to a common null geodesic  or $x=y$.  All that and everything follows
 also holds in  manifolds endowed with an Euclidean metric where $\sigma$ (defined as above)  is everywhere nonnegative.
It turns out   that  \cite{Friedlander} if $\gamma$ is the unique geodesic from $p$ to $q$ in a convex normal neighborhood containing
$p,q$, with affine parameter $\lambda\in [0,l]$
\begin{align}
&\uparrow \spa d_y\sigma(x,y)|_{y=\gamma(\lambda)} = \lambda \dot{\gamma}(\lambda) \label{AA}\:,\\
&2\sigma(x,y) = {\bf g}_x(d_x \sigma(x,y), d_x \sigma(x,y))= {\bf g}_y(d_y \sigma(x,y), d_y \sigma(x,y)) \label{dsds}\:.
\end{align}
 {\bf A.1}. {\em Lorentzian manifold}. A (smooth) {\bf Lorentzian manifold} $(M,{\bf g})$ is a  $n\geq 2$-dimensional
smooth manifold $M$ with a smooth Lorentzian metric ${\bf g}$ (with signature $(-,+,\cdots,+$)). \\
{\bf A.2}.  {\em Signature of vectors}.  A vector
$T\in T_xM$, $T\neq 0$, is said to be {\bf space-like}, {\bf time-like} or {\bf null}
if, respectively, ${\bf g}_x(T,T)>0$,  ${\bf g}_x(T,T)<0$, ${\bf g}_x(T,T)=0$.   $T\neq 0$ is said
to be  {\bf causal} if it is either time-like or null.
The same nomenclature is used for
co-vectors $\omega \in T^*_xM$ referring to $\uparrow\spa \omega \in T_xM$, where ${\bf g}_x(\uparrow\spa \omega, \:\cdot\:) = \omega$.
If $T\in T_pM$, $|T| := \sqrt{|{\bf g}_p(T,T)|}$, similarly, if $\omega \in T^*_pM$,   $|\omega| := \sqrt{|{\bf g}_p(\uparrow\spa \omega, \uparrow\spa
 \omega)|}$.\\
{\bf A.3}.  {\em Time orientation}.
A  Lorentzian manifold $(M,{\bf g})$ is said to be  {\bf time orientable} if it admits a smooth non vanishing vector field $Z\in TM$ which is
everywhere time-like. A {\bf time orientation}, ${\cal O}_t$, on a time-orientable Lorentz manifold, $(M, {\bf g})$, is one of the two
equivalence classes of smooth time-like vector fields $Z$ with respect to the equivalence relation
$Z\sim Z'$ iff ${\bf g}(Z,Z') <0$ everywhere.
For each point $p\in M$, an orientations determines an analogous equivalence  class of time-like vectors of $T_pM$, ${\cal O}_{tp}$.
In a orientable Lorentz manifold, to assign a time orientation it is sufficient to single out a timelike
vector in  $T_pM$ for a $p\in M$.   With the given definitions, a causal vector (co-vector )
 $T\in T_pM$ ($\omega\in T^*_pM$) is said to be {\bf future directed} if
${\bf g}_{p}(Z(p),X) <0$ (${\bf g}_{p}(Z(p),\uparrow\spa\omega) <0$).
A causal vector (resp. covector )
 $T\in T_pM$ ($\omega\in T^*_pM$) is said to be {\bf past directed} if 
${\bf g}_{p}(Z(p),X) <0$ (${\bf g}_{p}(Z(p),\uparrow\spa\omega) >0$)\\
{\bf A.4}.  {\em Spacetime}.
  A {\bf spacetime} $(M, {\bf g}, {\cal O}_t)$ is a  Lorentzian manifold $(M, {\bf g})$ which is time orientable and  equipped with
 a time orientation ${\cal O}_t$; the points of $M$  are also called {\bf events}. \\
{\bf A.5}.  {\em Regularity of  curves and causal curves}.  
In a spacetime $M$,   a {\bf piecewise $C^k$}  {\bf curve} defined in  a (open, closed, semi-closed) non-empty interval in
$\bR$, $I$, is a continuous map   $\gamma: I \to M$ with  a finite partition  of $I$  such that each subcurve obtained by
restricting $\gamma$ to each subinterval of the partition ({\em including its boundary}) is $C^k$.
If the partition coincides with $I$ it-self, the curve is said to be  $C^k$.
A piecewise $C^1$ curve $\gamma$ is said to be 
 {\bf time-like}, {\bf space-like}, {\bf null}, {\bf causal} if its tangent vector  $\dot\gamma$  is
respectively  time-like, space-like, null, causal, {\em everywhere} in each subinterval $I$ of the associated  partition.
A piecewise $C^1$ causal curve in a spacetime $\gamma: I \to M$
is said to be  {\bf  future} ({\bf past}) {\bf directed} if
if its tangent vector  $\dot\gamma$  is {\bf  future} ({\bf past}) {\bf directed}  {\em everywhere} in each subinterval $I$ of the associated  partition.
In a spacetime $M$, if $p,q \in M$,  a  curve $\gamma : [a,b] \to M$ is said to be {\bf from $p$ to $q$} if
$\gamma(a)=p$ and $\gamma(b)=q$.    \\
{\bf A.6}. {\em Continuous causal curves}.
It is possible to extend the notion of causal  future directed curves, considering
{\bf continuous future-directed causal curves} $\gamma : I \to M$. That is by requiring that, for each $t\in I$ there is a neighborhood of $t$, $I_t$ and
a  convex normal neighborhood of $\gamma(t)$, $U_t$, such that, for $t'\in I_t\setminus \{t\}$, one has $\gamma(t')\neq \gamma(t)$ and
there is a  future-directed causal (smooth) geodesic segment  $\gamma' \subset U_t$  from $\gamma(t)$ to $\gamma(t')$
if $t'>t$
there is a  future-directed causal (smooth) geodesic segment $\gamma' \subset U_t$  from $\gamma(t')$ to $\gamma(t)$
if $t'<t$.       Similar definitions hold concerning  {\bf continuous future-directed timelike curves}, by replacing
``causal'' with ``timelike'' in the definitions above.
In this work a causal curve is supposed to be a continuous causal curve, moreover  continuous curves $\gamma :I\to M$
and $\gamma' : I' \to M$ are identified if there is an increasing  homomorphism $h : I\to I'$
and $\gamma'\circ h =\gamma$.\\
{\bf A.7}. {\em Causal relations of events}.
In a spacetime $(M, {\bf g}, {\cal O}_t)$, if $p,q\in M$, \\
    (i) $p\preceq q$ means that either $p=q$ or
 there is a  future-directed causal  curve from $p$ to $q$, \\
    (ii) $p\prec q$   means that $p\preceq q$
 and $p\neq q$,   \\
 (iii) $p\past q$ means that there is a  future-directed time-like curve from $p$ to $q$. \\
$\past$ and $\preceq$ are clearly transitive.\\
 {\bf Remark}. In a spacetime $(M,{\bf g},{\cal O}_t)$, if $p,q,r\in M$,
$p\past q$ and $q\preceq r$ entail $p\past r$, and similarly $p\preceq q$ and $q\past r$ entail
$p\past r$  \cite{Penrose72}. \\
\noindent {\bf A.8}. {\em Causal sets}.
  We make use the following notations.  Consider a spacetime $(M, {\bf g}, {\cal O}_t)$ and  $S\subset M$, then
\begin{align}
J^{+}(S) := &\{ q\in M\:|\: p \preceq q \mbox{ for some $p\in S$}\}   \mbox{  the {\bf causal future} of $S$ } \nonumber \:,   \\
J^{+}(S) := &\{ q\in M\:|\: q \preceq p \mbox{ for some $p\in S$}\}   \mbox{  the {\bf causal past} of $S$ }\nonumber\:,\\
I^{+}(S)  :=  &\{ q\in M\:|\: p \past q \mbox{ for some $p\in S$}\}  \mbox{  the {\bf chronological future} of $S$ } \nonumber\:,   \\
I^{-}(S)  :=  &\{ q\in M\:|\: q \past p \mbox{ for some $p\in S$}\}   \mbox{  the {\bf chronological past} of $S$ }  \nonumber \:.
\end{align}
Moreover $I(p,q) := I^+(p) \cap I^-(q)$ (which is not empty iff $p \past q$) and   $J(p,q) := J^+(p) \cap J^-(q)$.
If $\emptyset \neq S\subset M$,  $I^+(S)$ and $I^-(S)$ are  open,
$S\subset J^{\pm}(S) \subset \overline{I^\pm(S)}$,
$I^{\pm}(S) =  \mbox{Int}(J^\pm(S))$  \cite{O'Neill}.\\
{\bf A.9}. {\em Properties of $I^\pm(p)$ and $J^\pm(p)$}.
 (Theorem 8.1.2 in \cite{Wald84}.) In a spacetime $(M,{\bf g},{\cal O}_t)$, taking  a sufficiently small normal convex neighborhood of  $p\in M$, $U$,
$exp_p^{-1}$ defines a local diffeomorphism, $\phi: U \to \bR^n$ with $\phi(p)= 0$,
and $\phi(U\cap I^{\pm}(p))= B \cap C$, where $B\subset \bR^n$ is an open ball centred in
$0$ and $C$ the open convex cone, with vertex $0$, made of all the future directed timelike vectors. This result implies that  both $I^\pm(p)$ and $J^\pm(p)$ are nonempty, connected by paths and connected. \\
{\bf A.10}. {\em Causal relations of events again}.
$p,q\in M$ are said to be  {\bf time related}, if either $I^+(p) \cap I^-(q)\neq \emptyset $ or  $I^-(p) \cap I^+(q) \neq \emptyset$,
 {\bf causally related} if either $J^+(p) \cap J^-(q) \neq \emptyset$ or $J^-(p) \cap J^+(q)\neq \emptyset$.
  Causally related events $p,q\in M$, $p\neq q$, which are not time related are called {\bf null-related}.
$S,S' \subset M$ are said to be {\bf spatially separated} if  $(J^+(S) \cup J^-(S)) \cap S' =\emptyset$
(which is equivalent to $(J^+(S') \cup J^-(S')) \cap S =\emptyset$).\\
{\bf A.11}. {\em Causally convex sets, strongly causal spacetimes, Alexandrov topology}.
 In a spacetime $M$,  we say that a set
 $S\subset M$ is  {\bf causally convex}  when
 $J(p,q) \subset S$ if $p,q\in S$.
 It can be proven that an open set $U\subset M$ is causally convex iff
 for any future-directed causal  curve $\gamma$ and any choice of  (continuous) parametrization
 $\gamma^{-1}(U)$ is  open  and connected in $\bR$. The transitivity of $\preceq$ implies that
  $J^+(S)$, $J^-(S)$, $J(r,s)$
are causally convex  for $\emptyset \neq S \subset M$ and  $r \preceq s$. Also using  the remark in A.7 one directly shows that
 $I^+(S)$, $I^-(S)$, $I(r,s)$ are causally convex for $\emptyset \neq S \subset M$ and $r\past s$. A  spacetime
is {\bf strongly  causal} when every event admits a fundamental set of open neighborhoods
consisting of {\bf causally convex} sets.
It is known that a  spacetime $M$ is strongly causal  iff the {\bf Alexandrov topology}, i.e.,   that generated by all the
 sets $I(p,q)$, $p,q\in M$, is  the topology of $M$ \cite{Penrose72,BEE}.\\
{\bf A.12}.  {\em Globally hyperbolic spacetimes}.
A {\bf globally hyperbolic} spacetime (see the end of 8.3 in \cite{Wald84} about possible equivalent definitions)
is a strongly-causal spacetime $(M, {\bf g}, {\cal O}_t)$ such that every  $J(p,q)$ is either empty or compact  for each pair $p,q\in M$.\\
If the spacetime $M$ is globally hyperbolic and $S\subset M$ is compact,  $J^{\pm}(S) = \overline{I^\pm(S)}$ \cite{Wald84}
 and thus, using $I^\pm(S) = \mbox{Int}(\overline{S})$ (A.8), $J^\pm(S)\setminus I^\pm(S) = \partial I^\pm(S)= \partial J^\pm(S)$.   In particular
 $J^\pm(p) = \overline{I^\pm(p)}$.\\
{\bf A.13}. {\em Stably causal spacetimes, global time functions}.
A spacetime $(M, {\bf g}, {\cal O}_t)$ is said to be {\bf stably causal} if there is a smooth map $f: M\to \bR$
with $d f$ everywhere timelike (other equivalent definitions are possible \cite{BEE}).
A continuous map $t: M \to  \bR$ is said to be a {\bf global time function}  if strictly increases along every future-directed
causal curve.
A stably causal spacetime admits a global time function given by either  $+f$ or $-f$, $f$ being defined above.
 Remarkably, also the converse is true \cite{Hawking68,Seifert77}: If a spacetime admits a (global) time function, it admits
 a smooth map $f : M \to \bR$ with $d f$ everywhere time-like.\\
{\bf A.14}. {\em Causal spacetimes}.
A spacetime is said to be  {\bf causal} if there are no events $p,q$ such that $p\prec q\prec p$ (equivalently, it  does not contain any closed causal curve).
It is trivially proven that in a causal spacetime $\preceq$ is a {\em partial-ordering relation} in causal spacetimes.
A spacetime is called  {\bf chronological} if  there are no events $p,q$ such that $p\past q\past p$ (equivalently, it  does not contain any closed timelike curve).\\
{\bf A.15}. {\em Implications of causal conditions}.
 It is known that \cite{BEE,O'Neill,HE}\\
{\bf globally hyperbolic} $\Rightarrow$ {\bf stably causal} $\Rightarrow$ {\bf strongly causal} $\Rightarrow$ {\bf causal} $\Rightarrow$ {\bf chronological}.\\
In particular $\preceq$ is a partial-ordering relation in globally hyperbolic spacetimes too.\\
{\bf A.16}. {\em Inextendible curves}.
A  causal curve $\gamma : I \to M$ is said to be future (past) {\bf inextendible} if it  admits no future (past)
{\bf endpoint}, i.e., $e\in M$ such that, for every neighborhood $O$ of $e$,
there is $t'\in I$ with $\gamma(t)\in O$ for $t>t'$ ($t<t'$). Any causal curve which admits an endpoint can be extended beyond that endpoint
into a larger causal curve (only continuous in general). Hausdorff's maximality  theorem
implies that every (causal, timelike) curve can be extended up to a inextendible (causal, timelike) curve.\\
{\bf A.17}. {\em Cauchy developments}.  
Let $S\subset M$ be any set in the spacetime $(M,{\bf g}, {\cal O}_t)$, $D^+(S)$ ($D^-(S)$)
indicates the {\bf future} ({\bf past}) {\bf Cauchy development of} $S$, i.e.,
the set of points $p$ of the spacetime, such that every past (future) inextendible causal curve through $p$ intersects $S$.
(In particular $S\subset D^\pm(S)$.)
$D(S) := D^+(S)\cup D^-(S)$ is the  {\bf Cauchy development of} $S$.\\
{\bf A.18}. {\em Achronal and acausal sets}.
A set  $S\subset M$ is said to be  {\bf achronal} if  $S\cap I^\pm(S) =\emptyset$ and {\bf acausal} if
$S\cap J^\pm(S) =\emptyset$.   An achronal smooth spacelike embedded  submanifold with dimension  dim$(M)-1$ turns out to be also acausal
(\cite{O'Neill} p. 425).\\
{\bf A.19}. {\em Cauchy surfaces}.  
A {\bf Cauchy surface} (for $M$ it-self),  $S\subset M$, is a closed achronal set such that ${D}(S) = M$.
There are different,
also inequivalent definitions, of Cauchy surfaces, we use the definition of \cite{Wald84} which is equivalent to that
given in \cite{O'Neill} as stated in lemma 29 in chapter 14 therein.  \\
{\bf A.20}. {\em Globally hyperbolic spacetimes and Cauchy surfaces}.
An important results states that: {\em  a spacetime $(M, {\bf g}, {\cal O}_t)$ is globally hyperbolic iff it admits a
Cauchy surface}. This statement  can be adopted as an equivalent definition of a globally hyperbolic spacetime
(see remark in the end of 8.3 in \cite{Wald84} for a proof of  equivalence of the various definitions of globally hyperbolicity).\\
{\bf A.21}. {\em Cauchy surfaces and global time functions in globally hyperbolic spacetimes}.
All Cauchy surfaces of  a globally hyperbolic spacetime $M$  are connected and homeomorphic. $M$ it-self
is homeomorphic to $\bR\times S$, $S$ being a Cauchy surface of $M$ and the projection map from $M$ onto
$\bR$ can be fixed to be a smooth global time function \cite{Penrose72,BEE,Wald84,O'Neill}. \\
{\bf A.22}. {\em Smooth Cauchy surfaces}.
The existence of spacelike smooth Cauchy surfaces in any globally hyperbolic spacetime is 
a very subtle issue. A a first,  not complete,  proof of existence of smooth Cauchy surfaces in general globally hyperbolic spacetimes  is due to Dieckmann \cite{Dieckmann88}, however the complete proof, by  Bernal and Sanch\'ez, is much more recent  \cite{BS03}.
In (quantum) field theories, those are used to give  initial data for  hyperbolic field equations
determining the dynamics of the fields everywhere in the spacetime \cite{Wald84,Wald94}.\\
{\bf A.23}.  {\em Sets $I(S,p)$ and $J(S,p)$ and their properties}.
In a globally hyperbolic spacetime  $(M,{\bf g}, {\cal O}_t)$, if $S\subset M$ is a smooth Cauchy surface
and $p\in J^+(S)$, $I(S,p)$ and $J(S,p)$ respectively denote $I^{-}(p) \cap I^+(S)$ and  $J^{-}(p) \cap J^+(S)$.  One can straightforwardly  prove that 
 $I(s,p)$  is not empty iff $p\in I^+(p)$.  It is not so difficult to show that $I(S,p)$ and $J(S,p)$ are causally convex.
A.8 implies that $I(S,p)$ is open and  $I(S,p) \subset J(S,p)$. Finally, $J(S,p)$ is compact (theorem 8.3.12 in \cite{Wald84})  and
$J(S,p)= \overline{I(S,p)}$ (Proposition 2.4).  Analogous properties hold for the analogously defined regions $I(p,S)$ and $J(p,S)$.

\section*{Appendix B.}

{\bf B.1}. {\em Proof of Lemma 2.1}.
In our hypotheses,  $f\in  {\cal C}_{[\mu_{\bf g}]}(\overline{I})$ is continuous on $\overline{I}$ and  smooth in an open set
$J:= \overline{I}\setminus C = I\setminus C $ where $C\subset \overline{I}$  is  closed  with measure zero and $\partial I \subset C$.
Notice that  $\mu_{\bf g}(J) = \mu_{\bf g}(I) =\mu_{\bf g}(\overline{I})$ by construction. Therefore,
concerning the first part of the thesis it is sufficient to show that, in  $I\setminus C$,  ${\bf g}(\uparrow\spa df,\uparrow\spa df)\leq 0$
and $\uparrow \spa df$ is past direct if  $\uparrow\spa df\neq 0$.
To this end, suppose  ${\bf g}(\uparrow\spa df,\uparrow\spa df)> 0$ in $p\in I\setminus C$,   then
there must be an open neighborhood of $p$, $U$, where the same inequality holds.
So define a  smooth vector field $T'$ which is  timelike, future directed and  orthogonal to
$\uparrow \spa df$ in $U$ and $|T'|= 2|df|$, then define $T:= T'- \uparrow \spa df$. $T$ is timelike and future-directed in $U$.
If $\gamma :[0,1] \to U$ is a smooth integral curve of $T$, $\gamma$ is timelike and future directed  and   it
trivially holds $f(\gamma(1))-f(\gamma(0)) = \int_0^1 {\bf g}_{\gamma(s)}(\uparrow\spa df, \dot{\gamma}) ds < 0$. This is not
allowed if $f$ is a causal function in $I$. Similarly if $\uparrow \spa df \neq 0$ is future directed at $p\in I\setminus C$, the same
fact  must hold in a neighborhood $U$ of $p$. Take a local coordinate frame $x^1,\ldots x^n$ in $U$ where $\partial_{x^1}$ is timelike,
future directed and orthogonal to the spacelike vectors $\partial_{x^k}$, $k=2,\ldots n$. Obviously ${\bf g}(\partial_{x^1}, \uparrow \sp df) < 0$.
Let $\gamma$ be an integral curve of $\partial_{x^1}$ in $U$. $\gamma$ is causal and future directed by construction and one gets the contradiction
$f(\gamma(1))-f(\gamma(0)) = \int_0^1 {\bf g}_{\gamma(s)}(\uparrow\sp df, \dot{\gamma}) ds < 0$.
Concerning  (\ref{lemma1}) it is sufficient to prove it in $I$.  Indeed, the thesis in $\overline{I}= I\cup \partial I$ is a direct consequence of the continuity
of $f$ and ${\bf d}$  in $\overline{I}$ and the fact that $\partial I$ has measure zero.  (In particular, if $x$ or  $y$ or both belong to $\partial I$
and $x\past y$   there are two sequences $\{y_n\}\subset I$, $\{x_n\}\subset I$ with $x_n\to x$ and $y_n\to y$ as $n\to \infty$.
The continuity of ${\bf d}$ implies that $x_n \past y_m$ if $n,m$ are sufficiently large and thus the right-hand side of (\ref{lemma1}) can be computed
restricting to $I$.)
Let us pass to prove that
   \begin{eqnarray}
   \frac{f(y)-f(x)}{{\bf d}(x,y)} \geq {\mbox{ess}\inf}_{I} |d f| \:\:\:\:\:\mbox{  for each pair  $x,y \in I$
with $x \past y$} \label{lemma1<}\:.
  \end{eqnarray}
    To this end, fix $x,y \in I$ with $x\past y$.
    Since the spacetime is globally hyperbolic there is a time-like future-directed segment geodesic $\gamma_0 :[0,1] \to M$ from $x$ to $y$.
   This geodesic completely belongs to $I$ because $I$ is causally convex.
    Using normal coordinates (see lemma 2.5 in Chapter 7 of \cite{O'Neill}) about  a  geodesic segment $\gamma'_0$,   with $\gamma_0 \subset \gamma'_0$,
    it is possible to define a smooth variation of $\gamma_0$, $(t,s) \to \gamma_s(t)$ with $t\in [0,1]$, $\delta >0$,  $s\in D_1$, $\gamma_{s=0}=\gamma_0$,
  $D_\delta$ being the open disk in
    $\bR^{\mbox{dim} M -1}$ with
    radius $\delta >0$ and centred in $0$. It is possible to arrange $(t,s) \to \gamma_s(t)$ in order that  (1)
    $(t,s) \to \gamma_s(t)$ with $(t,s) \in (0,1) \times D_1$ defines an admissible  local coordinate map, (2) each curve $\gamma_s$ is time-like and future-directed
    for $t\in (0,1)$ and
    admits $t$-limits towards $0^+$ and $1^-$ defining smooth future directed causal curves from $x$ to $y$. Notice that for every $s\in D_1$, $\gamma_s\spa \rest_{(0,1)}\subset I$  by construction.
    Take $s\in D_\delta$, $0<\delta<1$ and consider, for $t\in [0,1]$, $t\mapsto h_s(t) = f(\gamma_s(t))$. This function
    is non-decreasing  and hence must admit derivative almost everywhere, such derivative is sommable and
    $f(y)-f(x) = h_s(1)-h_s(0) \geq \int_0^1 h'_s(\tau) d\tau$. The derivative is nonnegative and thus we may also write, using Fubini's theorem,
    $$f(y)-f(x) \geq   \frac{1}{\mbox{vol}(D_\delta)} \int_{D_\delta} ds \int_0^1 d\tau h'_s(\tau) =  \frac{1}{\mbox{vol}(D_\delta)}  \int_{[0,1]\times D_\delta}
    \frac{h'_s(\tau)}{\sqrt{|g(t,s)|}} \:\:d\mu_{\bf g}(t,s)\:.$$
    $\mbox{vol}(D_\delta)$  is the $R^{\mbox{dim} M-1}$ volume of $D_\delta$.  In other words
    $$f(y)-f(x) \geq  \frac{1}{\mbox{vol}(D_\delta)}  \int_{[0,1]\times D_\delta} \frac{|{\bf g}_{\gamma_s(\tau)}(\uparrow \spa d f, \dot{\gamma}_s)|}{\sqrt{|g(t,s)|}}
    \:\:d\mu_{\bf g}(t,s)\:.$$
     Barring vanishing measure sets, $\uparrow \spa d f$  is causal and past-directed or vanishes.
    Referring to an orthonormal basis of $T_{\gamma_s(t)}M$, $e_1,\ldots, e_D$, ($D=\mbox{dim} M$)
     where $e_1 = \dot{\gamma}_s(t)/|\dot{\gamma_s}(t)|$  is timelike, one straightforwardly proves that  if $T\in T_{\gamma_s(t)}M$ is causal and future directed
    or vanishes then
     $|{\bf g}_{\gamma_s(t)}(T, \dot{\gamma}_s(t))| \geq |T| |\dot{\gamma}_s(t)|$. Hence, posing $T = \uparrow \spa d_{\gamma_s(t)} f $, we have
         \begin{eqnarray}
         f(y)-f(x) \geq
         \frac{1}{\mbox{vol}(D_\delta)}\int_{[0,1]\times D_\delta} \frac{|{\bf g}_{\gamma_s(\tau)}(\uparrow \spa d f, \dot{\gamma}_s)|}{\sqrt{|g(t,s)|}} \:\:d\mu_{\bf g}(t,s)
    \geq \frac{ {\mbox{ess} \inf}_{[0,1]\times D_\delta} |d f|  }{\mbox{vol}(D_\delta)} \int_{D_\delta}ds \int_0^1 |\dot{\gamma}_s(t)|\: dt      \nonumber
    \end{eqnarray}
    and thus,
    $$  f(y)-f(x)  \geq \frac{\left({\mbox{ess}\inf}_{I}  |d f|\right) }{\mbox{vol}(D_\delta)} \int_{D_\delta}ds L(\gamma_s)\:.$$
      Changing variables $s\to \delta \sigma$
       $$  f(y)-f(x)  \geq \frac{\left({\mbox{ess}\inf}_{I}  |d f|\right) }{\mbox{vol}(D_1)} \int_{D_1}d\sigma
 L(\gamma_{\delta\sigma})\:.$$
 Notice that  $\sigma \mapsto L(\gamma_{\delta\sigma}) \leq  L(\gamma_{0}) = {\bf d}(x,y)$ is continuous in $D_1$.
 Taking the limit as $\delta \to 0^+$ (using Lebesgue's dominate convergence theorem) we have
        $$  f(y)-f(x)  \geq \frac{\left({\mbox{ess}\inf}_{I}  |d f|\right) }{\mbox{vol}(D_1)} \int_{D_1}d\sigma L(\gamma_0) =
        \frac{\left({\mbox{ess}\inf}_{I}  |d f|\right) }{\mbox{vol}(D_1)}
\: {\mbox{vol}(D_1)} {\bf d}(x,y)\:.$$
  As  $x\past y$, ${\bf d}(x,y)>0$ and thus
        \begin{eqnarray}\inf\left\{\left.\frac{f(y)-f(x)}{{\bf d}(x,y)} \:\right|\: x,y\in I,\:\: x\past y\right\}
     \geq {\mbox{ess}\inf}_{I}  |d f| \:.\label{in} \end{eqnarray}
      To conclude the proof it would be sufficient to  show that,   if $f$ is a time function,
      for every $\epsilon>0$ there are $x_\epsilon \past y_\epsilon$
      in $I$ such that
      \begin{eqnarray}
      \left|\frac{f(y_\epsilon)-f(x_\epsilon)}{{\bf d}(x_\epsilon,y_\epsilon)} -  {\mbox{ess}\inf}_{I} |d f|\right| <\epsilon         \label{inter} \:.
      \end{eqnarray}
       To this end notice that, if  ${\mbox{ess}\inf}_{I} |d f|>0$  there  must be  sequence
      $\{z_n\}_n \subset I$ such that
      each $\uparrow \spa d_{z_n} f$ is time-like (and thus  past-directed as we said above) and
       $|d_{z_n} f| \to  {\mbox{ess}\inf}_{I} |d f|$ as $n\to \infty$.
       In that case
      (\ref{inter}) is a consequence of  the statement
      "for each $z_n$, and each $\mu>0$   there are $x_{n,\mu} \past y_{n,\mu}$
      in $I$ such that
      \begin{eqnarray}
      \left|\frac{f(y_{n,\mu})-f(x_{n,\mu})}{{\bf d}(x_{n,\mu},y_{n,\mu})} -    |d_{z_n} f| \right| <\mu         \label{inter2} \:.
      \end{eqnarray}
        Let us prove the statement above.   Let  $d_{z_n} f$ be timelike and past-directed.
        Define the normalized vector  $e_1 = -\uparrow \spa d_{z_n} f/ |d_{z_n} f|$ and complete the basis of $T_{z_n}M$ with $D-1$ space-like vectors
        normalized  and orthogonal to $e_1$. Finally consider the Riemannian normal coordinate system    $\xi^1,\ldots,\xi^D$
       centred on $z_n$ generated
        by the basis $e_1,\ldots, e_D$.   We restrict such a coordinate system in a sufficiently small convex normal neighborhood of $z_n$.
        By (d) of Proposition 2.1  if $y$ has coordinates  $\xi^1,\ldots, \xi^D$, $\xi^1= {\bf d}(z_n, y)$.
          Then define $x_{n,\mu} = z_n$, and for every $y\equiv (t,0,\ldots,0)$ one has, by Lagrange's theorem
          (where $y'\equiv  (t',0,\ldots,0)$ with ${t}'\in (0,t)$)
             $$\frac{f(y)-f(x_{n,\mu})}{{\bf d}(x_{n,\mu},y)} =  \left|\frac{\partial f}{\partial \xi^1}\right|_{y'} \to
             \left|\frac{\partial f}{\partial \xi^1}\right|_{z_n} = |d_{z_n} f|\:,$$
             as $t \to 0^+$, i.e., $y\to z_n$. For every $\mu>0$, the existence of $x_{n,\mu}  \past y_{n,\mu}$ in $I$
             such that (\ref{inter2}) is fulfilled  follows trivially.
The same proof can be used for the case
 ${\mbox{ess}\inf}_{I} |d f|=0$ 
 provided that a  sequence  $\{z_n\}_n \subset I$ exists such that
      each $\uparrow \spa d_{z_n} f$ is  timelike  and
       $|d_{z_n} f| \to  0$  as $n\to \infty$. Let us prove that such a sequence do exist if $f$ is a time function. Suppose it is not the case and  ${\mbox{ess}\inf}_{I} |d f|=0$.
So it must happen that $|df|\equiv 0$ in some $E \subset I\setminus C$ with $\mu_{\bf g}(E) \neq 0$. In turn
it  implies that there is some $q\in I\setminus C$ where $df_q$ is a null vector or vanishes.
So,  take a sequence of  open neighborhoods of $q$, $U_i \subset I\setminus C$, where $df$ is smoothly defined,
such that $U_{i+1} \subset U_i$ and  $\cap_i U_i = \{q\}$.
If  $df_{q_i}$ is timelike for some  $q_i\in U_i\setminus \{q\}$ for every $i$,
the wanted sequence exists  and this is assumed to be impossible by hypotheses.
 So it must be $|df| = 0$ in some  $U_{i_0}$. But this is not
possible too because, if $df_r\neq 0$ for some $r\in U_{i_0}$,  the time-function $f$ would be constant along
 a future-directed causal curve given by an integral curve
of $\uparrow \sp df$ in a neighborhood of  $r$.  Conversely, if $df \equiv 0$ in $U_{i_0}$, $f$ would be constant along any timelike
future directed curve in $U_{i_0}$.   $\Box$\\

\noindent {\bf B.2}. {\em Proof of Proposition 2.4}.
It is convenient to prove {\bf (d)} before  because most of {\bf (a)}  is a straightforward consequence of the former.
{\bf (d)} If $S\subset M$ is a smooth Cauchy surface for $M$ and $p\in I^+(S)$, the set $I(S,p)$  is
open and  causally convex  (A.23). Since $\partial I^+(S) = S$, one has  $\partial I(S,p) \subset S \cup \partial I^-(p)$
and hence  $\partial I^+(S)$ has measure zero since $S$ is a smooth hypersurface and   (a) of Theorem 2.1 holds.
To conclude that $I(S,p)\in {\cal X}$, it is sufficient to show that $\overline{I(S,p)}= J(S,p)$ noticing that the latter is compact and causally
convex (A.23).    Obviously,
$\overline{I(S,p)} \subset J(S,p) = \overline{I^+(S)}\cap \overline{I^+(S)}$, so we have to show that
$\overline{I^+(S)}\cap \overline{I^-(p)} \subset \overline{I(S,p)}$.  Using the decomposition $\overline{I^+(S)}\cap \overline{I^-(p)}=
(\partial {I^+(S)}\cap \partial{I^-(p)})\cup  (\partial{I^+(S)}\cap {I^-(p)})\cup ({I^+(S)}\cap \partial{I^-(p)})\cup({I^+(S)}\cap {I^-(p)})$,
one finds that the only thing to be shown is that $x\in \partial {I^+(S)}\cap \partial{I^-(p)}$ implies $x\in \overline{I^+(S) \cap I^-(p)}$.
Take such an $x$. Notice that $x\in \partial I^+(S) = S$.  Let $B_x$ be an open neighborhood of $x$ and $\gamma$ a maximal
causal geodesic segment
from $x=\gamma(0)$ to $p=\gamma(1)$ which exists by (i) of Proposition 2.1.  Extend   $\gamma$ into an inextendible
causal  geodesic $\gamma'$. Notice that $\gamma$ (and $\gamma'$) must be null, because ${\bf d}(x,p)=0$ as a consequence
of (a), (f)  of Proposition 2.1 and A.12.
Since $p\in I^+(S)$ and $\gamma'$ intersects $S=\partial I^+(S)$ in $x$ only ($S$ being a Cauchy surface), $\gamma(t)\in I^+(S)$ for $t>0$.
It is possible to fix  $t_0>0$, $t<1$, such that $x':= \gamma(t_0) \in B_x$.
 Subsegments of a  maximal geodesic segment are maximal and hence $x'\in \partial I^-(p)$. Therefore there is a  sequence of points $\{x_n\}\subset I^-(p)$ with
$x_n\to x'$ as $n\to \infty$.  As $x'\in I^-(p), I^+(S), B_x$ and these sets are open,  for some $N\in \bN$ it must hold $x_n \in I^-(p),  I^+(S), B_x$
if $n>N$. We have proven that for every open neighborhood $B_x$ of $x$ there is some $x_n \in  I^-(p)\cap I^+(S) \cap B_x$. In other words,
$x\in  \overline{I^-(p)\cap I^+(S)}$.
Let us pass to consider the open diamond regions $I(r,s)$. We want to show that, if $p\in M$, there is a fundamental set of open neighborhoods of $p$,
 $\{ I(r_n,s_n) \}\subset  {\cal X}$ and in particular $\overline{I(r_n,s_n)}= J(r_n,s_n)$.  From proposition 4.12  in \cite{Penrose72} one finds
that\footnote{The reader should pay attention to the fact that the definition of causally convex sets
 given in \cite{Penrose72} is different from that used in this paper.},  in a globally hyperbolic
spacetime $M$ (it is sufficient the strongly causal condition),
 each point $p$  admits  a convex normal neighborhood $U_p$  and an open neighborhood, $A_p$,
such that (1) $A_p\subset U_p$ and (2) if $r,s\in A_p$, $I(r,s)\subset A_p$.   In $A_p$, take a future directed geodesic segment
through $p$, $\gamma$, and a two sequences of points on $\gamma$, $\{r_n\}$, $\{s_n\}$ such that  $r_n \past r_{n+1} \past p \past s_{n+1}\past s_n$,
 and $r_n,s_n \to p$ as $n\to \infty$.  As the spacetime is strongly causal (A.11), using the remark in A.7
one proves that $\{I(r_n,s_n)\}$ is a fundamental set of neighborhoods
of $p$, and $J(r_{n+1},s_{n+1}) \subset I(r_n,s_n) \subset U_p$. It is clear that each $I(r_n,s_n)$ is open, causally convex and $\partial I(r_n,s_n)
\subset \partial I^+(r_n) \cup \partial I^-(s_n)$ has measure zero. As $J(r_n,s_n)$ is causally convex (A.11) and compact (A.12), to conclude it is sufficient to show that
$J(r_n,s_n) = \overline{I(r_n, s_n)}$. Suppose this is not the case and thus there is $x\in J(r_n,s_n)$ with $x\not \in \overline{I(r_n, s_n)}$.
As is known, causal curves from $r_n$ to $s_n$ which are not  smooth geodesic segments from $r_n$ to $s_n$ can be approximated by
timelike curves from $r_n$ to $s_n$ \cite{O'Neill}. This means that there must be a smooth null geodesic segment $\eta\subset J(r_n,s_n)$ from
$r_n$ to $s_n$ with  $x\in \eta$. Let us show that this is impossible. Indeed, since $r_n\past s_n$,  by (i) of Proposition 2.1  there is a timelike (and thus $\neq \eta$) geodesic
segment $\eta'$ from
$r_n$ to $s_n$ and both $\eta,\eta'$ must belong to the same geodesically convex neighborhood $U_p$.\\
{\bf (a)}  (ii) in  (d) implies $\cup {\cal X} = M$. Let us pass to show that ${\cal X}$  is a direct set. We want
to show that if $A,B\in {\cal X}$, there is $C\in {\cal X}$ with $A,B\subset D$. From now on $D:= A\cup B$.
As the spacetime is globally hyperbolic, it is homeomorphic
to $\bR \times S$, where $S\subset M$ is a smooth Cauchy surface (A.19-A.21). Then consider the natural smooth time function $t:M \to \bR$ (which exists by
A.13, A.15, A.21) associated to the former Cartesian factor. Take $t_1< \min t\spa \rest_{\overline{D}}$  and $t_2> \max t\spa \rest_{\overline{D}}$.
The Cauchy surface $S_1=\{x\in M \:|\: t(x)=t_1\}$ is in the past (with respect to $t$)
of $\overline{D}$ and the Cauchy surface $S_2:=\{x\in M \:|\: t(x)=t_2\}$ is in the future  of $\overline{D}$ and $S_1$.
By definition of Cauchy surface, if $p\in \overline{D}$, every inextendible future-directed timelike curve $\gamma$ through $p$ must intersect both $S_1$ and $S_2$.
Let $q$ the intersection of $\gamma$ and $S_2$ and $F$ the set of such points $q$. By construction, it holds $\overline{D} \subset \cup_{q\in F} I(S_1,q)$. Since $\overline{D}$ is
compact we can extract a finite covering from that found above. In particular we have  $\overline{D} \subset \cup_{i=1}^n I(S_1,q_i) $.
To conclude define $C:= \cup_{i= 1}^n I(S_1,q_j)$. $C$ is open because union of open sets, causally convex (if $p,q\in C$ satisfy $q\preceq p$, $p\in I(S_1,q_k)$
for some $k$ and thus $q$ and every causal curve from $p$ to $q$ belong to $I(S_1,q_k)\subset C$),  $\partial C \subset  \cup_{i= 1}^n \partial I(S_1,q_j)$
has measure zero because every $I(S_1,q_j)\in {\cal X}$ and  thus $\partial I(S_1,q_j)$ has measure zero, $\overline{C}=\cup_{i=1}^n J(S_1,q_i)$   is compact
(because union of compacts) and causally convex (the proof is similar to that for $C$). We conclude that  $C\in {\cal X}$ and $A,B\subset C$.
{\bf (b)} If $f$ is a global smooth time function, which exists by A.13 and A.15 in globally hyperbolic spacetimes, and   $A\in {\cal X}$,
then, trivially,  $f\spa\rest_{{A}} \in {\cal T}_{[\mu_{\bf g}]}(A)$ and $f\spa\rest_{\overline{A}} \in {\cal C}_{[\mu_{\bf g}]}(\overline{A})$.
{\bf (c)} If $p,q\in \overline{I}$ and  $p\preceq q$, it holds $f(p)\leq f(q)$ for all $f\in {\cal C}_{[\mu_{\bf g}]}(I)$, ${I}\in {\cal X}$ because
  $\overline{I}$ is causally convex.
Let us prove that if  $f(p)\leq f(q)$ for all essentially smooth
causal functions $f$ defined in any $\overline{I}\in {\cal X}$ with $p,q\in \overline{I}$, then  $p\preceq q$.   Suppose that the implication is false.
If $p$ and $q$ are spatially separated, as in the proof of Theorem 2.2
one finds two spatially separated, sufficiently small, regions $I(p',q')$ and $I(p'',q'')$ which respectively contain $p$ and $q$ and have spatially separated
closures. By (c) of Proposition 2.3, $f_c: z \mapsto {\bf d}(p',z) + c {\bf d}(p'',z)$,
$c>0$ is an element of ${\cal C}_{[\mu_{\bf g}]}(\overline{I})$ 
with $I= I(p',q')\cup I(p'',q'')\in {\cal X}$. Moreover $f_c(q) = c{\bf d}(p'',q) < {\bf d}(p',p) = f_c(p)$ for $c$
sufficiently small and this is a contradiction.
If $q \prec p$, te map $f: z\mapsto d(x,z)$, defined on $J(x,y)$ with $x\past q$ and $p\past y$, produces a contradiction once again.
$\Box$\\

\noindent {\bf B.3}. {\em Proof of Theorem 4.1}.  In the following we take advantage of  Riesz' representation theorem \cite{Rudin70}  which
proves that there is a bijective map $L\mapsto \mu_L$ between the set of positive linear functionals $L$ on $C_c(\Omega)$, $\Omega$
being a locally compact  topological space, and the set of regular Borel measures on $\Omega$,   such that   $L(f) = \int_\Omega f \:d\mu_L$ for
every $f\in C_c(\Omega)$.  \\
 {\bf (a)} By (a) of Proposition 2.4, for every $\mu\in \gP$,  there is $I\in {\cal X}$ with $\mbox{supp}(\mu) \subset \overline{I}$.
 It is trivially proven that $\lambda^{(\mu)}_{I}$ is a state on $\overline{{\cal A}_I}$. Moreover, varying $I$, one obtains equivalent  states
 since, by trivial properties of measures,  $\mbox{supp}(\mu)\subset  \overline{I},\overline{J}$ entails
$J_{K,I}(\lambda^{(\mu)}_{I}) = J_{{K},J}(\lambda^{(\mu)}_{J})$
for $I,J\leq K$. We only have to show that if $\Lambda_\mu$ is the locus generated by some $\lambda^{(\mu)}_{I}$,
every element $\lambda_{J} \in \Lambda_\mu$ must belong to $F(\mu)$. To this end assume that $\lambda_{J} \in \Lambda_\mu$,
that is  $\lambda_{J}\sim \lambda^{(\mu)}_{I}$.
That equivalence relationship can be re-written as follows: for every $K\in {\cal X}$ with $\overline{I}\cup \overline{J} \subset \overline{K}$ and
  for every  $f\in C(\overline{K})$, it holds
$\int_{\overline{K}} f d\mu = \lambda_{J}(f\spa\rest_{\overline{J}})$.
Using the fact that $\mbox{supp}(\mu) \subset \overline{K}$ and (a) of Proposition 2.4,
the obtained identity implies that  if $h\in C_c(M)$ and $\mbox{supp}(h) \subset M\setminus \overline{J}$,
$\int_M h \:d\mu=0$.  Uryshon's lemma \cite{Rudin70} implies that $\mu(R)=0$ for every compact $R\subset M\setminus \overline{J}$, then
 the regularity of $\mu$ implies that $\mbox{supp}(\mu)\subset \overline{J}$ and thus
$\lambda_J \in F(\mu)$.  We have proven that $F(\mu)$ is a locus, but also that  $F(\mu)\ine I$ implies $\mbox{supp}(\mu)\subset \overline{I}$.
To conclude notice that if $\mbox{supp}(\mu)\subset \overline{I}$
then $F(\mu)\ine I$ because $\lambda_I(\cdot) := \int_M \cdot \: d\mu \in F(\mu)$.
{\bf (b)} Injectivity: if $\mu\neq \mu'$, by Riesz' theorem  there is $f\in C_c(M)$ with $\int_M f\: d\mu \neq \int_M f\: d\mu'$. Taking
$K\in {\cal X}$ with $\mbox{supp}(\mu),\mbox{supp}(\mu)'\subset K$  one gets $\lambda_K^{(\mu)}(f\spa\rest_{\overline{K}})\neq
\lambda_K^{(\mu')}(f\spa\rest_{\overline{K}})$ and thus $F(\mu)\neq F(\mu')$. Surjectivity: if $\Lambda$ is a locus, take $I\in {\cal X}$
with $\Lambda \ine I$. Define $L_\Lambda(f) :=  \Lambda(f\spa\rest_{\overline{I}})$ for every $f\in C_c(M)$. $L_\Lambda$ turns out to be
a positive linear functional, therefore, Riesz' theorem proves that $L_\Lambda(f) = \int_M f \: d\mu_\Lambda$ for some
regular Borel probability measure $\mu_\Lambda$ with $\mbox{supp}(\mu_\Lambda) \subset \overline{I}$. Then consider the locus
$F(\mu_\Lambda)$ and some $\lambda^{(\mu_\Lambda)}_J \in F(\mu_\Lambda)$.
Tietze's extension theorem \cite{Rudin70} entails $\lambda^{(\mu_\Lambda)}_J \in  \Lambda$ and thus $F(\mu_\Lambda)= \Lambda$ by (a).
 {\bf (c)} First we prove that the elements of $F(\delta_x)$ are pure states.
$f\in \overline{{\cal A}_{I}}= C(\overline{I})$
and $x\in \overline{I}$  imply that $\lambda^{(\delta_x)}_{I}:  f \mapsto f(x)$ defines a pure state because $\langle {\cal H}, \pi,\Omega \rangle$ is
a GNS triple for $\lambda^{(\delta_x)}_{I}$ if ${\cal H}=\bC$, $\Omega=1\in \bC$ and $\pi : f\mapsto f(x)$, and $\pi$ is trivially irreducible. So
$F{(\delta_x)}$ is a pointwise locus for every $x\in M$.
$F$  restricted to the space of Dirac measures $\{\delta_x\}_{x\in M}$ is surjective
 onto $\gL_p$. Indeed, take $\Lambda \in \gL_p$ and let $\lambda_{I} \in \Lambda \cap {\cal S}_{pI}$.
As irreducible representations of a commutative $C^*$-algebras are unidimensional,
a pure state $\omega$ on a commutative $C^*$-algebra admits a GNS representation on $\bC$.
As the cyclic vector is $1\in \bC$, one sees that $\omega$ is also multiplicative:
$\omega(ab)=\omega(a)\omega(b)$. In other words
$\omega$ it-self is  an irreducible $\bC$-representation of the $C^*$-algebra. Therefore $\lambda_I$ is
an irreducible representation of the commutative $C^*$-algebra $C(\overline{I})$.
A known theorem in commutative $C^*$-algebras theory
(e.g., see proposition 2.2.2 in \cite{Landi}) implies that  there is $x_\Lambda\in \overline{I}$ such that $\lambda_{I}(f) = f(x_\Lambda)$
for all $f\in C(\overline{I})$. (Precisely, the theorem states that  $h_I : x \mapsto
\lambda^{(\delta_x)}_{I}$ is a homeomorphism from $\overline{I}$ onto ${\cal S}_{pI}$ equipped with the Gel'fand topology.)
Then consider $F(\delta_{x_\Lambda})$. It is clear that  $\lambda_{I}\in F(\delta_{x_\Lambda})$ and thus $\Lambda = F(\delta_{x_\Lambda})$ by (a). 
 Up to now we have proven that the map $H : M \to \gL_p$ with $H(x) = F(\delta_x)$ is a bijection from $M$ onto $\gL_p$. It remains to show that
$H$ is a homeomorphism when $\gL_p$ is equipped by the inductive-limit topology obtained by equipping each ${\cal S}_I$ by the weak $*$-topology
(Gel'fand topology).
To prove that $H$ is a homeomorphism, notice that $M$ can be naturally identified to  $M'$,  the inductive limit of the
class of compact sets $\{\overline{I}\}_{I\in{\cal X}}$
equipped with a class of maps $F_{I,J} : \overline{J} \to \overline{I}$, when $J\subset I$,
$F_{I,J}$ being  the inclusion map. As $M\equiv M'$, the  injective inclusion maps $F_I \mapsto M'$
($F_I : x \mapsto [x]$ where $x\in \overline{I}$, $I\in {\cal X}$ and
$[x]\in M' \equiv M$ being the equivalence class of $x$ in the inductive limit) coincide with the usual
inclusion maps of each $\overline{I}$ in $M$ it-self.
By definition the  inductive-limit topology is the finest topology on the inductive limit set
which makes continuous all the inclusion maps $F_I$.  In other words a set $A\subset M'\equiv M$ is open iff
$A\cap \overline{I}$ is open in the topology of $\overline{I}$, for all $I\in {\cal X}$. As the sets $I$ are open and
$\cup {\cal X} = M$, the inductive-limit topology on $M'\equiv M$ coincides to the original topology of $M$.
To conclude,  consider the following ingredients: the  space $\gL_p$ realized as the inductive limit of the family $\{{\cal S}_{pI}\}_{I\in {\cal X}}$
(with maps $J_{I,J}$) equipped with the inductive-limit topology induced by the
Gel'fand topology in the spaces ${\cal S}_{pI}$, the
injective inclusion  maps $G_{I} : {\cal S}_{pI}
 \to \gL_p$ and the homeomorphisms $h_{I}: \overline{I}  \to {\cal S}_{pI}$ said above. Using (b) above, it is a trivial task to show
that, for every $I\in {\cal X}$, $G_I\circ h_I = H\circ F_I$.    As every  $G_I\circ h_I$  is continuous, $H$ turns out to be continuous.
Conversely, since  it also holds $H^{-1} \circ G_I  =  F_I \circ h_I^{-1}$  and  every $F_I \circ h_I^{-1}$ is continuous,
$H^{-1}$ turns out to be continuous  too.
We have obtained that $H$ is a homeomorphism.
  $\Box$  \\

 \noindent {\bf B.4}. {\em Proof of Theorem 4.2}.  {\bf (a)} In the given hypotheses, take $I\in {\cal X}$ with $\Lambda,\Lambda'\ine I$.
 If  $\lambda_I\in \Lambda\cap {\cal S}_I$ and $\lambda'_I\in \Lambda'\cap {\cal S}_I$, one has
 $\lambda_I(t)=\lambda'_I(t)$, for every $t\in Co_I$. By (b) of Proposition 4.3, the linearity and the continuity of the states
 $\lambda_I,\lambda'_I$, one gets $\lambda_I = \lambda'_I$ and thus $\Lambda=\Lambda'$. \\
 The proof of {\bf (b)} and {\bf (c)} is based on the following lemma.\\

\noindent  {\bf Lemma B.1}.
 {\em In the hypotheses of Theorem 4.2, $\Lambda \unlhd \Lambda'$ implies both $\mbox{supp}(\mu')\subset J^+(\mbox{supp}(\mu))$
 and $\mbox{supp}(\mu)\subset J^-(\mbox{supp}(\mu'))$, where $F(\mu)=\Lambda$ and $F(\mu')=\Lambda'$ and $F$ is defined in Theorem 4.1.}\\

 \noindent {\em Proof of Lemma B.1}.   We prove      $\mbox{supp}(\mu)\subset J^-(\mbox{supp}(\mu'))$, the other inclusion  is analogous  taking
 $p\past q$ and $-{\bf d}(\cdot,q)$ in place of ${\bf d}(q,\cdot)$ in the following proof. Suppose that  $\Lambda \unlhd \Lambda'$ and there is
 $p\in \mbox{supp}(\mu)$ with
 $p\not \in J^-(\mbox{supp}(\mu'))$. Since $J^-(\mbox{supp}(\mu'))$ is closed as $\mbox{supp}(\mu')$ is compact (see A.12), there is
 an open neighborhood of $p$ which have no intersection with $J^-(\mbox{supp}(\mu'))$. Take $q\past p$ in such a neighborhood.
 $J^+(q) \cap J^-(\mbox{supp}(\mu')) =\emptyset$  by construction. Let $I\in {\cal X}$ be  such that $\mbox{supp}(\mu), \mbox{supp}(\mu') \subset I$,
 such a set exists because of (a) of Proposition 2.4 and let $t\in Co_I$.
 By (c) of Proposition 2.3, $t_{\alpha}= t+  \alpha {\bf d}(q,\cdot))\spa \rest_{\overline{I}} \in Co_I$ for all
 $\alpha >0$. Therefore $\Lambda \unlhd\Lambda'$ entails, making use of  Theorem 4.1,
  $\int_{\overline{I}} t_{\alpha} \: d\mu \leq \int_{\overline{I}} t_{\alpha} \:d\mu'$. Since $\alpha {\bf d}(q,\cdot)$ vanishes on
 $\mbox{supp}(\mu') \subset J^-(\mbox{supp}(\mu'))$, the same inequality can be re-written  $\int_{\overline{I}} t \: d\mu + \alpha \int_{\overline{I}} {\bf d}(q,\cdot)
  \: d\mu  \leq \int_{\overline{I}}t \:d\mu'$ for every $\alpha >0$. On the other hand it holds 
 $\int_{\overline{I}} {\bf d}(q,\cdot) \: d\mu> 0$. Indeed
  (1) if $U_p \subset \overline{U_p}\subset I^+(q)\cap I$ is an open neighborhood
  of $p$, it must be $\mu(U_p)\neq 0$
 because $p\in \mbox{supp}(\mu)$, moreover (2) ${\bf d}(q,\cdot)\spa\rest_{\overline{U_p}}\geq \gamma >0$ because of (a) of Proposition 2.1
  and the continuity of ${\bf d}$. By consequence
   there must be some sufficiently large $\alpha >0$ which
  produces a contradiction in  $\int_{\overline{I}} t \: d\mu + \alpha \int_{\overline{I}} {\bf d}(q,\cdot)
  \: d\mu  \leq \int_{\overline{I}}t \:d\mu'$.   $\Box$

  Let us come back to the main proof. Concerning {\bf (c)}, the proof of the statement   "$x\preceq y$ implies $F(\delta_x) \unlhd F(\delta_y)$"
   is obvious by the given definitions.
  Using the proven lemma, the proof of the statement  "$F(\delta_x) \unlhd F(\delta_y)$ implies $x\preceq y$" is straightforward.
  Concerning {\bf (b)}, notice that by the  lemma and (a) of Theorem 4.1,  $\Lambda \unlhd\Lambda' \unlhd\Lambda''$
  implies, with obvious notations, $\mbox{supp}(\mu') \subset  J^+(\mbox{supp}(\mu))\cap J^-(\mbox{supp}(\mu''))$. Every open set $I\in {\cal X}$ such that
  $\overline{I}$ contains both   $\mbox{supp}(\mu)$  and   $\mbox{supp}(\mu'')$  must contain $J^+(\mbox{supp}(\mu))\cap J^-(\mbox{supp}(\mu''))$ because
  $\overline{I}$ is causally convex.
  Hence $\mbox{supp}(\mu')\subset J^+(\mbox{supp}(\mu))\cap J^-(\mbox{supp}(\mu'')) \subset \overline{I}$. In other words, using (a) of Theorem 4.1,
  if $\Lambda \unlhd \Lambda' \unlhd \Lambda''$ and $\Lambda,\Lambda''\ine I\in {\cal X}$, then
  $\Lambda'\ine I$. $\Box$

\section*{Appendix C.}

\noindent {\em Proof of Theorem 2.1}. 
The proof of Theorem 2.1   is based on two lemmata.\\

\noindent{\bf Lemma C.1}.
{\em If $(M, {\bf g}, {\cal O}_t)$ is globally hyperbolic and $p\in M$, referring to the definitions above,
$C^+(p)$ and  $\partial J^+(p)=\partial I^+(p) = J^+(p)\setminus I^+(p)$ are  closed without internal points and have measure zero,
finally $J^+(p)\setminus \left(C^+(p) \cup \partial J^+(p)\right) = I^+(p) \setminus C^+(p)$  is homeomorphic to $\bR^{\mbox{dim}(M)}$.}\\

\noindent {\em Proof.}
 From now on $n:= \mbox{dim}(M)$ and $V^+_p\subset T_pM$ is the cone made of  future-directed causal vectors and $0$.
  First consider  $\partial J^+(p)=\partial I^+(p) = J^+(p)\setminus I^+(p)$, these identities being given in A.12.   It is obvious that $\partial J^+(p)$
is closed, let us prove that it has measure zero.
$J^+(p) \setminus I^+(p) \subset \exp_p(U_p\cap \partial V_p^+)$
where  $U_p$ is the open domain of the exponential map at $p$ (see the Appendix A).
Indeed if $q\in J^+(p) \setminus I^+(p)$, either $q=p$ or, by (i) of Proposition 2.1, there is a geodesic from $p$ to $q$ which  must be null-like it being maximal and $q\not \in I^+(p)$.
Therefore (re-scaling the vector if necessary) there must be a vector $v\in \partial V_p^+\cap U_p$ with $\exp_pv=q$.
The Lebesgue measure of $\partial V^+_p\subset \bR^n$ vanishes and thus, since $\exp_p$ is
smooth and thus locally Lipschitz,  $\partial I^+(p)$ must have measure zero. Indeed  one has that the part of
 $\partial I^+(p)$ contained in the domain $V$ of any local coordinate chart $(V,\psi)$ has measure zero, with respect to the Lebesgue coordinate measure and thus
 $\mu_{\bf g}$, because $\psi \circ \exp_p$ is locally Lipschitz on $(\exp_p^{-1}(V))$ for all $k\in \bN$. Then  the countable measurability of $\mu_{\bf g}$ and
 the existence of a countable atlas of the manifold entails the thesis for the whole set $\partial I^+(p)$.  
The closure of $C^+(p)$ was proven in theorem 9.35 of \cite{BEE}, the absence of internal points
is a trivial consequence of the measure zero (since nonempty open sets have positive measure $\mu_{\bf g}$).
The last statement in the thesis is a consequence of proposition 9.36 in \cite{BEE}, due to Galloway. 
It remains to show
that $C^+(p)$ has measure zero.
Similarly to the proof for $\partial J^+(p)$, it is sufficient to  prove that the Lebesgue measure in $\bR^n$ of $\Gamma^+_{ns}(p)$ is zero: since
$\Gamma^+_{ns}(p)\subset U_p$ and  $C^+(p) = \exp_p (\Gamma^+_{ns}(p))$, the latter has measure zero if $\Gamma^+_{ns}(p)$ has measure zero.
To this end  notice that $UM_p$ can be thought to be embedded in $\bR^n$ and diffeomorphic to
the intersection of the sphere $S^{n-1}$, $\sum_{i=1}^{n} (X^i)^2=1$ and the cone $V^+$, $X^1\geq \sqrt{\sum_{i=2}^{n} (X^i)^2}$.
$UM_p$ is compact by construction. Fix $N\in \bN$ and consider  set
 $K_N := \{ v\in UM_p\:|\: s_1(v) \leq N\}.$ As $s_1$ is lower semicontinuous,
$K_N$ is closed and thus compact (since $K_N\subset UM_p$ which is compact and the topology being Hausdorff), moreover $\cup_N K_N = UM_p$. As a second step
we define $S_N = \{ v\in K_N\:|\: s_1(v) v \in U_p\}$.
It is clear that, by the countability of the Lebesgue measure,
  the thesis is proven if one shows that, for every $N\in \bN$, the image of the  map $v\mapsto s_1(v) v$ with  $v\in S_N$
has measure zero.
The map $v \mapsto s_1(v)$, $v\in  S_N$ is continuous (see the beginning of this appendix).  Using the continuity of
$v \mapsto s_1(v) v$ and the  fact that $U_p$ is open, it arises that $S_N$ is open with respect to the topology of $A_N$.
We conclude that $S_N= K_N \cap B_N$,  where $K_N$ is compact  and $B_N$ is an open set in $S^{n-1}$. If $B_N$ is not connected we shall refer to each
connected component of $B_N$ in the following.  The open set $B_N$ admits a finite or  countable class of components
because the topology of $S^{n-1}$ is second countable.
Consider a countable
class of compact sets $H_n\subset B_N$ such that $U_n H_n = B_N$ (they do exist because $B_N$ is a connected manifold or it holds for each connected
component). $H_n\cap K_N$ is compact
(since the topology of $S^{n-1}$ is Hausdorff), $K_N\cap H_n \subset S_N$ and $\cup_n (K_N\cap H_n) = S_N$.
The function $v\mapsto s_1(v)v$ is continuous on each compact $K_N\cap H_n$ and thus its image has measure zero in $\bR^n$. By countability
the image of $v\mapsto s_1(v)v$, $v\in S_N$ has measure zero as required.
$\Box$\\

\noindent The proof of the theorem ends by proving the lemma below.   In the proof  we make use of  the
sets $$A_p := \{ \lambda v\:|\: v\in UM_p,\: \lambda \in
  [0, s_1(v)) \} \:\:\: \mbox{and}\:\:\:
{\cal A}_p := A_p\setminus \{v\in T_pM \:|\: {\bf g}_p(v,v)=0\}\:.$$
$A_p\subset U_p$ as one can straightforwardly prove using the definition of $s_1$ and the domain of the exponential map at $p$, $U_p$,
moreover  ${\cal A}_p$ is open as a consequence  of the  lower semicontinuity  of $s_1$. \\

\noindent{\bf Lemma C.2}.
{\em If $(M, {\bf g}, {\cal O}_t)$ is globally hyperbolic and $p\in M$, referring to the definitions above the
following statements hold true.   \\
{\bf (a)} $\exp_p\spa\rest_{A_p}$ is surjective onto $J^+(p)\setminus C^+(p)$,
$\exp_p\spa\rest_{{\cal A}_p}$ is surjective onto $I^+(p)\setminus C^+(p)$; \\
{\bf (b)}  $\exp_p\spa\rest_{A_p}$ and $\exp_p\spa\rest_{{\cal A}_p}$
are smooth and injective;\\
{\bf (c)} $(\exp_p\spa\rest_{A_p})^{-1}\in C^\infty(J^+(p)\setminus C^+(p))$
and $(\exp_p\spa\rest_{{\cal A}_p})^{-1}\in C^\infty(I^+(p)\setminus C^+(p))$; \\
{\bf (d)} The map $q \mapsto {\bf d}(p,q)^2$ belongs to $C^\infty(J^+(p)\setminus C^+(p))$;\\
{\bf (e)} The map $q \mapsto {\bf d}(p,q)$ belongs to $C^\infty(I^+(p)\setminus C^+(p))$\\
{\bf (f)}  ${\bf g}_q(\uparrow \spa d_q {\bf d}(p,q),\uparrow \spa d_q {\bf d}(p,q)) = -1$ for $q\in I^+(p)\setminus C^+(p)$.}  \\

\noindent {\em Proof.} {\bf (a)}  Take  $q\in J^+(p)\setminus C^+(p)$. If $q=p$, $q=\exp_p(0)$ and $0\in A_p$. If $q\neq p$,
 by (i) of Proposition   2.1 (since the spacetime is globally hyperbolic)
 there is a future directed causal geodesic $\gamma : [0,b) \to M$ with  $\gamma(0)=p$ and $\gamma(a)=q$, $a<b$ and $\gamma$ is maximal from
 $p$ to $q$.
 Rescaling the affine parameter of $\gamma$, we can assume that $v:= \dot{\gamma}(0)\in UM_p$. It must hold $a\leq  s_1(v)$ by maximality
and $a \neq s_1(v)$ because it would imply $q\in C^+(p)$ by definition.
Therefore
 $q = \exp_p(\lambda v)$ with $\lambda \in [0, s_1(v))$, namely, $q\in \exp_p(A_p)$. If $q \in J^+(p)\setminus C^+(p)$ but 
$q\not \in \partial J^+(p)$ then (the spacetime being globally hyperbolic) $q\in I^+(p)\setminus C^+(p)$ and so $v$
above must belong to ${\cal A}_p$.
{\bf (b)} As is  known (see the Appendix A), the exponential map is smooth where it is defined. Let us consider the injectivity.
Suppose there are $u,v \in A_p$ with $exp_p(u)=exp_p(v)$. This is equivalent to say that
$exp_p(\lambda v_0) = exp_p(\mu u_0) = q$ for some $v_0,u_0 \in UM_p$ and $0<\lambda <s_1(v_0)$, $0<\mu <s_1(u_0)$.
In other words $q$ is contained in a maximal future-directed causal geodesic from $p$ to some $q'$
(after $q$), and thus the subsegment from $p$ to $q$ is a maximal geodesic, too.
Moreover there is another maximal future-directed causal geodesic from $p$ to $q$ it-self. Lemmata 9.1 and 9.12 in \cite{BEE}
imply that $q$ cannot be the image of a point in $A_p$ and this is impossible.
{\bf (c)}
 It is a trivial consequence of (a), (b) and the fact that $\exp$ is a local diffeomorphism about 
every point of $A_p$. This is  because there are no conjugate points with $p$ along each future-directed causal geodesic 
starting from $p$ before the corresponding cut point as stated in theorems 9.12 and 9.15 of \cite{BEE}. 
{\bf (d)} If $q\in J^+(p)\setminus C^{+}(p)$, there is a  causal
 future-directed geodesic, $\gamma$, from $p$
to $q$ whose length coincides with ${\bf d}(p,q)$ and whose initial tangent vector is nothing but
$\left(\exp\spa\rest_{A_p}\right)^{-1}(q) \in A_p$. Therefore
${\bf d}(p,q)^2 = L(\gamma)^2 = -{\bf g}_p\left(\left(\exp\spa\rest_{A_p}\right)^{-1}(q),\left(\exp\spa\rest_{A_p}\right)^{-1}(q)
\right) $ from trivial  properties of geodesics. From now on $-2\sigma(p,q)$ indicates the right-hand side of the obtained identity.
{\bf (e)} ${\bf d}(p,\cdot) = \sqrt{{\bf d}(p,\cdot)^2}$ and $x\mapsto \sqrt{x}$ is smooth for $x>0$. ${\bf d}(p,\cdot)^2$
cannot vanish in the open set $I^+(p)\setminus C^+(p).$
{\bf (f)} $d_q {\bf d}(p,q) = d_q \sqrt{-2 \sigma(p,q)}$. Thus $d_q {\bf d}(p,q) = -(-2 \sigma(p,q))^{-1/2} d_q \sigma(p,q)$
and by consequence one gets  ${\bf g}_q(\uparrow \spa d_q {\bf d}(p,q),
\uparrow \spa d_q {\bf d}(p,q))
 = (-2\sigma)^{-1}{\bf g}_{q}(\uparrow \spa d_q \sigma(p,q),\uparrow \spa d_q \sigma(p,q))$. (\ref{dsds}) holds in geodesically convex
 neighborhoods. However it can also be proven in our hypotheses following the proof of theorem 1.2.3, items (iii) and (iv), in
\cite{Friedlander} which only employs the variational definition of (timelike) geodesics. Using (\ref{dsds}) one has
${\bf g}_q(\uparrow \spa d_q {\bf d}(p,q),\uparrow \spa d_q {\bf d}(p,q))= -1$.
 $\Box$ $\Box$

 \end{document}